\documentclass[12pt]{article}

\usepackage{graphicx,natbib} %,epsf
\usepackage{enumerate}

\usepackage{url} % not crucial - just used below for the URL 

\newcommand{\blind}{0}

\usepackage{tikz}
\usetikzlibrary{shapes.geometric, arrows}
\usepackage[doublespacing]{setspace} % Double spacing
\usepackage{lineno} % For numbered lines
\usepackage{authblk} % Manages authors' affiliation
\usepackage{amsmath} % Math symbols
\usepackage{bm} % Bold math symbol
\usepackage{amssymb} % For real number like symbols
\usepackage{natbib} % For citation without parenthesis
\usepackage[inline]{enumitem} % For lists in the text
\usepackage{float}
\usepackage[labelsep=period,font=scriptsize]{caption} % "Figure 1." instead of "Figure 1:"
\usepackage[utf8]{inputenc} % For accented characters
\usepackage[T1]{fontenc} % For accented characters
\usepackage[title]{appendix}%appendix titles
\usepackage{multirow}%multirow
\usepackage{booktabs}
\usepackage{array}
\usepackage{booktabs,subcaption,amsfonts,dcolumn}
\usepackage{soul}
\newcolumntype{M}[1]{>{\centering\arraybackslash}m{#1}}
\usepackage{changes}

\newcommand{\stkout}[1]{\ifmmode\text{\sout{\ensuremath{#1}}}\else\sout{#1}\fi}
\setdeletedmarkup{\stkout{#1}}

\DeclareMathOperator{\Var}{Var}

\DeclareMathOperator*{\argmax}{argmax} 
\DeclareMathOperator*{\arginf}{arginf}

\tikzstyle{process} = [rectangle, minimum width=3cm, minimum height=1cm, text centered, draw=black ]
\tikzstyle{arrow} = [thick,->,>=stealth]
\begin{document}

\def\spacingset#1{\renewcommand{\baselinestretch}%
{#1}\small\normalsize} \spacingset{1}

%%%%%%%%%%%%%%%%%%%%%%%%%%%%%%%%%%%%%%%%%%%%%%%%%%%%%%%%%%%%%%%%%%%%%%%%%%%%%%

\if0\blind
{
  
\title{Real-time Monitoring of Functional Data}
\author[1]{Fabio Centofanti}
\author[1]{Antonio Lepore \thanks{Corresponding author. e-mail: \texttt{antonio.lepore@unina.it}}}
\author[2,3]{ Murat Kulahci }
\author[2]{Max Peter Spooner}

\affil[1]{Department of Industrial Engineering, University of Naples Federico II, Piazzale Tecchio 80, 80125, Naples, Italy}
\affil[2]{Technical University of Denmark, Department of Applied Mathematics and Computer Science, Kongens Lyngby, Denmark}
\affil[3]{Luleå University of Technology, Department of Business Administration, Technology and Social Sciences, Luleå, Sweden}

\setcounter{Maxaffil}{0}
\renewcommand\Affilfont{\itshape\small}
\date{}
\maketitle
} \fi

\if1\blind
{
  \bigskip
  \bigskip
  \bigskip
  \begin{center}
    {\LARGE\bf Real-time Monitoring of Functional Data}
\end{center}
  \medskip
} \fi

\bigskip
\begin{abstract}
	With the rise of Industry 4.0, huge amounts of data are now generated that are apt to be modelled as \textit{functional data}.
%	The new Industry 4.0 paradigm has produced  factories able to acquire  a huge amount of data that are apt to be modelled as \textit{functional data}.
		In this setting, standard \textit{profile monitoring} methods aim to assess the stability over time of a completely observed functional quality characteristic.
		However, in some practical situations, evaluating the process state in  real-time, i.e., as the process is running, could be of great interest to significantly improve the  effectiveness of monitoring.
		To this aim, we propose a new method, referred
		to as functional real-time monitoring  (FRTM), that is able to  account for both phase and amplitude variation through the following steps: (i) registration;
		(ii) dimensionality reduction; (iii) monitoring of a partially observed functional quality characteristic.
An extensive Monte Carlo simulation study is performed to quantify the  performance of  FRTM with respect  to two  competing  methods.
Finally, an example is presented where  the proposed method is used to  monitor batches from a  penicillin production process  in real-time.
\end{abstract}

\noindent%
{\it Keywords:}  Functional Data Analysis, Profile Monitoring, Statistical Process Control, Curve Registration
\vfill
%\hfill {\tiny technometrics tex template (do not remove)}

\newpage
\spacingset{1.45} % DON'T change the spacing!
\section{Introduction}
\label{sec_intro}
Within the modern Industry 4.0 framework,  factories  have been increasingly equipped with  advanced acquisition systems that allow huge amounts of data to be recorded at a high rate.
Particularly relevant is the case where data are apt to be modelled as  functions defined on multidimensional domains, i.e.,  \textit{functional data} \citep{ramsay2005functional,ferraty2006nonparametric,horvath2012inference,kokoszka2017introduction}.
A typical problem in industrial applications  is evaluating the stability over time of a process through  some characteristics of interest in the form of functional data, hereinafter referred to as functional quality characteristics.
This problem stimulated a new statistical process control (SPC) \citep{montgomery2007introduction} branch, named \textit{profile monitoring}, whose aim is to  continuously monitor a functional quality characteristic  in order to assess if  only normal sources of variation (also known as common causes) apply to the process, or if assignable sources of variations (also known as special causes) are present. In the first case  the process is said to be  in control (IC) otherwise is said to be an out-of-control (OC) process.
Relevant works on profile monitoring include \cite{woodall2004using,noorossana2012statistical,grasso2016using,grasso2017phase,menafoglio2018profile,capezza2020control,centofanti2020functional}.

Standard profile monitoring methods consider the case of completely observed functional data, that is, they assess the presence of special causes once the functional quality characteristic has been completely observed.
In practice,  assessing the presence of assignable sources of variations in real-time, i.e., before the process completion, is also of great interest to significantly improve the monitoring effectiveness.
%For instance, this is the case for long processes, where identifying anomalous behaviours while the process is still running could provide useful process insights.
%% trigger corrective actions as retune  process parameters or interrupt the production. 
% However, also short processes could benefit from a   monitoring strategy in real-time,
%  because it could allow   implementing automatic feedback systems  to effectively recalibrate the process parameters to deal with specific OC conditions.
  Although some multivariate SPC frameworks have been adapted for real-time applications \citep{wold1998modelling,gonzalez2011real,gonzalez2014multisynchro,spooner2018monitoring}, none of these methods tries to take advantage   of the functional nature of the data.
  To the best of the authors' knowledge, the only work that briefly sketches a possible real-time monitoring strategy for functional data is proposed by \cite{capezza2020control}  who monitors CO\textsubscript{2} emissions in a maritime application.
However, this method lacks  generality being specifically designed for the considered application.

To implement a more general and effective  monitoring strategy in real-time, the first issue to consider is related  to the registration or alignment of  incomplete functional data.
During the real-time monitoring phase, the functional quality characteristic, which is observed up to a given domain point in time, should be compared to an appropriate reference distribution that describes the process IC performance up to such point. However, the identification of the appropriate reference  distribution is a non-trivial problem, because processes could exhibit different temporal dynamics. Phase variation, which refers to the lateral displacement of the curve features, is a well-known issue in functional data analysis \citep{gasser1995searching,ramsay2005functional,marron2014statistics,marron2015functional}. The presence of phase variation often inflates data variance, distorts latent structures and, thus, could badly affect the effectiveness of the monitoring strategy by masking the effects of unnatural deviations associated with special causes. Different strategies have been proposed to separate phase and amplitude components, where the former is represented by  a warping function and the latter refers to the size of the curve features. Landmark registration, where a few anchor points form the basis of the registration, was studied by \cite{kneip1992statistical,gasser1995searching}. Different nonparametric strategies based on a regularization penalties were proposed by \cite{ramsay1998curve} and \cite{ramsay2005functional}.
Among relevant works on this subject,   \cite{srivastava2011registration}  proposed  a novel geometric framework for registration based on the use of the Fisher-Rao Riemannian metric, whereas \cite{james2007curve}  developed an alignment method based on equating the moments of a given set of curves. To frame the method present below it is also worth mentioning  the dynamic time warping technique for functional data  proposed by \cite{wang1997alignment} and \cite{gasser1999synchronizing}. 

%Thus, the registration step allows  identifying the appropriate reference IC conditions through the separation of the phase and amplitude variations.
%A straightforward approach to work around the real-time monitoring is to consider  the aligned partially observed functional data  as complete on the corresponding partial domain, after having identified the   phase and amplitude components, and apply standard profile monitoring methods  at each domain point. 
The mainstream profile monitoring literature \citep{woodall2004using,mosesova2006profile,colosimo2007use}  considers the registration process as a pre-processing step i.e.,  after the registration step  the amplitude component is monitored alone, whereas phase variation is regarded as a nuisance effect that may mask the true data structure and, thus, it may reduce the monitoring performance.
However, this approach could be risky  because it may overlook potentially useful  information to assess the process state contained in the phase component.
On the other hand, the registration step may have the drawback to mitigate OC behaviours of 
new curves by forcing them to resemble the observations in the reference sample and transferring the OC features over the phase component.
The first work in the direction of combining curve registration  within the profile monitoring framework was proposed by \cite{grasso2016using}, who introduced a novel approach to jointly monitor the stability over time of both the amplitude and phase components.

Along this line, we present a new method, referred to as functional real-time monitoring  (FRTM),  to monitor a functional quality characteristic in real-time by taking into account both the phase and amplitude components identified through a real-time alignment procedure. 
In particular,  FRTM  is designed for Phase II monitoring that refers to the perspective monitoring of new observations,  given a clean set of  observations, which will be hereinafter referred to as IC or reference sample and used to characterize the IC operating conditions of the process (Phase I).
To this end, FRTM  applies  a real-time procedure consisting of (i) registering the partially
observed functional data to the appropriate reference curve;
(ii) performing a dimensionality reduction by proposing a novel version of mixed functional principal
component analysis (Happ, 2018; Ramsay and Silverman, 2005) specifically designed
to  account for both the amplitude and phase components; (iii) monitoring the functional quality characteristic in the reduced space through an appropriate monitoring strategy.
The first step (i) is based on the functional dynamic time warping (FDTW) method proposed by \cite{wang1997alignment}, which is the functional extension of the well-known dynamic time warping (DTW) designed to align two signals with different dynamics \citep{sakoe1978dynamic,itakura1975minimum}.
In particular, we consider a modification of the FDTW that takes into account  partial signals matches and is referred to as open-end/open-begin FDTW (OEB-FDTW).
The second step (ii) aims at reducing the intrinsically infinite dimensionality of the phase and amplitude components obtained through the registration step (i).
However, as standard dimensionality reduction methods \citep{ramsay2005functional,happ2017mfpca} are not able to successfully capture the constrained nature of the warping functions, we propose a modification of the standard functional principal
component analysis \citep{happ2018multivariate}, hereinafter referred to as mFPCA,  where warping functions are transformed through an isometric isomorphism \citep{happ2019general}.
In the last step (iii), both phase and amplitude components are monitored through the profile monitoring approach introduced by \cite{woodall2004using} and, then, used in \cite{noorossana2012statistical,grasso2016using,pini2017domain,centofanti2020functional}, which is based on the simultaneous application of  Hotelling’s $T^{2}$  and the squared prediction error ($SPE$) control charts.

An extensive Monte Carlo simulation study is performed to quantify the monitoring performance of FRTM in the identification of mean shifts in the functional quality characteristic that may arise  in the amplitude or phase components. To this end, two competing monitoring strategies are considered, the first is a profile monitoring approach that does not take into account curve misalignment, whereas the other uses only  pointwise  information, without considering the functional nature of the data.
Finally, the practical applicability of the proposed method is demonstrated through an example   where batches from a  penicillin production process are  monitored in real-time.

The paper is structured as follows. Section \ref{sec_method} introduces FRTM through the definition of its founding elements. Section \ref{se_perfo} contains the Monte Carlo simulation to assess and compare the performance of  FRTM with competing methods, while Section \ref{se_casestudy} presents the data example. 
Section \ref{se_conclusions} concludes the paper.
Supplementary Materials for the article are available online.
All computations and plots have been obtained using the programming language \textsf{R} \citep{R2021}.

\section{Functional Real-time Monitoring}
\label{sec_method}
In the following sections, the key elements of  FRTM are introduced. The OEB-FDTW and mFPCA methods are introduced in Section \ref{sec_fdtw} and Section \ref{sec_mfpca}, respectively. The monitoring strategy is introduced in Section \ref{sec_mon}. Then, these elements are put together in Section  \ref{sec_prme}, where the proposed procedure is illustrated by detailing both Phase I and Phase II.
\subsection{Functional Dynamic Time Warping}
\label{sec_fdtw}
The curve registration problem is the first element of  FRTM method  and  can be formalized as follows.
Let $ N $ functional observations $ X_i $, $ i=1,\dots,N $ be a random  sample of functions whose realizations  belong to $ L^2\left(\mathcal{D}_{X_i}\right) $, the space of square integrable functions defined on  a compact domain $ \mathcal{D}_{X_i} $ in $ \mathbb{R} $. 
  Let $ X_i $ follow the general model
\begin{equation}\label{eq_tw}
X_i\left(x\right)=g_i\left(h_i^{-1}\left(x\right)\right) \quad x \in \mathcal{D}_{X_i},  i=1,\dots,n, 
\end{equation}
where $ h_i: \mathcal{D}_{h_i}\rightarrow \mathcal{D}_{X_i}$ are strictly  increasing square integrable random functions, named \textit{warping functions}, which are  defined on the compact domain $ \mathcal{D}_{X_i}\subset \mathbb{R} $ and capture the phase variation by mapping the absolute time $t$ to  the observation time $x$; whereas, $ g_i$ are functions in $   L^2\left(\mathcal{D}_{h_i}\right) $, named \textit{amplitude functions}  and capture the amplitude variation.
In this setting, the registration problem consists in estimating the warping functions  such that the similarity among the registered curves $ X_i(h_i\left(t\right)),t \in \mathcal{D}_{h_i} $ as a function of $h_i$, is maximized.
Several methods arise by different definitions of similarity among  curves. 
In this work,  we propose the  OEB-FDTW as an open-end/open-begin version of the FDTW of \cite{wang1997alignment}.
Given a reference or template curve $ Y \in L^2\left(\mathcal{D}_{Y}\right)$, the OEB-FDTW estimates the warping function $ h_i $ by solving the following variational problem
\begin{equation}\label{eq_fdtw}
\hat{h}_i,\hat{\alpha}=	\arginf_{h_i: \mathcal{D}_{h_i}\subseteq \mathcal{D}_{Y}\rightarrow\mathcal{D}_{X_i},\alpha\in[0,1]}\frac{1}{|\mathcal{D}_{h_i}|}\int_{\mathcal{D}_{h_i}}	\left[F(Y,X_i,Y',X'_i,h_i,\alpha)\left(t\right)+\lambda R\left(h'_i(t)-T_i/T_0\right)\right]dt,
\end{equation}
with 
\begin{multline*}
	F(Y,X_i,Y',X'_i,h_i,\alpha)(t)=\alpha^2\left(\frac{Y\left(t\right)}{||Y||}-\frac{X_i\left(h_i\left(t\right)\right)}{||X_i||}\right)^2\\+(1-\alpha)^2\left(\frac{Y'\left(t\right)}{||Y'||}-\frac{h'_i(t)X'_i\left(h_i\left(t\right)\right)}{||X'_i||}\right)^2\quad  t\in\mathcal{D}_{h_i},
\end{multline*}
where  $ R(u) $ is $u^2 $ if $ s^{min}\leq u+T_i/T_0\leq s^{max} $ or $ \infty $ otherwise; $ s^{min} $ and $ s^{max} $ are the  minimum and maximum allowed values of the first derivative of $ h_i $;  $ ||f|| $ is the sup-norm of $ f $ , and $ |\mathcal{D}_{h_i}| $ is the size of $ \mathcal{D}_{h_i} $. The parameter  $ \alpha $ tunes  the dependence of the alignment on the amplitude of the curves $ Y(t) $ and $ X_i(h_i(t)) $,  and their first derivatives $ Y'(t) $ and $ h'_i(t)X'_i(h_i(t)) $, $ t\in\mathcal{D}_{h_i} $. We are implicitly assuming that $ h'_i $ exists for  $ t\in\mathcal{D}_{h_i} $. The second right-hand term in Equation \eqref{eq_fdtw}, with the smoothing parameter $ \lambda\geq0 $, aims to penalize too irregular warping functions by constraining their first derivative to  be not too steep and to lie between $ s^{min} $ and $ s^{max} $. Indeed, large values of $ \lambda $ shrink the resulting warping function slope toward $ T_i/T_0 $, which corresponds to  the case of linear rescaling.

Usually, the values of  $ h_i $ at the boundaries of $ \mathcal{D}_{h_i}$ are constrained to be equal to the boundary values of $ \mathcal{D}_{X_i}$, corresponding to the case when the start and  end points of $ Y $ and $ X_i $ coincide.
However, this assumption could be unrealistic due to the lack of knowledge of the exact boundaries \citep{shallom1989dynamic,wang1997alignment,shallom1989dynamic} and, thus, could produce unsatisfactory warping function estimates.
To avoid this, the proposed OEB-FDTW does not consider boundary constraints by allowing the warping function to have values different from the boundary values of $ \mathcal{D}_{X_i}$ at the beginning and at the end of the process (which explains the OEB prefix). To further improve flexibility, a candidate solutions in Equation \eqref{eq_fdtw} is a warping function defined on the domain $ \mathcal{D}_{h_i}\subseteq \mathcal{D}_{Y}$. These assumptions allow considering both the case when  $ X_i $ reveals  temporal dynamics that partially agree with those of the template $ Y $ and  the case where some  $ X_i $ dynamics are not reflected in the template, i.e., partial matches \citep{tormene2009matching}.
The normalization factor $ \frac{1}{|\mathcal{D}_{h_i}|} $ is needed  to avoid that warping functions defined  on smaller domains are preferred in the optimization, which   implies the objective function to be simply an average distance.

The FDTW proposal of \cite{wang1997alignment} solves the variational problem by dynamic programming  at fixed $ \alpha $ and  grid searching for the best $ \alpha $.
However, the fact that $ \mathcal{D}_{h_i} $ is unknown strongly complicates the computation. The variational problem in Equation \eqref{eq_fdtw} is indeed not allowed to be solved by dynamic programming techniques because  local solutions depend on the size of $ \mathcal{D}_{h_i} $ and, thus, on the complete warping function trajectory \citep{shallom1989dynamic,kassidas1998synchronization}.
To still use the favourable features of the dynamic programming algorithm, we however consider an approximation of the optimal solution,  by modifying  the idea of \cite{shallom1989dynamic}, based on the current average distance  at each  algorithm iteration. Further details about the numerical solution of the variational problem in Equation \eqref{eq_fdtw} are provided in  Supplementary Materials A.

To define the OEB-FDTW solution, the smoothing parameter $ \lambda $ has to be chosen. The literature is lacking  model selection methods  for curve registration.
%A straightforward proposal to help  practitioners  in the selection of the appropriate value of $ \lambda $ may consist of the  study of 
We propose to select the optimal $ \lambda $ based on the behaviour of the \textit{average curve distance} ($ ACD $) defined as 
\begin{equation}\label{eq_acd}
ACD(\lambda)=	\sum_{i=1}^{N}\frac{1}{|\mathcal{D}_{\hat{h}_{i,\lambda}}|}\int_{\mathcal{D}_{\hat{h}_{i,\lambda}}}	F(Y,X_i,Y',X'_i,\hat{h}_{i,\lambda},\hat{\alpha}_{\hat{h}_{i,\lambda}})\left(t\right)dt
\end{equation} 
at different values of $ \lambda $ as the maximum value  such that the  increase in $ ACD $ is less than or equal to a given fraction $ \delta $ of the difference between the maximum and the minimum distance, i.e.,  $ACD(\infty)$ and $ ACD(0) $. The function   $ \hat{h}_{i,\lambda} $ and $ \hat{\alpha}_{\hat{h}_{i,\lambda}} $ are the solutions of the optimization problem in Equation \eqref{eq_fdtw} at a given value of $ \lambda $, being $ \mathcal{D}_{\hat{h}_{i,\lambda}} $  the corresponding domain. Although this criterion works well in both  the simulation study (Section \ref{se_perfo}) and in the data example (Section \ref{se_casestudy}),    whenever possible we recommend to directly plot and inspect the $ ACD$ curve and to use any information available from the specific application to select $ \lambda $. 

As detailed in \cite{wang1997alignment}, the template function should result similar to the typical  curve pattern and resemble the features of the sample curves.
A standard approach to choose the template function is through a \textit{Procrustes} fitting process, where the data are  used  both to define the template and to estimate the warping functions \citep{ramsay1991some,grasso2016using}. Specifically, the procedure  starts by choosing  an initial estimate of the template function   e.g., the average curve or a curve chosen based on prior knowledge of the process. Then, at each iteration, all the curves in the sample are  registered to the template curve, then, the average curve of the registered sample curves will be used as template in the subsequent iteration. Few iterations are usually sufficient to converge to a satisfactory template curve.

\subsection{Mixed Functional Principal Component Analysis}
\label{sec_mfpca}
For each functional observation $ X_i $, $ i=1,\dots,N $,  the  curve registration  procedure described in Section \ref{sec_fdtw} returns the registered curve $  X^*_i\left(t\right)=X_i(\hat{h}_i\left(t\right)),t \in \mathcal{D}_{\hat{h}_i} $ and the warping function $ \hat{h}_i $.   The OEB-FDTW solution is hereinafter denoted by $ h_i $.
% and the true warping function defined in Equation \eqref{eq_tw}.
%As mentioned in the introduction, the registration phase is often considered as a preprocessing step to obtain the registered functions $ X^*_i $, which become the main focus of the analysis. On the contrary, in this work, we follow a different approach where the warping functions $ h_i $ are considered as a crucial source of information.
Note that the pair $ \left(X^*_i,h_i\right) $ consists of two random functions both  defined on the domain $\mathcal{D}_{h_i} $. As  stated in the introduction,  we propose as the second element of the proposed FRTM procedure, a novel version of   the mixed functional principal
component analysis, named mFPCA and specifically designed to reduce the infinite dimensionality of the problem and take into account both the  amplitude and phase components.

Unfortunately, the space of warping functions has a complex geometric structure, as it is not closed under addition and is not equipped with a natural scalar product. Thus,  standard operations based
on Euclidean geometry can only be applied with great care \citep{lee2016combined}.
We circumvent this issue by considering a two-steps approach \citep{happ2019general,hadjipantelis2015unifying}, where each warping function is transformed to a square integrable function through an appropriate mapping, and, then the classical functional principal component analysis is applied. Results are then back transformed through the inverse map.
Following \cite{happ2019general},  the centred log-ratio ($clr$) transformation is applied to the first derivative of the warping function $h_i$. This transformation is  isometric  with respect to the geometry of the Bayes Hilbert space \citep{hron2016simplicial} when the warping functions have values in the same range.
However, this is not necessarily in our setting, where the ranges $ \mathcal{D}_{X_i} $ of each  warping function may not be the same.
This can be easily solved by considering in place of $h_i$, the following functions, denoted by $h^*_i$ and defined on the range $ \left[0,1\right] $
\begin{equation*}
	h^*_i\left(t\right)=\frac{	h_i\left(t\right)-F_{0i}}{F_{1i}-F_{0i}} \quad t\in \mathcal{D}_{h_i},
\end{equation*} 
where $ F_{0i} $ and $ F_{1i} $ are the values of $ h_i $ at the boundaries of $ \mathcal{D}_{h_i} $, with $ F_{1i}>F_{0i} $. Then, we propose to apply the $clr$ transformation  to the first derivative $ h^{*'}_i $ of $ h^{*}_i $, and obtain
\begin{equation*}
	v_i(t)=clr(h^{*'}_i)(t)=\log\left(h^{*'}_i(t)\right)-\frac{1}{|\mathcal{D}_{h_i}|}\int_{\mathcal{D}_{h_i}}\log\left(h^{*'}_i(t)\right)dt \quad t\in \mathcal{D}_{h_i}.
\end{equation*}
In this way, the mFPCA can be  performed by extending the approach of \cite{silverman1995incorporating, ramsay2005functional} to $ Z_i=(X^*_i,v_i,F_{0i}, \tilde{F}_{1i})\in \mathbb{Z}=L^2\left(\mathcal{D}_{h_i}\right)\times L^2\left(\mathcal{D}_{h_i}\right)\times\mathbb{R} \times\mathbb{R}$, where $ \tilde{F}_{1i}=log(F_{1i}-F_{0i}) $ is introduced to ensure the validity of the constraint $ F_{1i}>F_{0i} $.
That is, we jointly account for both amplitude and phase variations  by considering, in a multivariate fashion, the  registered function, the $clr$ transformation of $ h^{*'}_i $, and the values of the warping function at the domain boundaries. 
Without loss of generality, let us assume that the $ Z_i $ components have zero mean, otherwise   they can be opportunely
centred by subtracting either their functional or scalar  mean.
Then, the first \textit{principal component} is obtained as the element $ \psi_1\in \mathbb{Z} $ such that
\begin{equation}\label{eq_mapc}
	\psi_1=\argmax_{\psi_1\in \mathbb{Z},||\psi_1||^2=1}\Var\left(\xi_{1i}\right),
\end{equation}
where $\xi_{1i}=\langle\psi_1,Z_i\rangle_{w}$ is the \textit{principal component score}, or simply  \textit{score}, that is associated to $ \psi_1 $, with $ ||\psi_1||^2= \langle\psi_1,\psi_1\rangle_{w}$. The inner product $ \langle\cdot,\cdot\rangle_w $ of two elements $ f=(f_1,f_2,f_3,f_4)\in\mathbb{Z} $ and $ g=(g_1,g_2,g_3,g_4)\in\mathbb{Z} $ is defined as
\begin{equation}\label{eq_ip}
	\langle f,g\rangle_w=\int_{\mathcal{D}_{h_i}}w_1(t) f_1(t)g_1(t)dt+\int_{\mathcal{D}_{h_i}}w_2(t) f_2(t)g_2(t)dt + w_3 f_3g_3+w_4 f_4g_4,
\end{equation}
where $ w_1,w_2 $  are positive weight functions and  $ w_3,w_4 $ are positive weight constants.
Subsequent principal components $ \psi_2,\psi_3,\dots \in \mathbb{Z} $ are obtained by considering the same maximization as in Equation \eqref{eq_mapc} with the additional constraint that each solution is orthogonal with respect to the  previous principal components, i.e., $   \langle\psi_k,\psi_{k-1}\rangle_w=\dots=\langle\psi_k,\psi_{1}\rangle_w=0 $, $ k=2,3,\dots $. The scores corresponding to $ \psi_2,\psi_3,\dots $ are $\xi_{ki}=\langle\psi_k,Z_i\rangle_w$, $ k=2,3,\dots $.
In practice, the dimensionality reduction consists in returning a  finite number $ L $ of principal components. Where, as in the multivariate setting, $ L $  can be chosen such that $ \psi_1,\dots,\psi_L$ explain at least
a given percentage of the total variability, say $ 70 \text{-}90\% $.  More sophisticated methods could be used as well, see \cite{jolliffe2011principal} for further details. 
However, to allow the principal components to be  estimated  through a basis function expansion approach \citep{silverman1995incorporating, ramsay2005functional}, the functions returned by the registration phase must be defined on the same domain.
%However, the registration output could be  functions defined on different domains $ \mathcal{D}_{h_i}\subseteq \mathcal{D}_{Y} $ because of the open-begin/open-end feature of the registration method. This causes a difficulty in the estimation  of the mFPCA mode, which requires   the curve sample to be 
To overcome this issue, we reduce it to a missing data imputation problem by setting the template curve domain $ \mathcal{D}_{Y} $ as the common domain and impute  the missing parts of the functional observations $ (X^*_i, h_i) $ by using the template curve as a reference, opportunely shifted to ensure continuity. Details on the missing data procedure are provided in Supplementary Materials A.2.

The weights $w_1,w_2,w_3,w_4$ in the inner product defined in Equation \eqref{eq_ip} are introduced to  account for any different unit of measurement  of the $ Z_i=(X^*_i,v_i,F_{0i}, \tilde{F}_{1i}) $ components. Indeed, while $ X_i^* $ are measured on the scale of the centred observations, $ v_i, F_{0i} $ and $ \tilde{F}_{1i} $ bring information about the phase component. Therefore, the weights should reflect the different nature of the data in terms of both amplitude and phase variation, and functional and scalar components. 
We propose to set the weights, for $ t \in\mathcal{D}_{h_i} $, as
\begin{equation}
	\label{eq_weigh}
	w_1(t)=\frac{k_1}{\sigma_1^2(t)|\mathcal{D}_{h_i}|}, \quad  w_2(t)=\frac{k_2}{\sigma_2^2(t)|\mathcal{D}_{h_i}|},\quad
	w_3=\frac{k_3}{\sigma_3^2}, \quad
		w_4=\frac{k_4}{\sigma_4^2},
\end{equation}
where $\sigma_1^2,\sigma_2^2,\sigma_3^2,\sigma_4^2$ are the  variances of $ X_i^*, v_i,F_{0i} $ and $\tilde{F}_{1i} $, respectively, and $ k_1,k_2,k_3,k_4 $ are constants that should be chosen  to appropriately  weight  amplitude and phase variation in the inner product computation. The denominators of Equation \eqref{eq_weigh} aim to equalize the overall variability of the functional  and scalar parts. For instance, if $ k_1=k_2=k_3=k_4=1 $, then all  components show the same overall variation and  thus, they equally contribute to the principal component calculation.  Note that the first component $ X_i^* $ is related to the registered curve whereas the remaining ones $ v_i, F_{0i} $ and $ \tilde{F}_{1i} $ are related to the warping function, hence we find it suitable to choose $ k_1=0.5 $ and $ k_2=k_3=k_4=0.5/3 $ to equally weight phase and amplitude components.
Although this weight choice works well in the simulation study (Section \ref{se_perfo}) and  the data example (Section \ref{se_casestudy}), there may be  rationales for  different choices driven by  the  specific application or  question of interest.

\subsection{The Monitoring Strategy}
\label{sec_mon}
As stated in the introduction, the third element of the proposed FRTM procedure relies on the consolidated monitoring strategy of \cite{woodall2004using} applied to  $ Z_i=(X^*_i,v_i,F_{0i}, \tilde{F}_{1i}) $ associated  with the functional quality characteristic $ X_i $.
Specifically,  Hotelling's $ T^2 $ and $ SPE $ control charts  assess the stability of $ Z_i $   both in the finite dimensional space spanned by the first principal components identified through the mFPCA (Section \ref{sec_mfpca}) and in its orthogonal complement space, respectively.
Hotelling's  $ T^2 $ statistic corresponding to $ Z_i $ is defined as
\begin{equation*}
	T^2_i=\sum_{k=1}^{L}\frac{\xi_{ki}^2}{\lambda_k},
\end{equation*}
where $ \lambda_k $ are the variances of the scores $ \xi_{ki}$ obtained  through the mFPCA. The statistic $ T^2_i $ is the standardized squared distance from the centre of the orthogonal projection of all $ Z_i $'s onto the space spanned by  $ \psi_{1},\dots,\psi_L $, referred  to also as principal component space.
Whereas, the distance of $ Z_i $  from its orthogonal projection onto the principal component space is measured through the $ SPE $ statistic, defined  as 
\begin{equation}
	SPE_i=||Z_i-\hat{Z}_i||_{w}^2,
\end{equation}
where $ ||\cdot||_{w} $ is the norm induced by the inner product $ \langle \cdot,\cdot\rangle_{w} $ defined in Equation \eqref{eq_ip} and $ \hat{Z}_i=\sum_{k=1}^{L}\xi_{ki}\psi_k$. Note that  $ Z_i,\psi_k $ and so $ \hat{Z}_i$ are in $ \mathbb{Z} $. Therefore,   addition and multiplication operations should be intended component-wise. That is for $ f=(f_1,f_2,f_3,f_4)\in\mathbb{Z} $ and $ g=(g_1,g_2,g_3,g_4)\in\mathbb{Z} $, $ f+g=(f_1+g_1,f_2+g_2,f_3+g_3,f_4+g_4) $ and $ af=(af_1,af_2,af_3,af_4) $, $ a\in\mathbb{R} $.
The control limits of  Hotelling's $ T^2 $ and $ SPE $ control charts can be obtained by the $ (1-\alpha^*) $ quantiles of the empirical distributions of the two statistics \citep{centofanti2020functional}.
Note that, to control the family wise error rate (FWER), $ \alpha^* $ should be chosen appropriately. We propose to use the \v{S}id\'ak correction, that is $\alpha^{*}=1-\left(1-\alpha\right)^{1/2}$ \citep{lehmann2006testing}, where $\alpha$ is the overall type I error probability.
\subsection{The Proposed Methodology}

\label{sec_prme}

\begin{figure}[h]
	\caption{Scheme of FRTM approach. }
	\label{fi_diag}
	
	\centering
	
	\includegraphics[width=\textwidth]{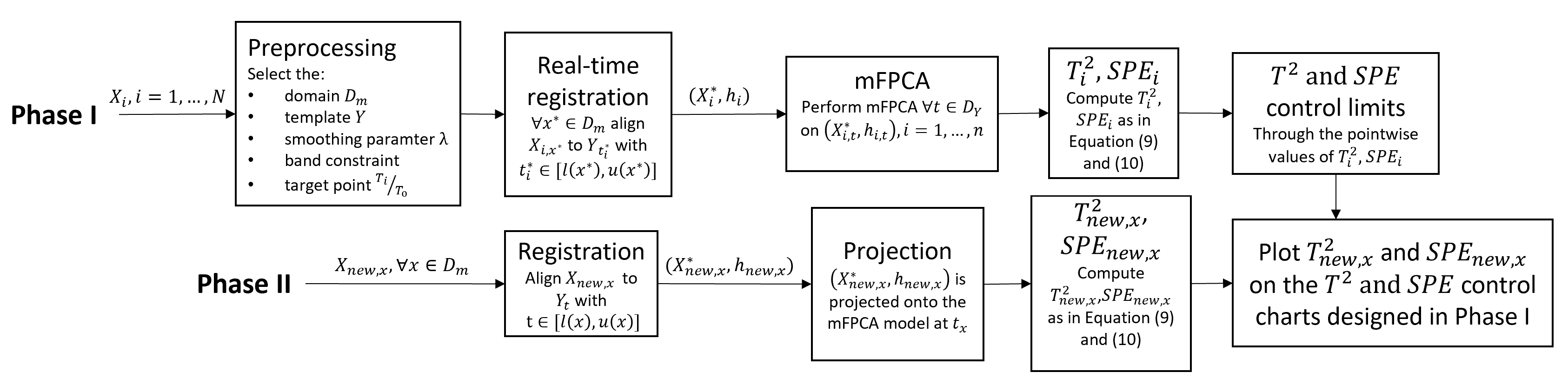}

\end{figure}

The proposed FRTM  collects  all the elements introduced in the previous sections.
As stated before, FRTM is designed for use in Phase II  where a set of  observations that represent the IC process are needed for the design of the control chart (Phase I).
Both phases are outlined in the scheme of Figure \ref{fi_diag} and detailed in the following.

 \subsubsection{Phase I}
 \label{sec_phaseI}
% \paragraph{Preprocessing}
In this phase, a reference sample denoted by $ X_i $, $ i=1,\dots,N $, is used to characterize the IC operating conditions of the process.

 A  \textit{preprocessing} step is needed to choose the common monitoring domain $ \mathcal{D}_{m} $, as the compact interval  over which the assessment of the process conditions could be of interest. Moreover,   based on the  domains   $ \mathcal{D}_{X_i} $ of the IC sample,  $ \mathcal{D}_{m} $ must be included in $ \cup_{i=1}^{N} \mathcal{D}_{X_i}  $, which implies  at least one observation $ X_i $ is defined for each $ x\in \mathcal{D}_{m}$. Reasonably, $ \mathcal{D}_{m}$ could be set  as $\cup_{i=1}^{N} \mathcal{D}_{X_i} $.  However,   different choices could be  driven by the specific application.
 The Procrustes fitting process and the $ ACD $ approach described in Section \ref{sec_fdtw} are used to identify an appropriate template function $ Y $  and an appropriate value of the smoothing parameter $ \lambda $. The subsequent real-time registration step considers  a band  constraint, applied to the terminal points of the partial warping functions and avoids  too extreme warping functions. The two extremes of the band are denoted by $ u,l:\mathcal{D}_{m}\rightarrow \mathcal{D}_{Y} $ and selected as the  $ (1-b/2) $ and $ (b/2) $ pointwise quantiles,  $ b\in\left[0,1\right] $, of the warping functions obtained on the completely observed functional data.
Note that, the band constraint serves also to set the maximum mismatch allowed range  at the beginning and the end of the template function domain $ \mathcal{D}_{Y} $ (as shown in Supplementary Materials A.1). Moreover,  the target point $ T_i/T_0 $ in Equation \eqref{eq_fdtw}  could be updated  by considering the mean slope of the  warping functions corresponding to $ X_i $, and avoid wrong alignments in the real-time registration step detailed below, especially in the first part of the process. 
%This modification is not necessary if the functional observations are not expected to show extreme open-end/open-begin behaviours, which means low temporal dynamic uncertainty over the domain boundaries.
 
 In the \textit{real-time registration} step, the  functional observations of the IC sample  are registered to the template function as they would have been observed until a specific point in $ \mathcal{D}_{m}$.
 Specifically, for each $ x^* \in \mathcal{D}_{m}$,  the partial functional observation up to $x^*$ is denoted by $ X_{i,x^*}:\mathcal{D}_{X_{i,x^*}}\subseteq\mathcal{D}_{X_{i}} $, where $ \mathcal{D}_{X_{i,x^*} }$ is a compact domain in $\mathbb{R}$ having the same $ \mathcal{D}_{X_{i}} $ left boundary  and right boundary equal to $ x^* $. 
Each partial observation $ X_{i,x^*} $ is aligned to the  template function truncated at $ t_{i}^*\in\left[l(x^*),u(x^*)\right]\subset\mathcal{D}_{Y} $ and denoted by $ Y_{t_{i}^*} $.
The point $ t_{i}^* $ is chosen such that the obtained warping function minimizes the   right-hand term in Equation \eqref{eq_fdtw} over the interval $ \left[l(x^*),u(x^*)\right] $. In addition, we do not allow  the partial warping function corresponding to $ X_{i,x^*} $, denoted by  $h_{i,x^*}  $,  to be open-end, i.e.,  $h_{i,x^*}(t_{i}^*)=x^*$. In this setting partial matches could be indeed easily handled by a different choice of $ t_{i}^* $. Open-end warping functions are allowed  for $ x^* $, such that $ u(x^*) $ is equal to the right boundary of $ \mathcal{D}_{Y} $.
The band constraint forces $h_{i,x^*}$ to be defined on a domain where the right boundary is inside the range $ \left[l(x^*),u(x^*)\right] $. However, this could be too restrictive, e.g., for the use in Phase II where curve dynamics may not be  well represented by the IC sample, and cause wrong alignments. To limit this issue, we consider an adaptive band constraint. The naive idea is based on the conjecture that  when the uncertainty of $h_{i,x^*}$ at the right domain boundary is sufficiently small for all $ x^* $  values  in a given portion of the domain, the available information up to $ x^* $  provides a reliable curve registration and   the band constraint  can be relaxed. 
%This adaptive band constraint approach takes advantage of the dependence of  registration results for close values of $ x^*\in \mathcal{D}_{m} $.
Details on the calculation of the adaptive band constraint are provided in  Supplementary Materials A.3.

Once all the IC observations are  aligned in real-time, they can be used to perform the  \textit{mFPCA} step, described in Section \ref{sec_mfpca}.
For each $ t\in\mathcal{D}_{Y} $, mFPCA is applied to the pairs $ \left(X^*_{i,t},h_{i,t}\right) $, $ i=1,\dots,N $, where both $X^*_{i,t}$ and $ h_{i,t} $ are defined on the compact domain $ \mathcal{D}_{Y,t} $, which has  the same left boundary of $ \mathcal{D}_{Y} $ and $ t $ as right boundary. 
However, it may exist $ i=1,\dots,N $ for which $ \left(X^*_{i,t},h_{i,t}\right) $ is not  available, because the   real-time registration step does not directly provide  aligned and warping functions defined on  $ \mathcal{D}_{Y,t} $. In this case,  the pair $ \left(X^*_{i,t_g},h_{i,t_g}\right) $ defined on the smallest $ \mathcal{D}_{Y,t_g}\supset \mathcal{D}_{Y,t} $, truncated at $ t $, is considered in place of $\left(X^*_{i,t},h_{i,t}\right)$.

In the last  step,  Hotelling's $ T^2 $ and $ SPE $ statistics are computed for each observation $ i=1,\dots,N $ and each $ x\in \mathcal{D}_{m} $. Specifically, for each $ x\in \mathcal{D}_{m} $ the observation $ X_{i,x} $, through the corresponding aligned and warping functions pair  $(X^*_{i,t_x},h_{i,t_x})$ on  $ \mathcal{D}_{Y,t_x} $, is projected on the   mFPCA model corresponding to $ t_x $. Therefore, the values of  $ T^2_{i} $ and $ SPE_{i} $ functions at $ x $ are computed as described in Section \ref{sec_mon}. Control limits are obtained  by considering the pointwise values of  $ T^2_{i} $ and $ SPE_{i} $ defined on $ \mathcal{D}_{m} $ for a given  overall type I error probability $\alpha$.
Note that this last step could be performed with a reference sample, referred to as tuning set, that is different from the one used in the mFPCA step. This possibly reduces possible overfitting issues and  increases the monitoring performance of FRTM.

In practice,  $ X_i $, and so the partial functional data $  X_{i,x^*} $, are usually observed through noisy discrete values, and functional observations are recovered through standard smoothing techniques.

\subsubsection{Phase II}
As in the traditional SPC, Phase II refers to the perspective monitoring of new observations, given a chosen reference sample in Phase I.
Let  $ X_{new,x} $ denote  a new observation  $ X_{new,x} $ of the quality characteristic observed until $ x\in \mathcal{D}_{m} $. Then, as described in  Section \ref{sec_phaseI}, we obtain the corresponding aligned and warping function pair  $(X^*_{new,t_x},h_{new,t_x})$, by projecting it  onto the  mFPCA space corresponding to $ t_x $, we compute the values of  $ T^2_{new} $ and $ SPE_{new} $  at $ x $  as described in Section \ref{sec_mon}.
An alarm signal is issued if it exists a $ x\in \mathcal{D}_{m} $ such that $ T^2_{new} $ or $ SPE_{new} $  violates the corresponding control limits.
Note that, although $ X_{new,x} $ can be defined also for $ x\notin  \mathcal{D}_{m} $, then $ X_{new,x} $ is monitored at $x\in \mathcal{D}_{m} $ alone. 

\section{Simulation Study}
\label{se_perfo}
%\subsection{Data Generation}
%\label{sec_data_generation}
An extensive Monte Carlo simulation study is performed to assess the performance of the proposed FRTM in identifying mean shifts of  the quality characteristic caused by perturbations of the amplitude and phase components. 
The data generation process is inspired by the works of \cite{wang1997alignment,gasser1999synchronizing} and is detailed in  Supplementary Materials B.
Two different scenarios are considered and  characterized by different  phase components. Specifically, Scenario 1 considers simple quadratic warping functions, whereas   Scenario 2 investigates a more complex temporal dynamics.
Each scenario explores three  models with decreasing levels of misalignment, quantified through the ratio  between  amplitude and phase variation and indicated through M1,  M2,  and M3.  A Low level of misalignment corresponds to small values of phase variation and mimics e.g., those settings where  a coarse registration is preliminarily applied to the data.
%Figure \ref{fi_dataicmain1} shows 100 realizations of IC curves for Scenario 1  for models, M1, M2 and M3.
%\begin{figure}[h]
%	\caption{A sample of $100$ randomly generated IC observations for each model M1,M2, and M3 in Scenario 1 of the simulation study.}
%	
%	\label{fi_dataicmain1}
%	
%	\centering
%	\includegraphics[width=.3\textwidth]{Figures/sim_data_Scenario_I_1.pdf}
%	\includegraphics[width=.3\textwidth]{Figures/sim_data_Scenario_I_2.pdf}
%	\includegraphics[width=.3\textwidth]{Figures/sim_data_Scenario_I_3.pdf}	
%	\vspace{-.5cm}
%\end{figure}
Three different types of OC conditions are analysed and correspond to  mean shifts  affecting   the phase component alone (Shift A),   the amplitude component alone (Shift B)  and both components simultaneously  (Shift C). For each shift 4 different severity levels $ d\in\lbrace 0.25,0.50,0.75,1.00\rbrace $ are explored, as well as to mimic  real-time monitoring OC scenarios, we simulate change points at  30\% and 60\% of the total duration of the process used to generate the data.
As results are analogous for both change point scenarios, in this section, we present only the former case, whereas the latter is presented in   Supplementary Materials C.
  
 FRTM method is compared with two simpler natural	 competing approaches, namely the  monitoring in real-time with \textit{no alignment}, referred to as NOAL, and the  \textit{pointwise} monitoring approach of the functional quality characteristic, referred to as PW.
The NOAL approach is based on the approach presented in \cite{capezza2020control}, where a dimensionality reduction based on functional principal components is followed by a monitoring strategy built on  the $ T^2$ and $ SPE $ control charts. The PW approach belongs to the class of methods	where a synthetic index for each curve is extracted to be used in a univariate control chart, where the pointwise value of the quality characteristic is monitored through a Shewhart-type control chart.
FRTM is implemented as described in Section \ref{sec_method},  the template function is obtained through the Procrustes fitting process with 2 iterations and a curve randomly chosen as initial estimate, $ s^{min}$ and  $ s^{max}$ are set equal to 0.01 and 1000, respectively,, and the value of $ \lambda $  in Equation \eqref{eq_fdtw} is chosen through the  \textit{average curve distance} (Equation \eqref{eq_acd}) with $ \delta=0.01 $. The band constraint is implemented with $ b=0.01 $. Moreover, in the mFPCA step, the number of retained principal components $ L $ is chosen such that the retained principal components explain at least $ 90\% $ of the total variability. The empirical quantiles of  $ T^2 $ and $ SPE $ statistics are obtained via the kernel density estimation approach \citep{chou2001control} with the Gaussian kernel, 1000 equally spaced points and bandwidth chosen by means of the  methods in \cite{sheather1991reliable}. While data are  observed through noisy discrete values, functional observations are obtained by considering the usual spline smoothing approach   based on cubic B-splines and a roughness penalty on the second derivative \citep{ramsay2005functional}. 50 simulation runs are performed for each scenario,  misalignment level, shift type and severity level. Each run considers an IC  sample of 500 observations, where one half is used in the mFPCA step, and the remaining  are used as tuning set to compute the control limits. The Phase II sample is composed of  500 OC observations.

FRTM and the competing method performance are assessed by means of the \textit{true detection rate} (TDR), which is the proportion of domain outside the control limits whilst the process is OC, and the \textit{false alarm rate} (FAR), which  is the proportion of domain outside the control limits whilst the process is IC. The FAR should be similar to the overall type I error probability $\alpha$ that is considered to obtain the control limits and, in this simulation study, is set equal to $ 0.05 $, whereas the  TDR  should be as close to one as possible.

\begin{figure}[h]
	\caption{Mean FAR ($ d=0 $) or TDR ($ d\neq 0 $) plus/minus one standard error achieved by FRTM, NOAL and PW, for each shift type (Shift A, B and C) and increasing misalignment level (M1, M2 and M3)   as a function of the severity level $ d $ in Scenario 1.}
	
	\label{fi_results_1}
	
	\centering
	\begin{tabular}{cM{0.28\textwidth}M{0.28\textwidth}M{0.28\textwidth}}
				\textbf{M1}&\includegraphics[width=.3\textwidth]{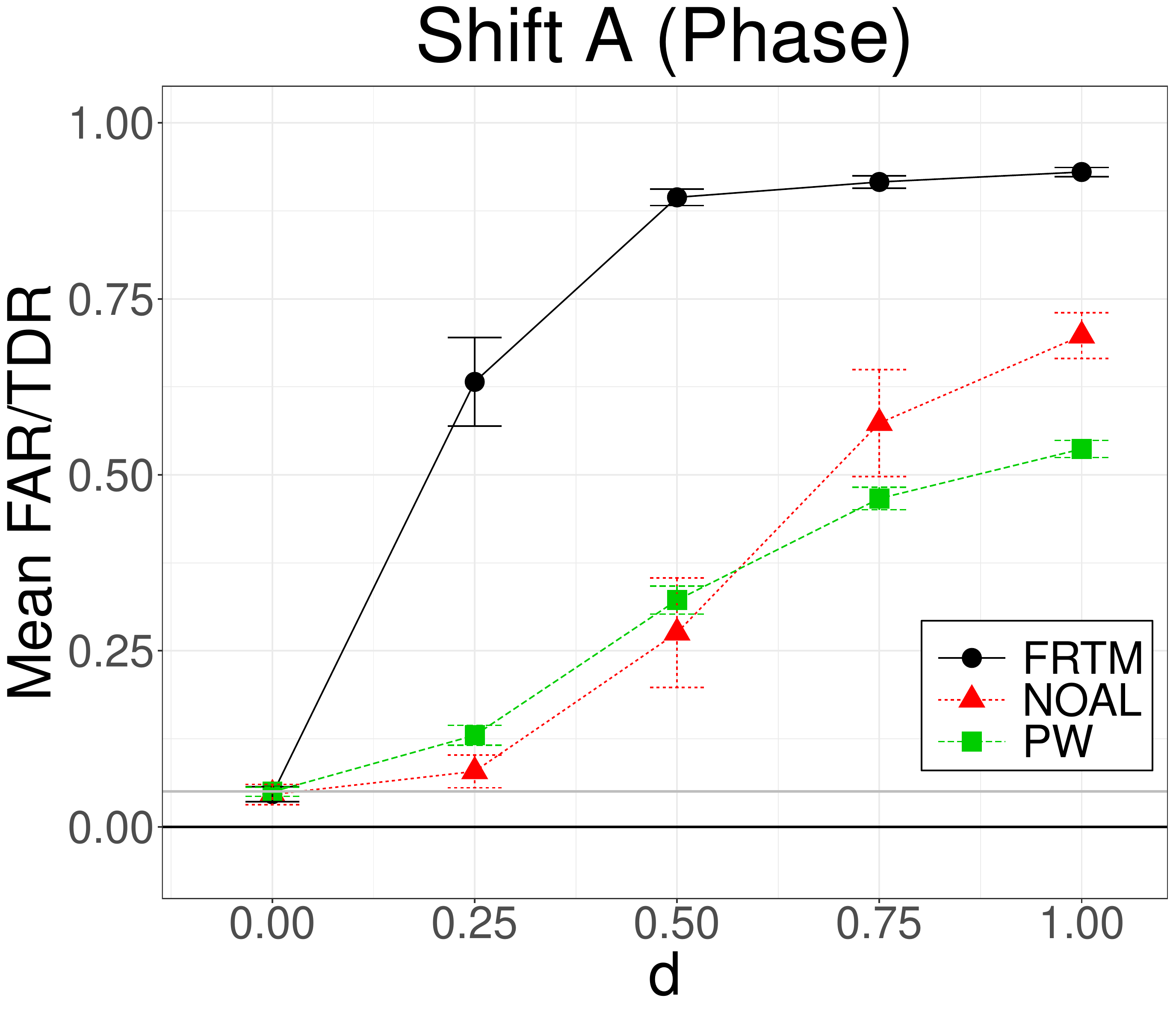}
		&\includegraphics[width=.3\textwidth]{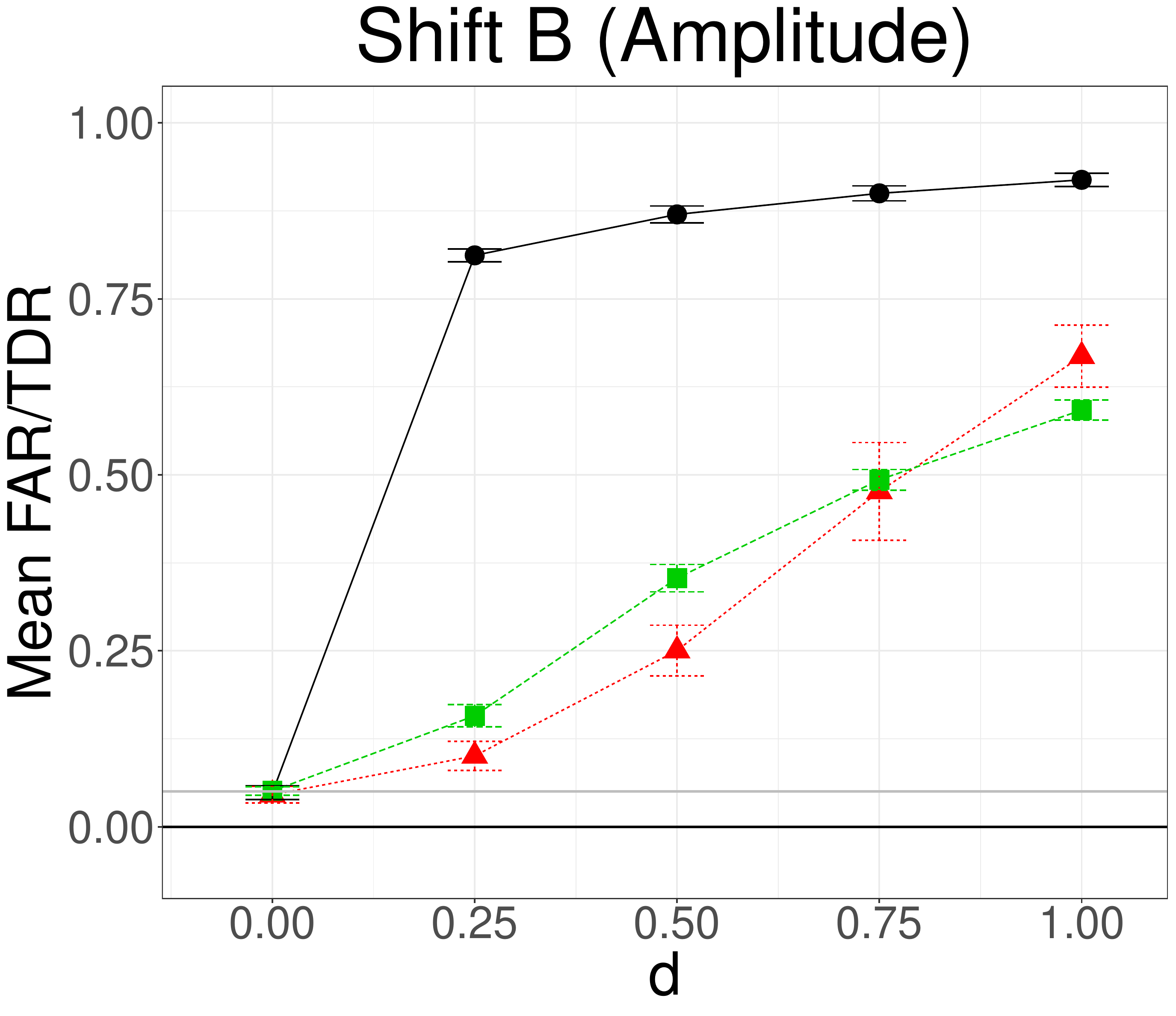}
		&\includegraphics[width=.3\textwidth]{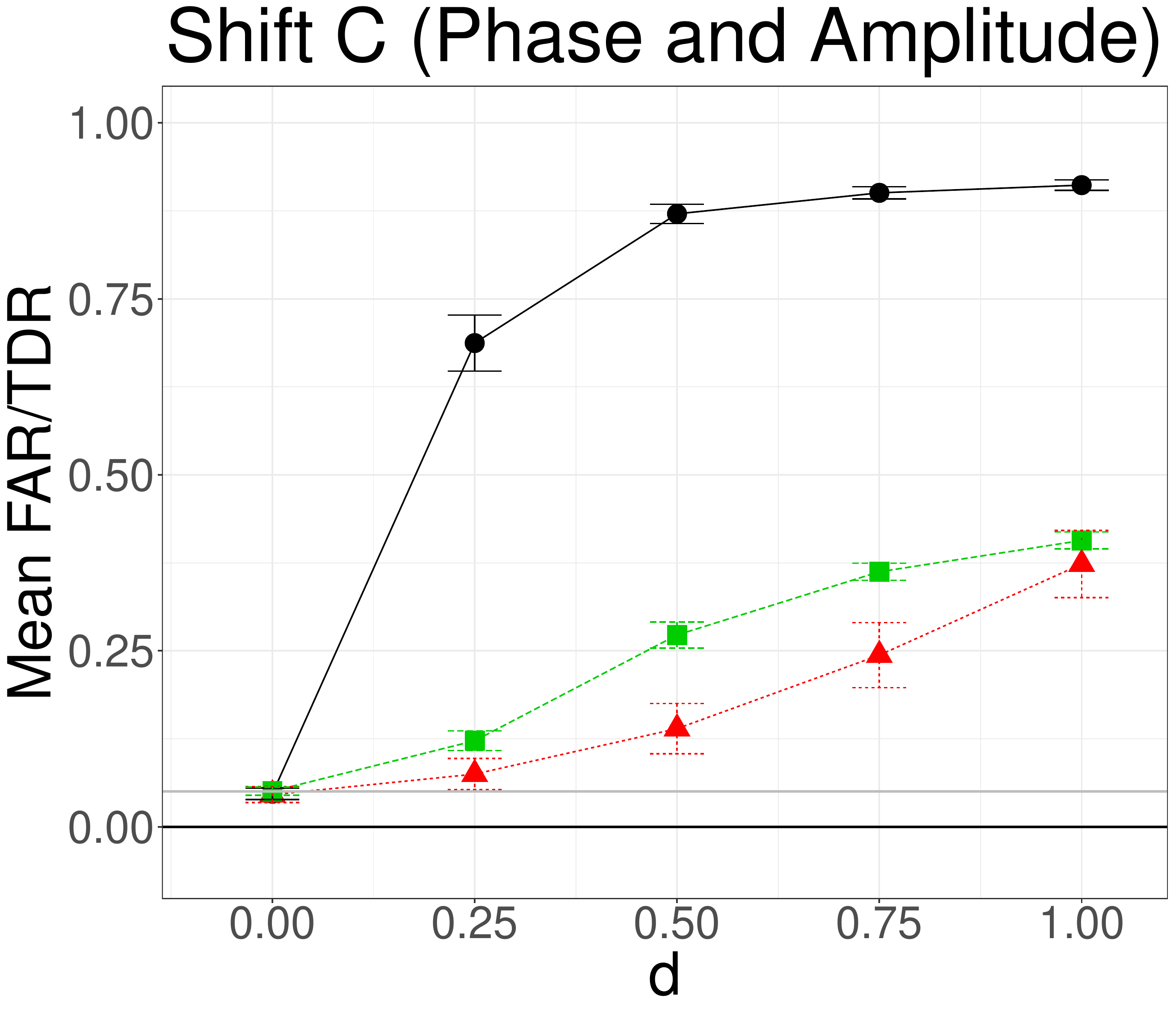}	\\
	\textbf{M2}&\includegraphics[width=.3\textwidth]{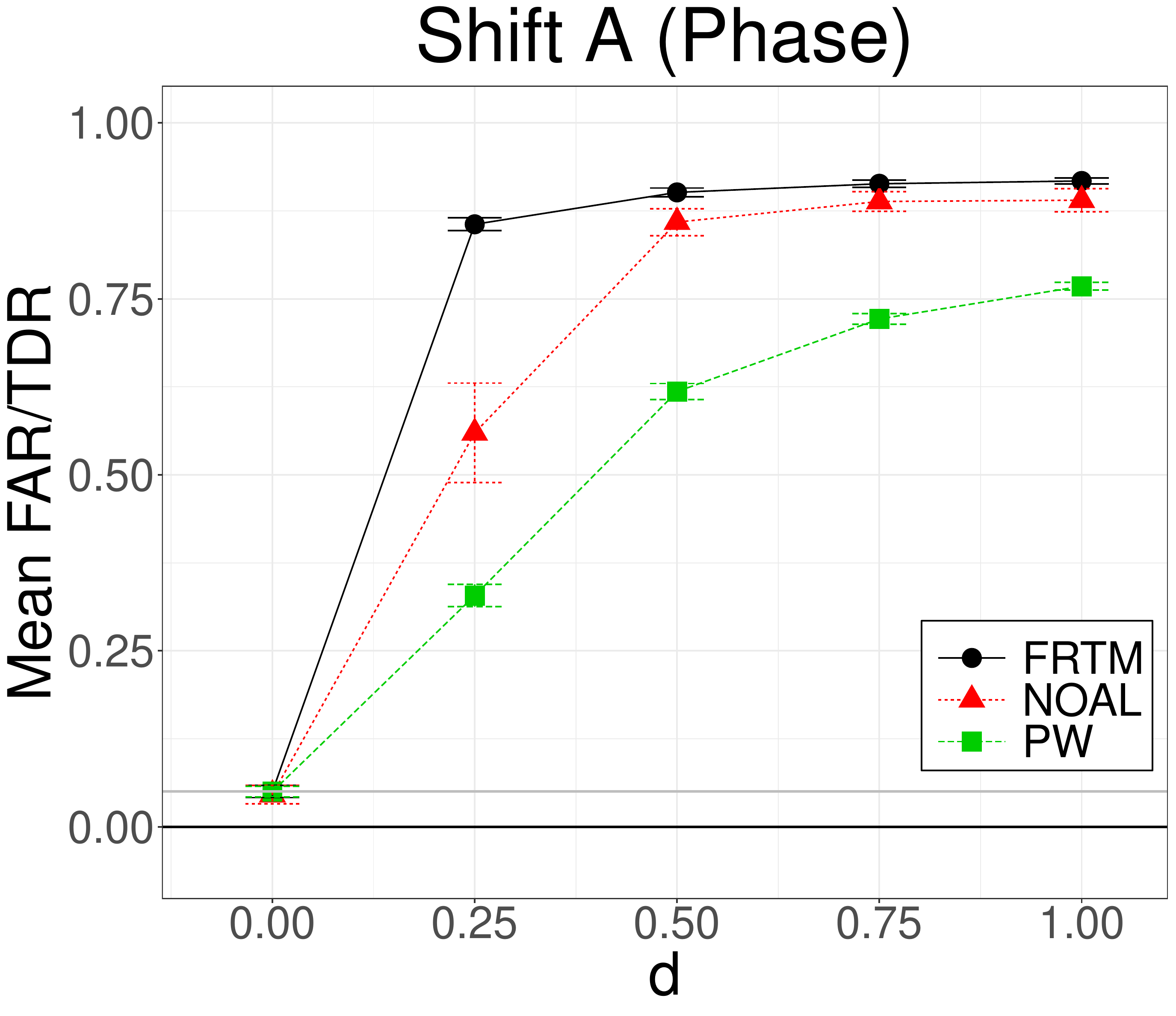}
	&\includegraphics[width=.3\textwidth]{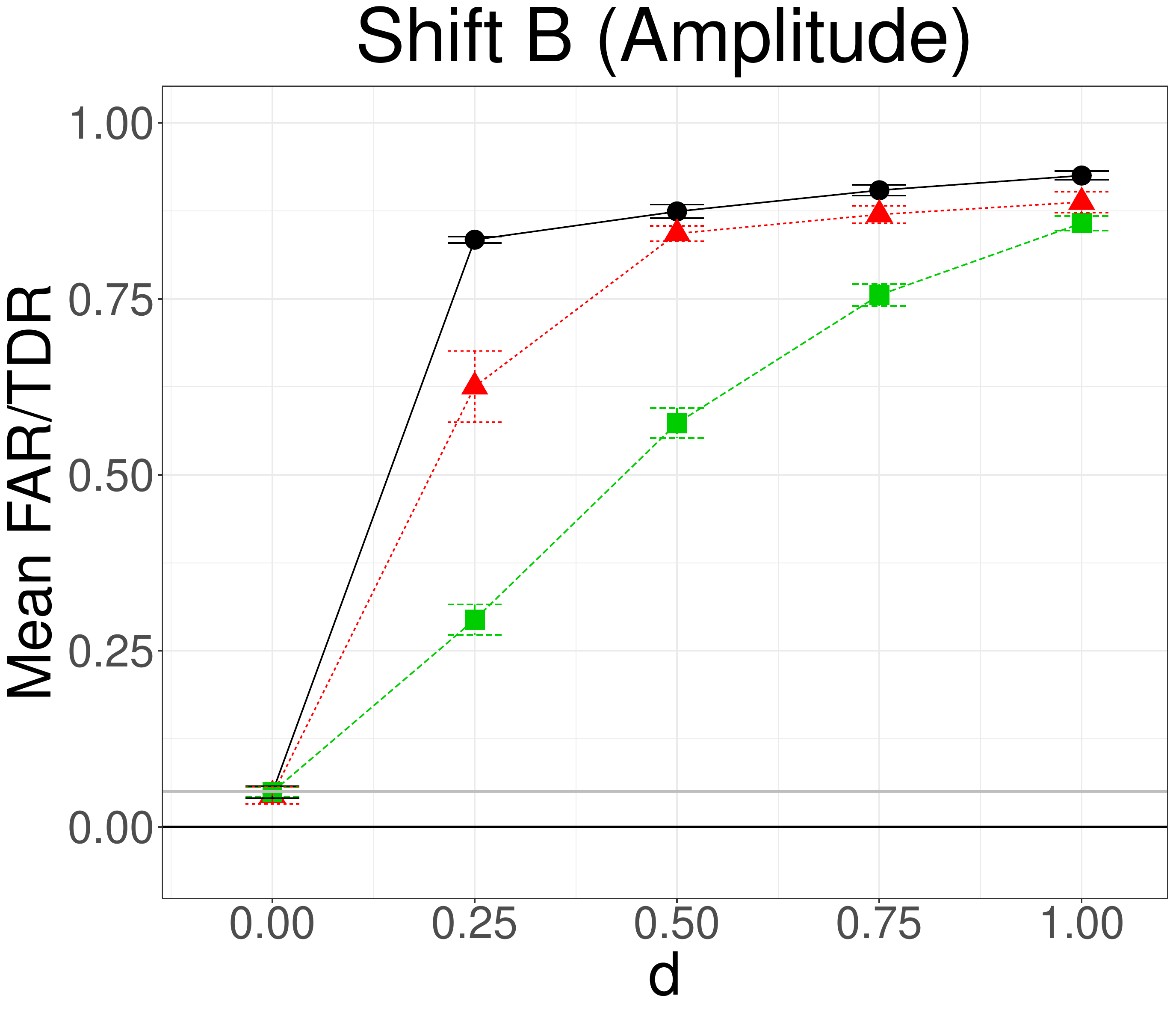}
	&\includegraphics[width=.3\textwidth]{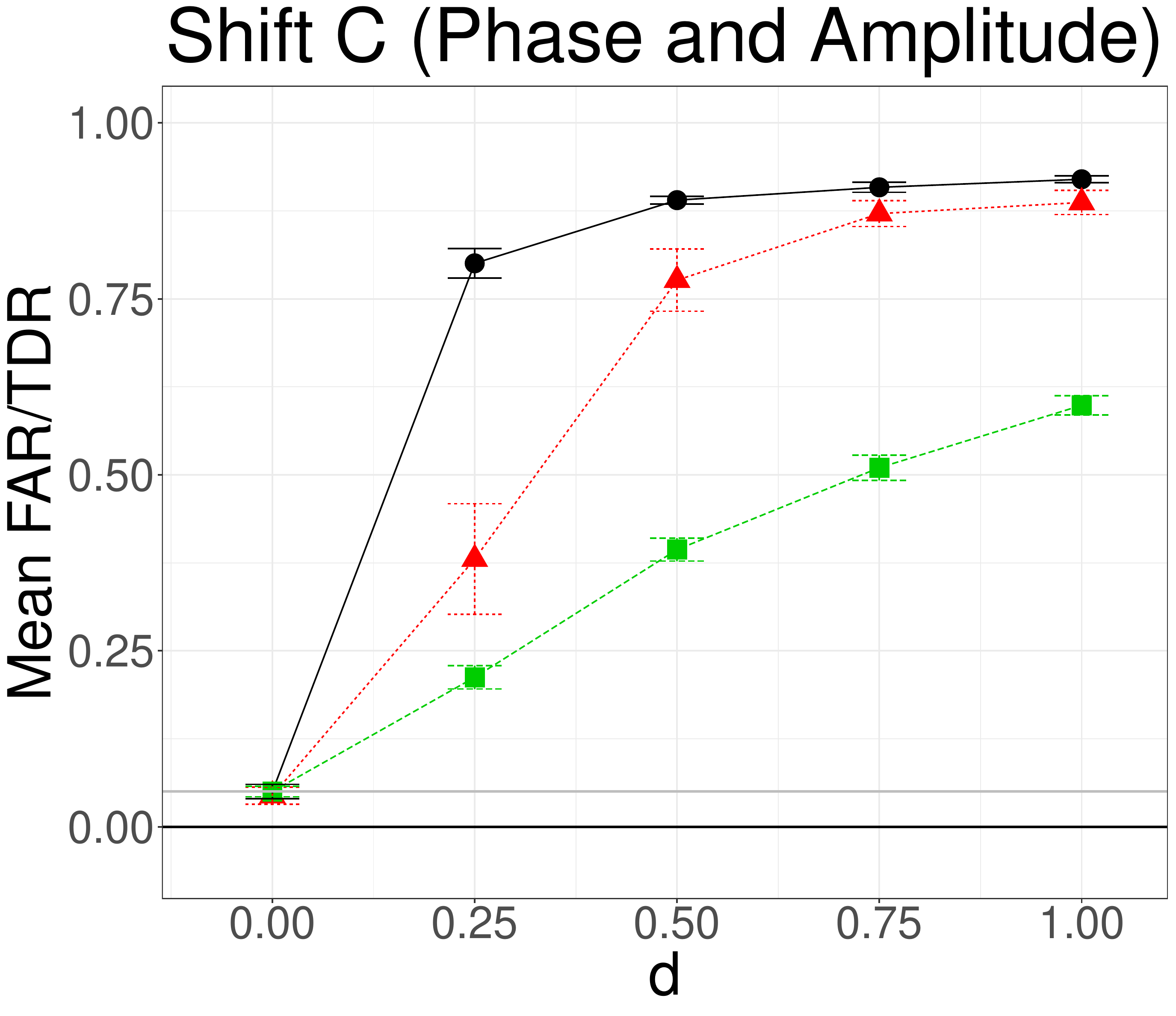}\\
		\textbf{M3}&\includegraphics[width=.3\textwidth]{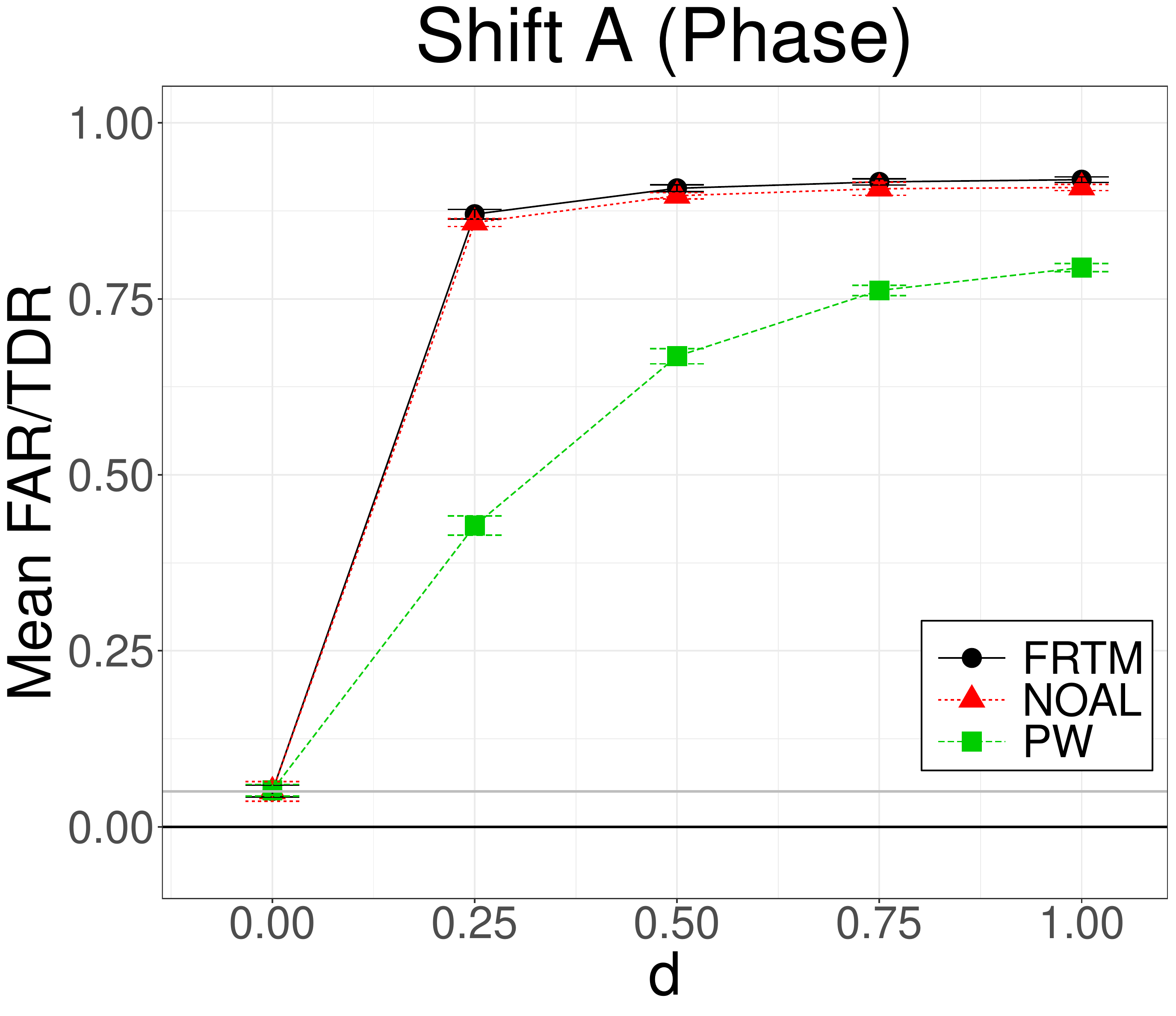}
	&\includegraphics[width=.3\textwidth]{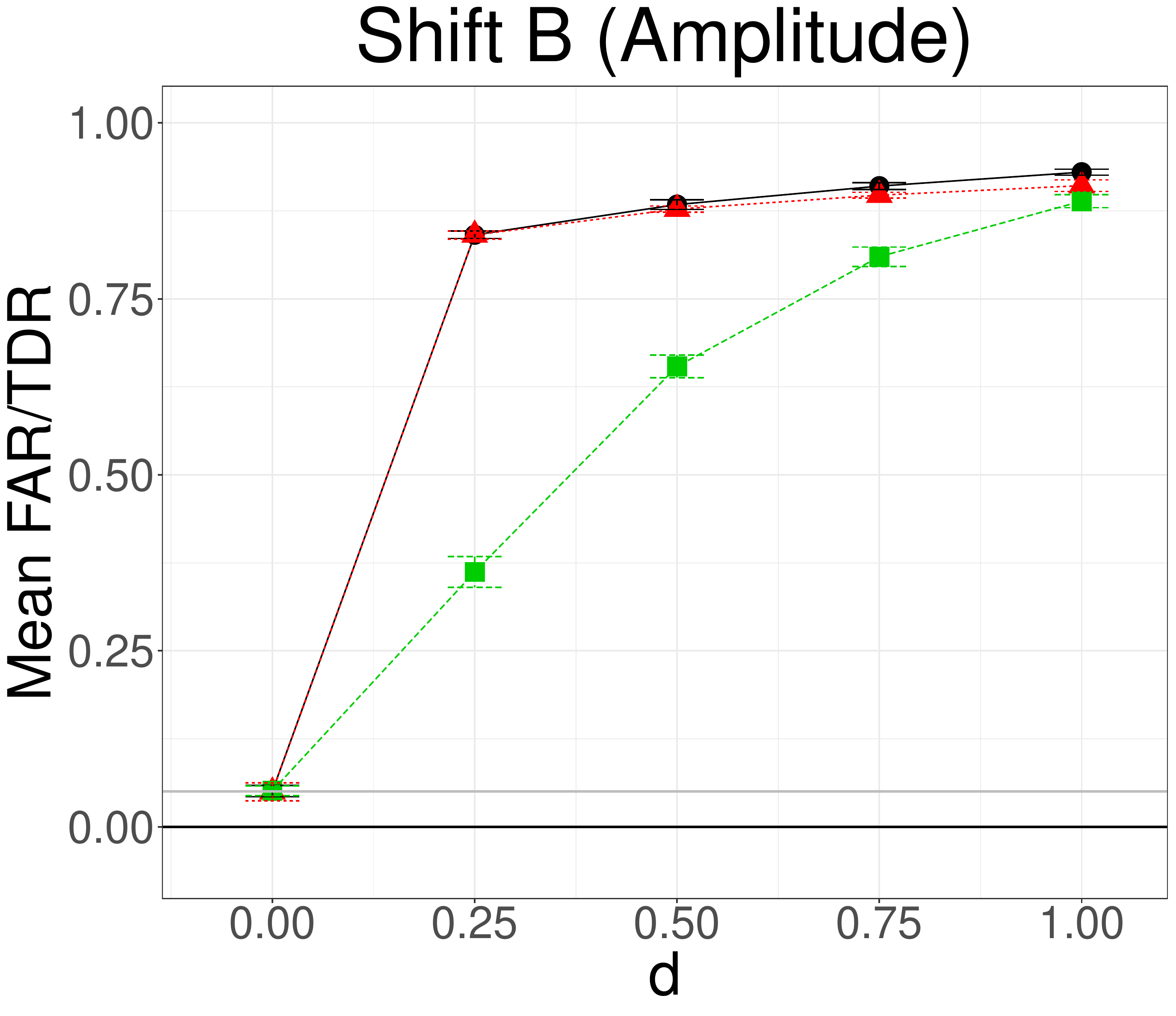}
	&\includegraphics[width=.3\textwidth]{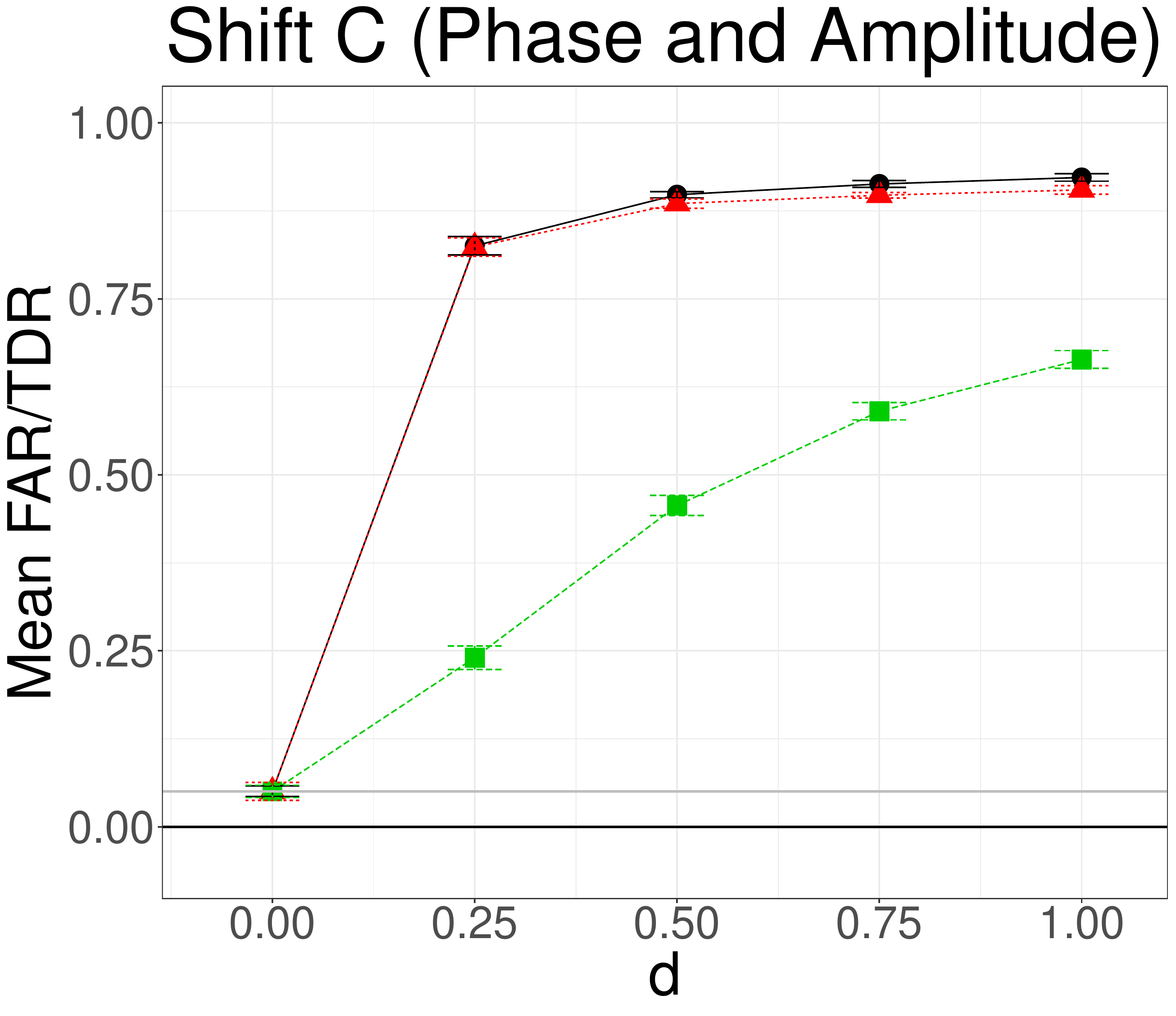}
	\end{tabular}
	\vspace{-.5cm}
\end{figure}

\begin{figure}[h]
		\caption{Mean FAR ($ d=0 $) or TDR ($ d\neq 0 $) plus/minus one standard error achieved by FRTM, NOAL  and PW, for each shift type (Shift A, B and C) and increasing misalignment level (M1, M2 and M3)   as a function of the severity level $ d $ in Scenario 2.}
	
	\label{fi_results_2}
	
	\centering
	\begin{tabular}{cM{0.28\textwidth}M{0.28\textwidth}M{0.28\textwidth}}
		\textbf{M1}&\includegraphics[width=.3\textwidth]{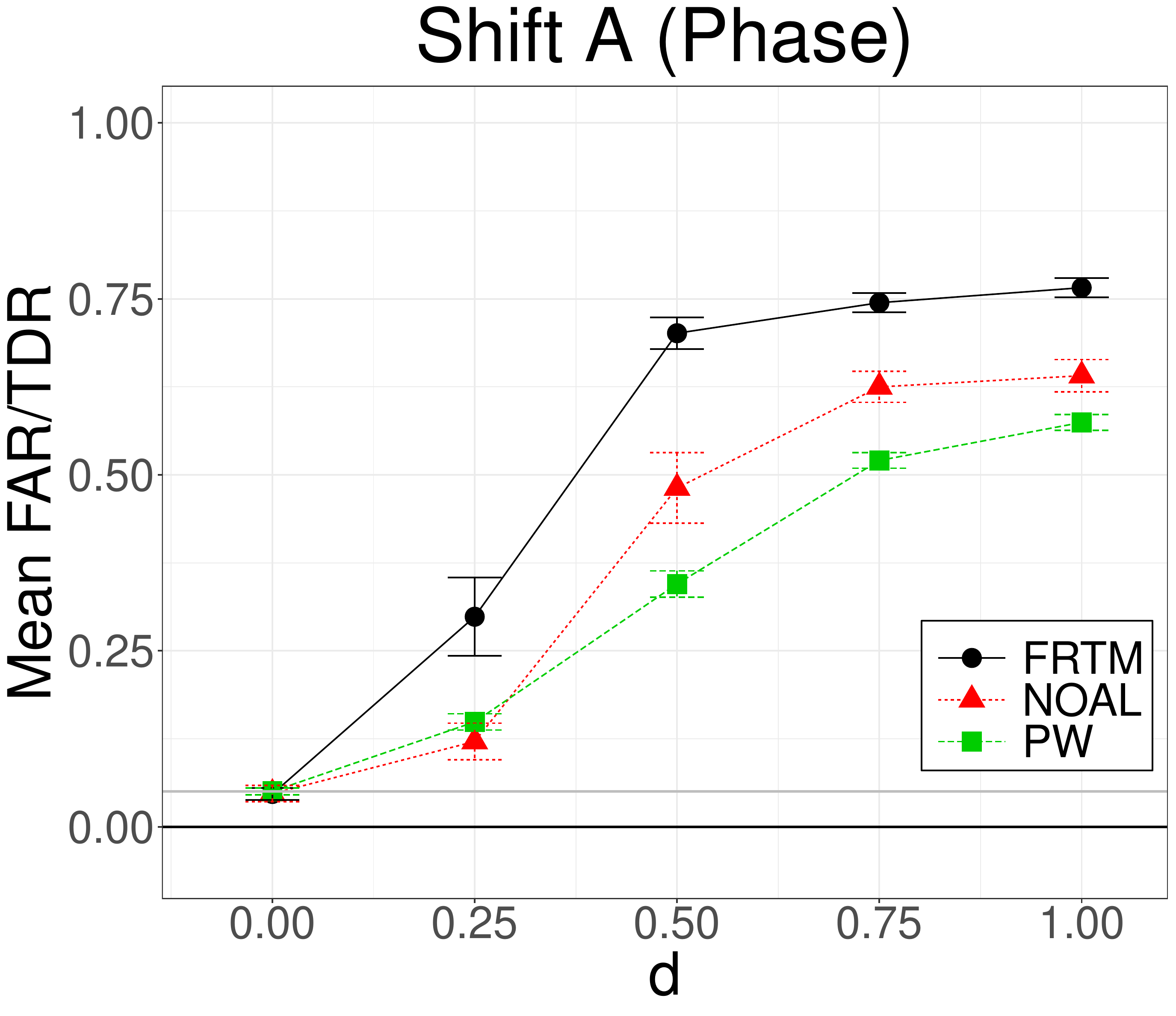}
		&\includegraphics[width=.3\textwidth]{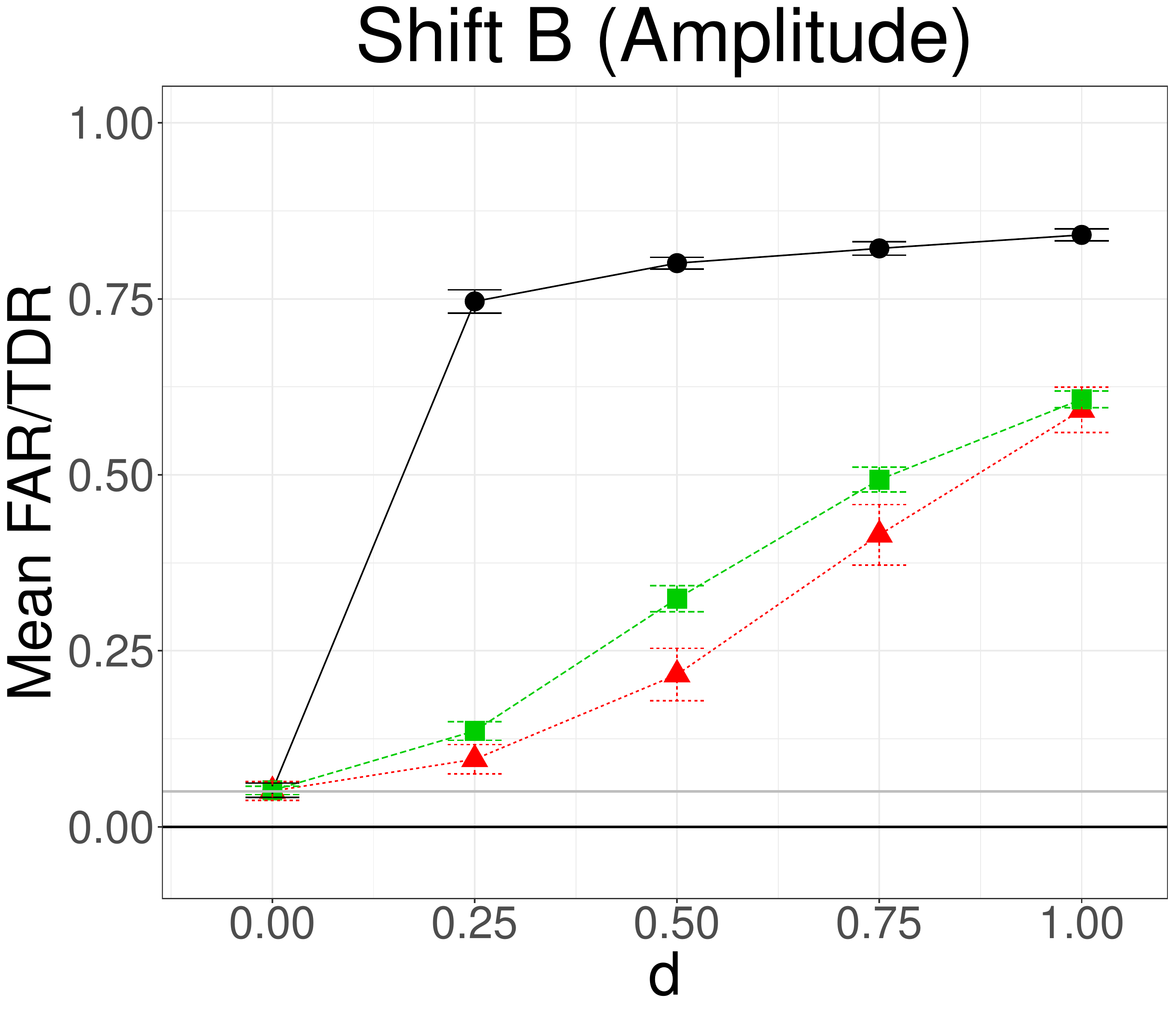}
		&\includegraphics[width=.3\textwidth]{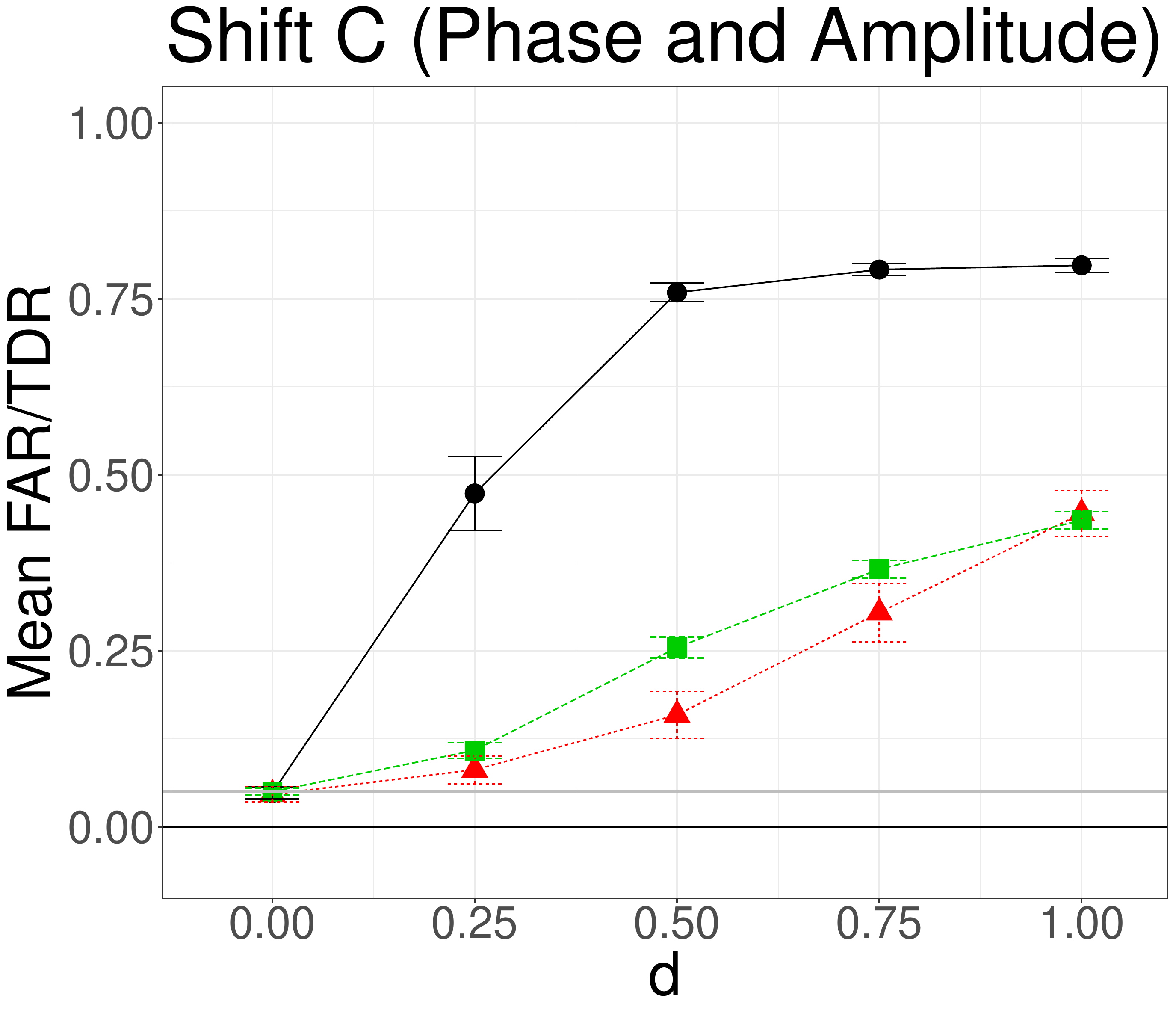}	\\
		\textbf{M2}&\includegraphics[width=.3\textwidth]{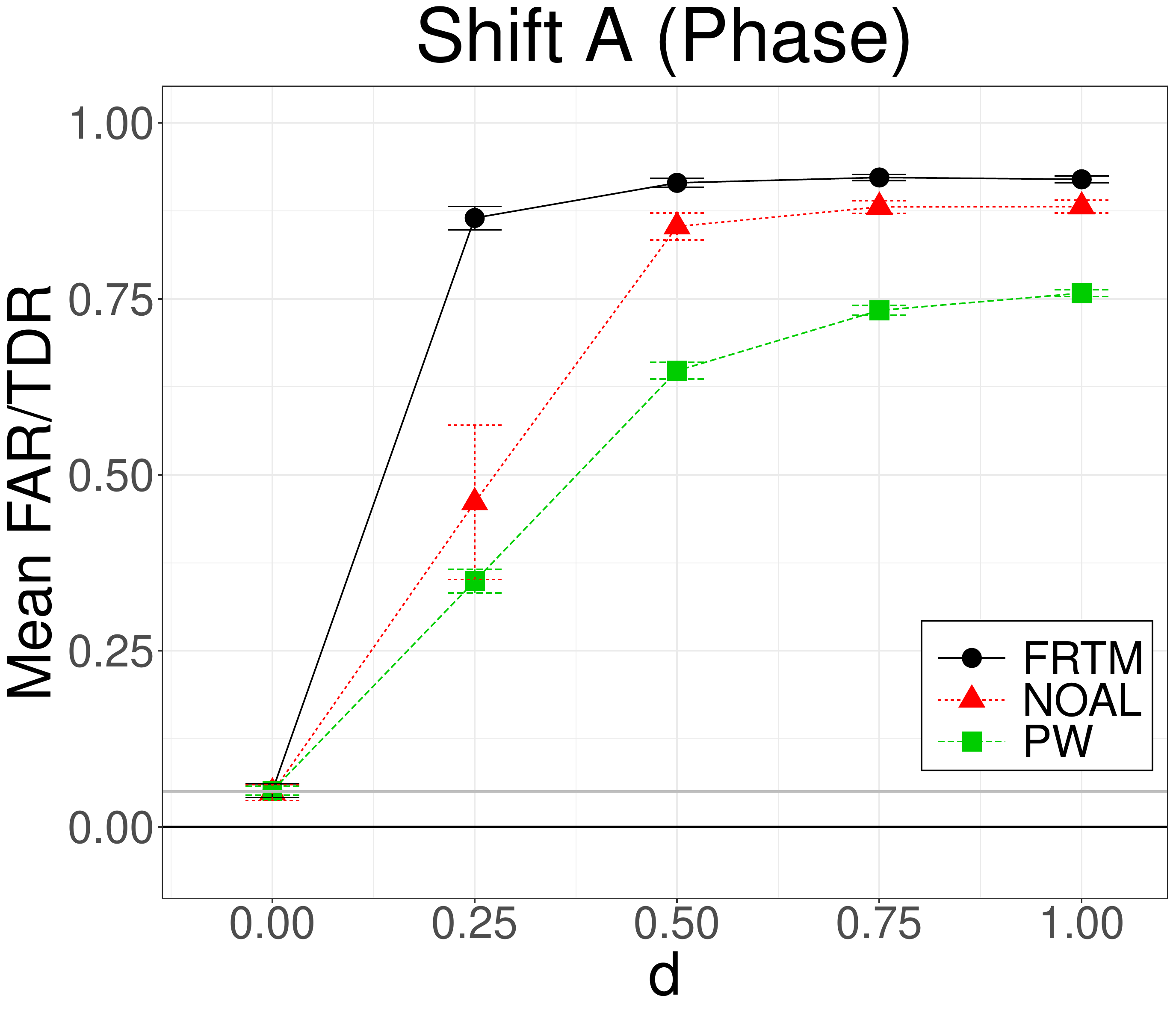}
		&\includegraphics[width=.3\textwidth]{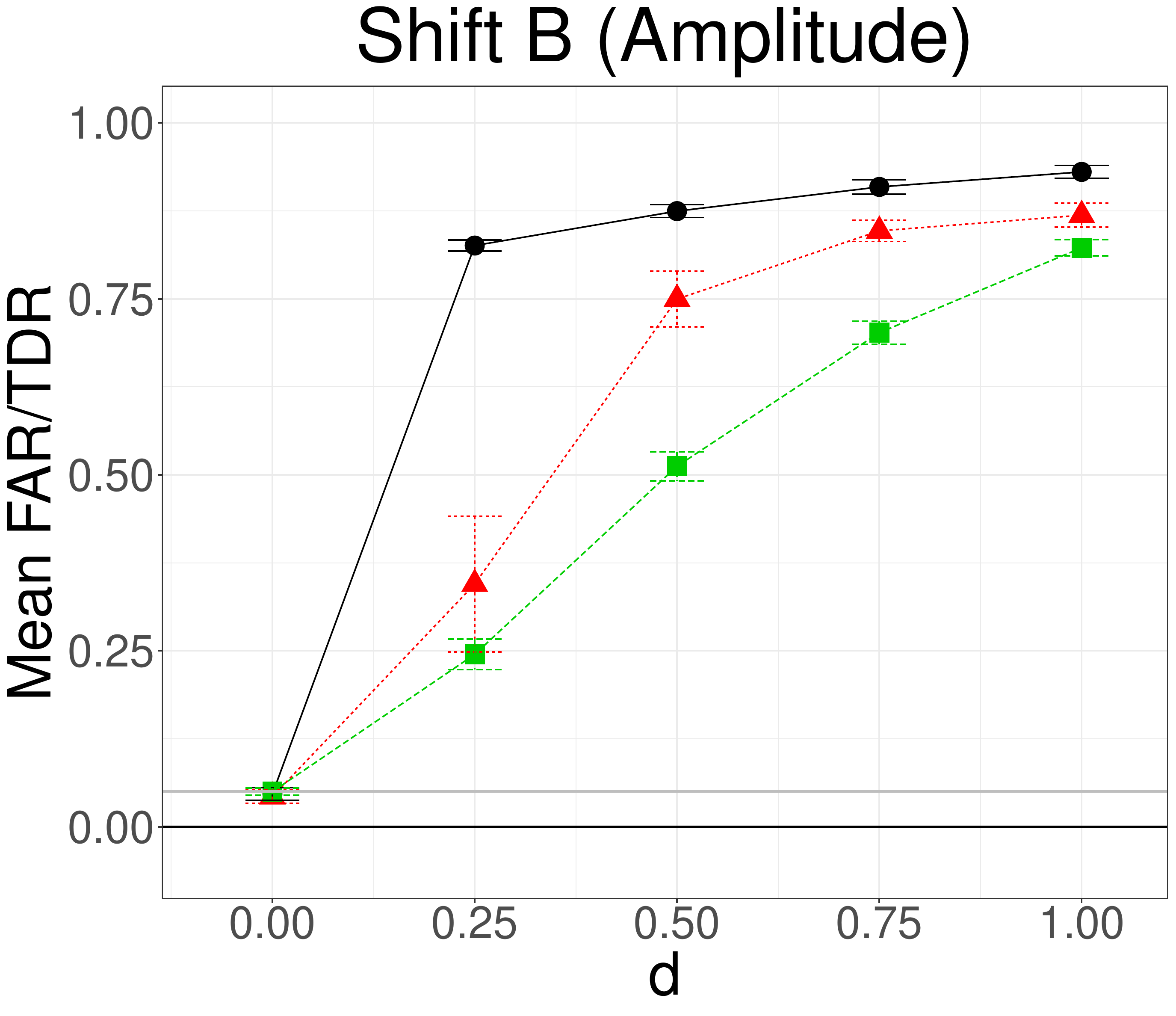}
		&\includegraphics[width=.3\textwidth]{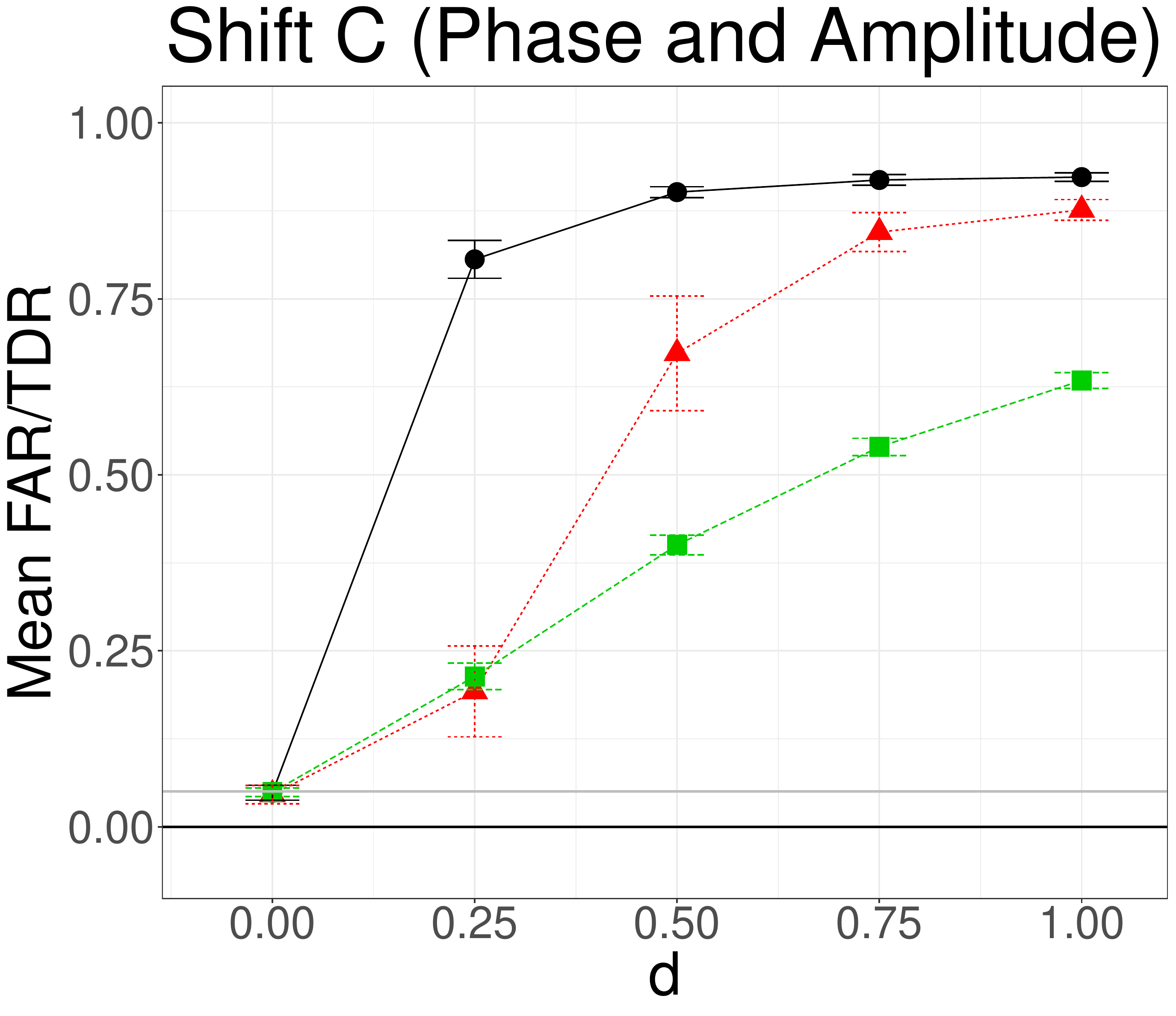}\\
		\textbf{M3}&\includegraphics[width=.3\textwidth]{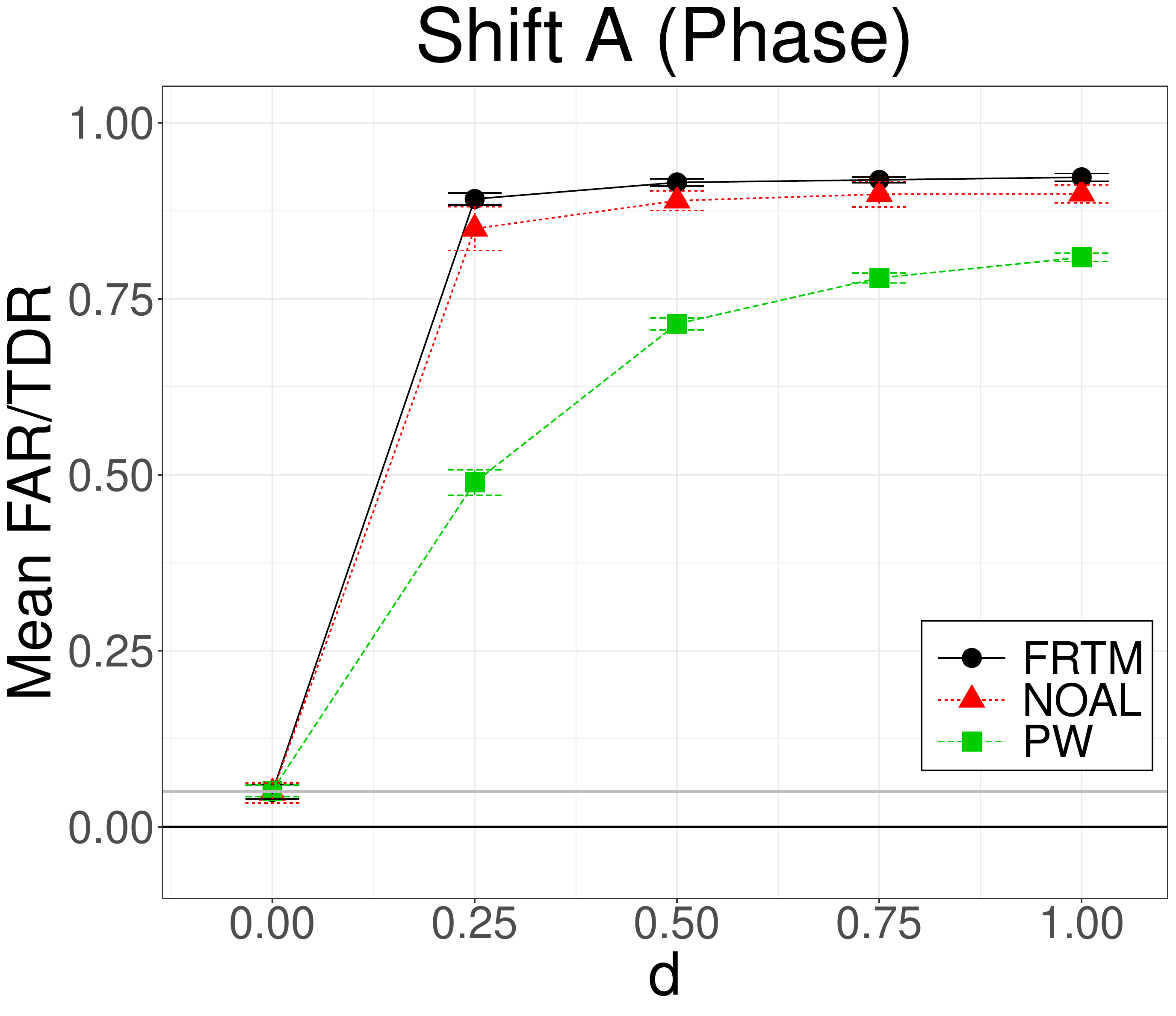}
		&\includegraphics[width=.3\textwidth]{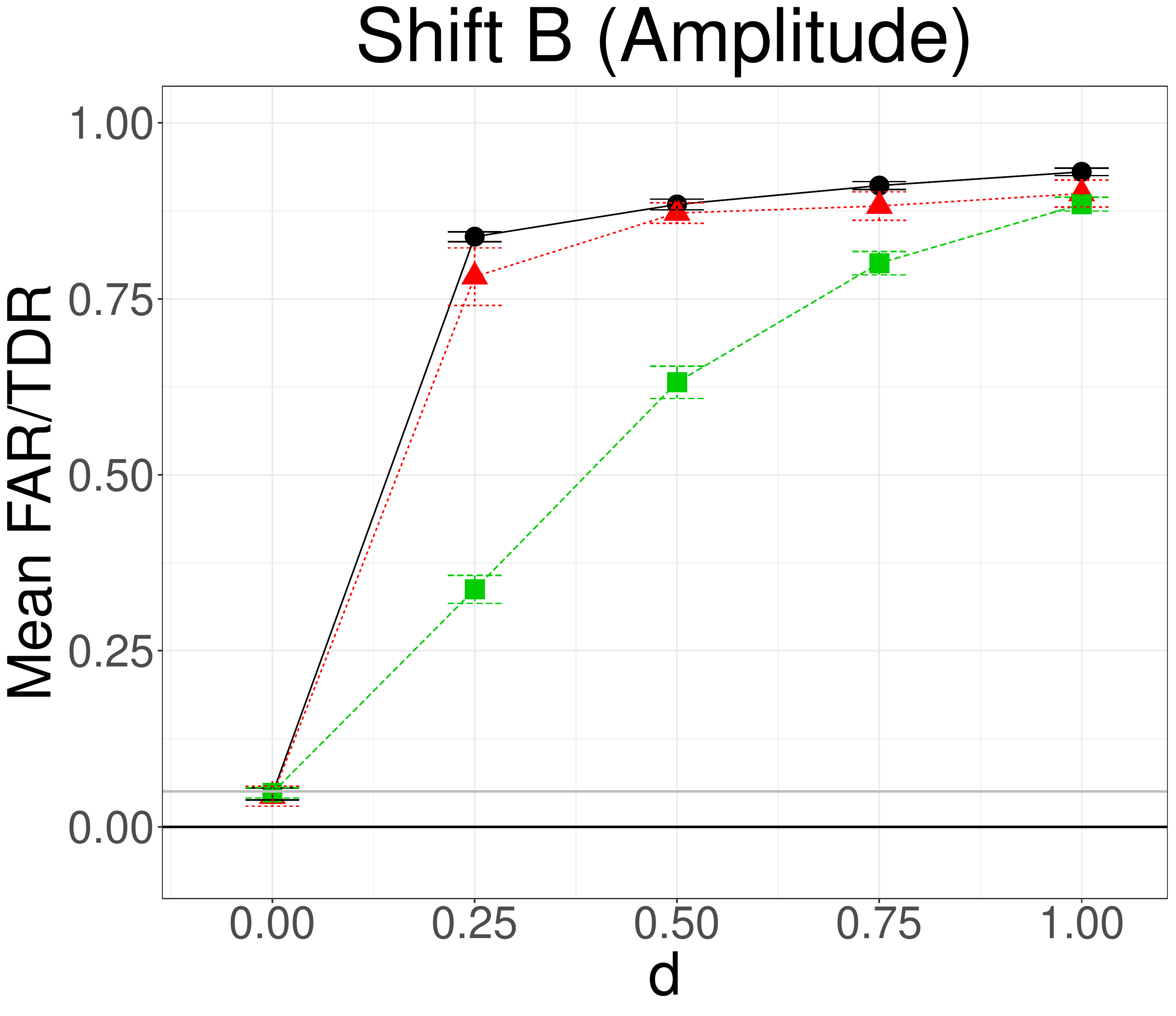}
		&\includegraphics[width=.3\textwidth]{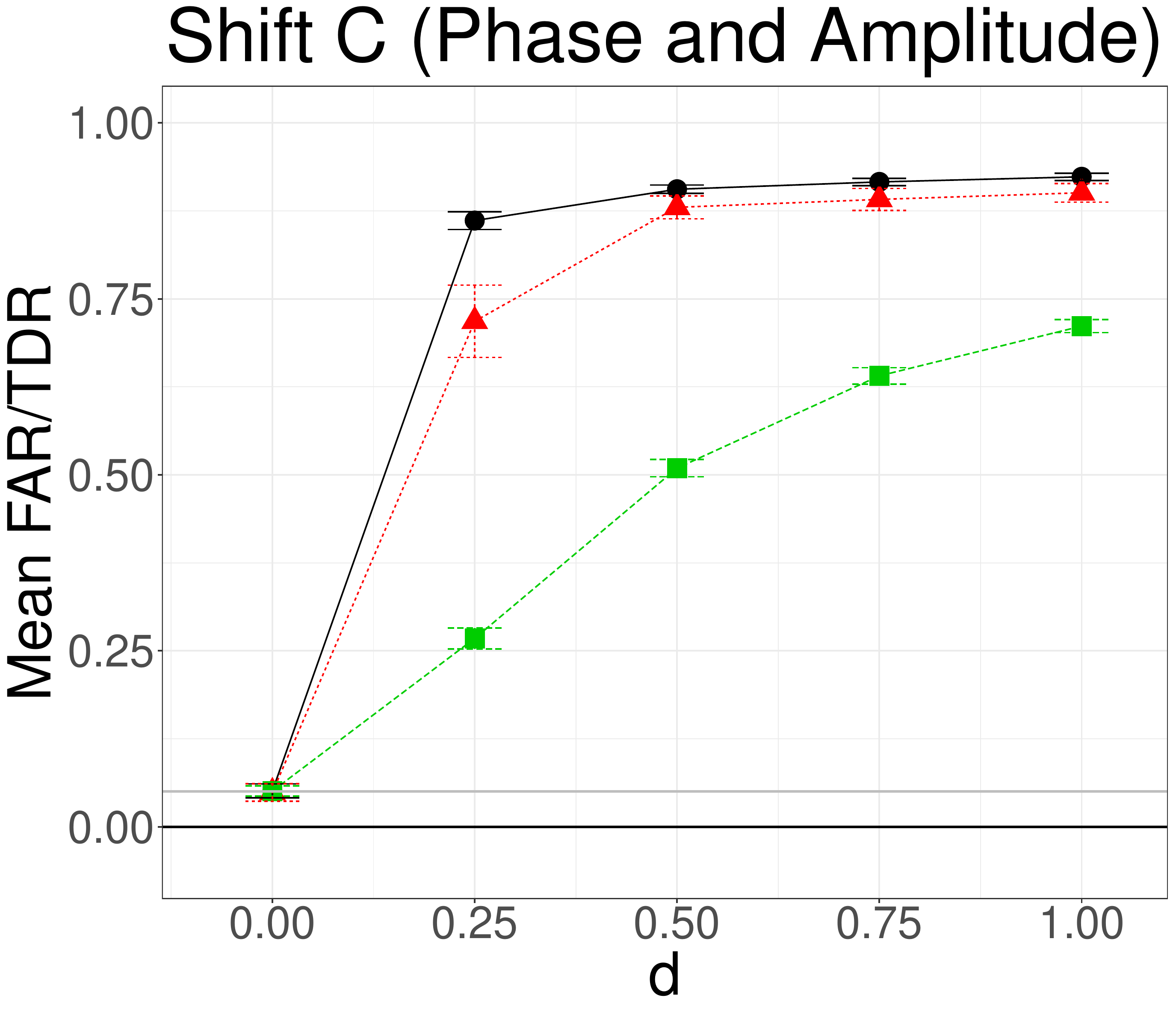}
	\end{tabular}
	\vspace{-.5cm}
\end{figure}

 Figure \ref{fi_results_1} and Figure \ref{fi_results_2}  display the mean FAR ($ d=0 $) or TDR ($ d\neq 0 $) for Scenario 1 and Scenario 2 respectively, as a function of  the  severity level $ d $     plus/minus one standard error  for each shift type and misalignment  M1, M2 and M3. Note that, the mean FAR  corresponding to IC in the first part of the process  for $ d\neq0 $ is not displayed here as it achieves  values comparable to those achieved for $ d=0 $.

 Figure \ref{fi_results_1} shows that all the methods successfully control the FAR  in Scenario 1  and  FRTM outperforms in terms of TDR both the NOAL and PW methods for all shift types and misalignments. As expected, the setting M1 with the largest phase variation  is the most favourable   for FRTM. In this case, for all   types of shifts, FRTM  outperforms the competing methods. The NOAL method is badly affected by large phase variation by showing particularly low detection rate performance, which is even comparable to that of the PW method. This comes from the fact that phase variation considerably influences the FPCA step   of the NOAL method by masking the presence of OC conditions.  On the contrary, FRTM is able to successfully deal with the function misalignment by separating  phase and amplitude components and considering both sources of variability in the monitoring procedure.
However, the performance difference between FRTM and the NOAL method decreases as the amplitude variation dominates the data variability. The NOAL method shows higher performance for misalignment M2 (i.e., mild amplitude/phase variation ratio)  and  TDR comparable with FRTM for misalignment M3 (i.e., low amplitude/phase variation ratio). It is interesting to note that the performance of FRTM is only slightly affected by the size of the phase variation, especially for large severity levels. This is particularly true for Shift B, where FRTM performance is almost the same for misalignment M1, M2, and M3. Moreover, for low values of misalignment, the higher complexity of FRTM  method is not reflected in a lower OC detection performance, which emphasizes the ability of FRTM to deal with a wide range of phase variation scenarios.

As noted in Figure \ref{fi_results_2}, also in Scenario 2, where the data  are generated with more complex phase components, FRTM performance is much higher than that of the competing methods.
However, the greater complexity of the phase component is reflected in a slightly lower detection power of FRTM,  mainly for misalignment M1.  Indeed, in Scenario 2, the largest mean TDRs  achieved by FRTM correspond to  Shift B and Shift C, where  OC condition affects the amplitude component. Whereas, for Shift A, where the OC condition arises   in the phase component alone, FRTM does not achieve  as good performance as  in Scenario 1.
The  ability of FRTM to successfully separate and then, combine the amplitude and phase components  makes it the best  performing real-time monitoring scheme for all the  considered settings.

\section{Data Example: Batch Monitoring of a Penicillin Production Process}
\label{se_casestudy}
To demonstrate the potential of FRTM in practical situations, in this section, we present an example  that addresses the issue of monitoring batches from a  penicillin production process.
	Batch  processes play an important role in the production and processing of high-quality speciality materials and products that appear  in the manufacturing, food and medicine industries among others \citep{nomikos1995multi,nomikos1995multivariate,spooner2017selecting,spooner2018monitoring}. A batch process is a finite-duration process where quantities of raw materials are subjected to a sequence of steps and conditions  that transform them into the final product.
Abnormal conditions that develop during these  operations can lead to  poor quality production.
Therefore, real-time SPC can be successfully employed     to detect and indicate OC conditions that can be possibly corrected prior to the batch completion. 
%However, batch processes present a great  challenge for monitoring, due to inherent non-stationarity, finite duration, non-linear response, and batch-to-batch variability.

The study considers the extensive dataset provided by \cite{van2015extensive}, generated from the morphological model of \cite{birol2002morphologically} for the  penicillin production in fed-batch cultivations. It describes the growth of biomass and production of penicillin in a fed-batch reactor, where the fermentation is operated in both batch and fed-batch modes. Further details about the penicillin production process can be found in \cite{birol2002modular, birol2002morphologically}.
Specifically, we focus on four datasets, named D1, D2, D3 and D4, that represent different assumptions about the production process. D1 represents the basic case where all initial conditions are well modeled by a normal distribution. D2 considers other non-Gaussian distributions for the initial conditions.  D3 includes also a batch-to-batch variation of the model parameters in addition to non-Gaussian initial conditions. Finally, D4 is similar to D3 with the added complexity of multi-modality due to two different strains in the simulation.
Each dataset consists of 400 NOC batches and several thousand  faulty batches where many  variables are measured online every 12 minutes throughout each batch. In this  example, we focus on the  monitoring of two  quality characteristics in real-time that are representative of the process state, considered in the first 100 hours of  production; namely,  the \textit{dissolved O\textsubscript{2} concentration}, indicated with O\textsubscript{2} and  measured in $mg/L$, and the \textit{base flow rate}, indicated with BFR and measured in $ mL/h $. 
 In  Supplementary Materials D, the plot of 400 IC observations of these two quality characteristics  is shown for each dataset, where   the presence of a significant proportion of phase variation is highlighted. Therefore, we expect  FRTM to perform better than the NOAL method, which has been shown in Section \ref{se_perfo} to be particularly sensitive  to the presence of phase variation. 
The faulty batches consist of 15 different fault types,  each with different fault magnitudes representative of a wide range of OC states (see \cite{van2015extensive} for further details on fault types). Specifically, 50 repetitions of each dataset/type/magnitude combination are available for a total of 3000 faulty batches, where the onset times of the OC conditions arise randomly during the process.
FRTM is implemented as in Section \ref{se_perfo}, with $ s^{min}=1$, $ s^{max}=1000 $, and the  value of $ \lambda $  in Equation \eqref{eq_fdtw}  chosen through the  average curve distance (Equation \eqref{eq_acd}) with $ \delta=0.05 $. The band constraint is implemented with $ b=0.01 $ and a  number of retained principal components $ L $ that explains at least $ 90\% $ of the total variability.  Moreover, the empirical quantile of the $ T^2 $ and $ SPE $ statistics are obtained via the kernel density estimation approach. As in the simulation study in Section \ref{se_perfo},  $\alpha$  is set equal to $ 0.05 $.

\begin{figure}[h]
	\caption{Mean TDR over the different fault magnitudes against fault type (1-15) achieved by FRTM, NOAL and PW for the O\textsubscript{2} and the BFR for  each dataset (D1, D2, D3, and D4).}
	
	\label{fi_results_cs}

	\hspace{-2.2cm}
	\begin{tabular}{cM{0.27\textwidth}M{0.27\textwidth}M{0.27\textwidth}M{0.27\textwidth}}
		\textbf{O\textsubscript{2}}&\includegraphics[width=.280\textwidth]{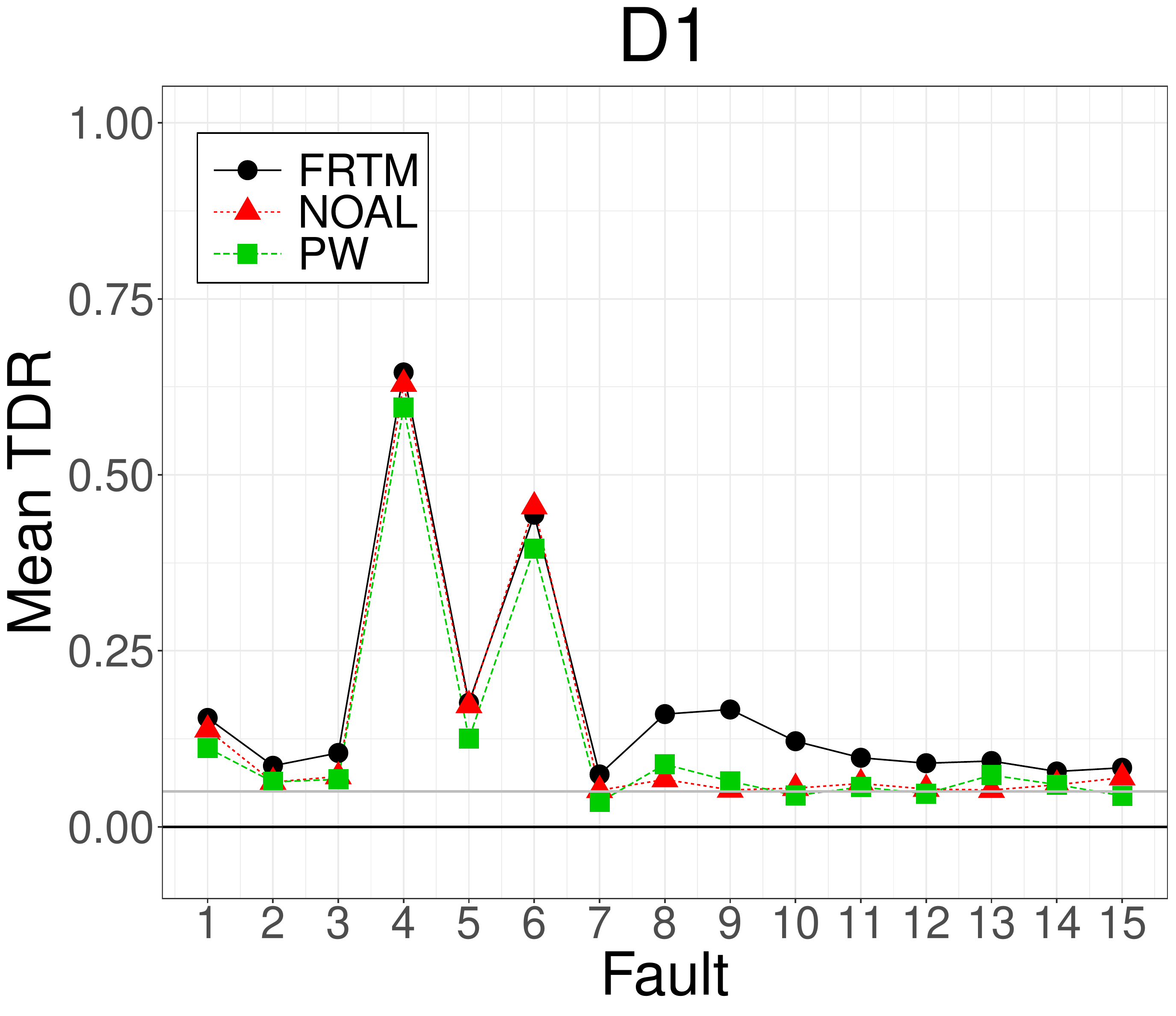}
		&\includegraphics[width=.280\textwidth]{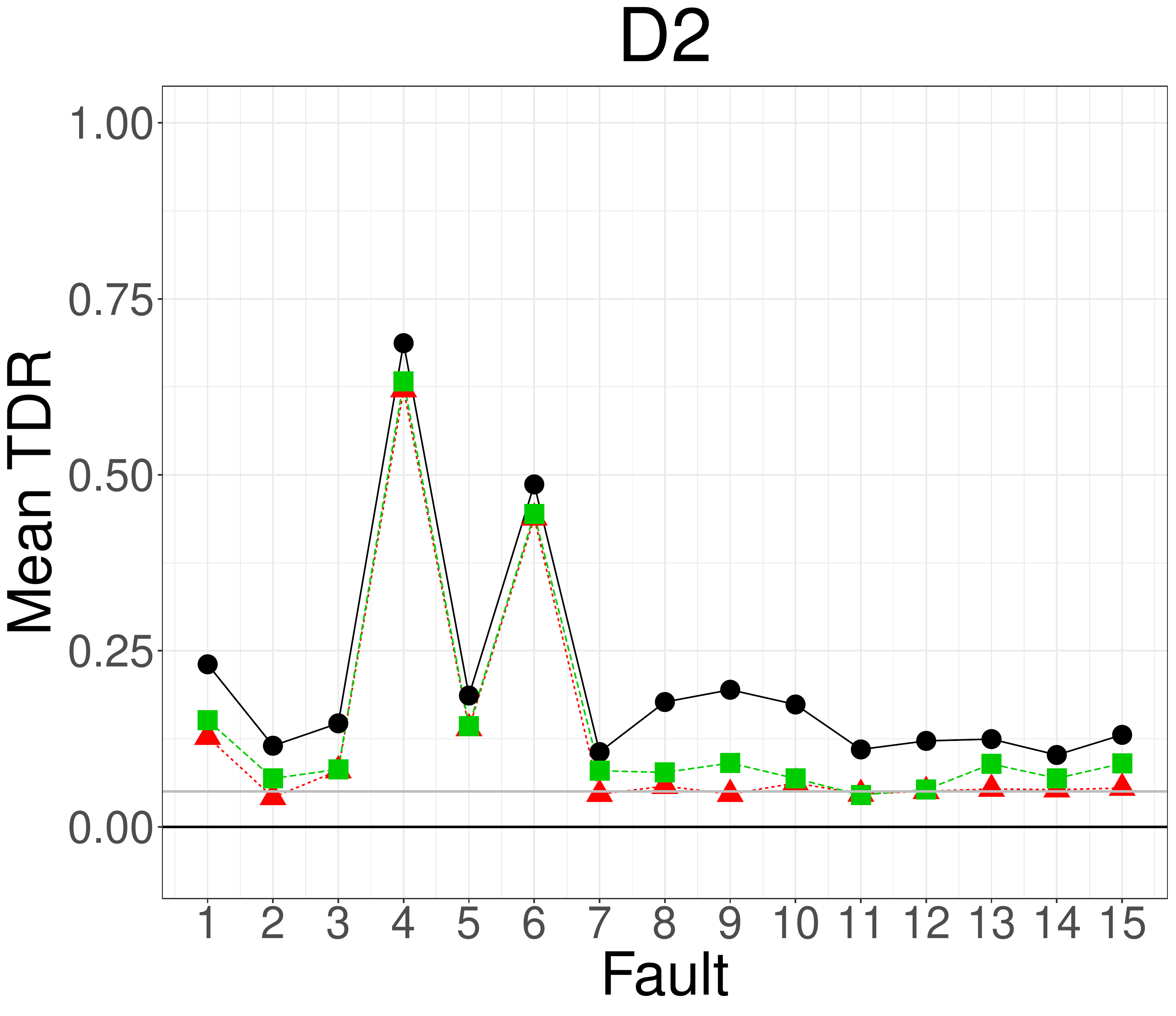}&\includegraphics[width=.280\textwidth]{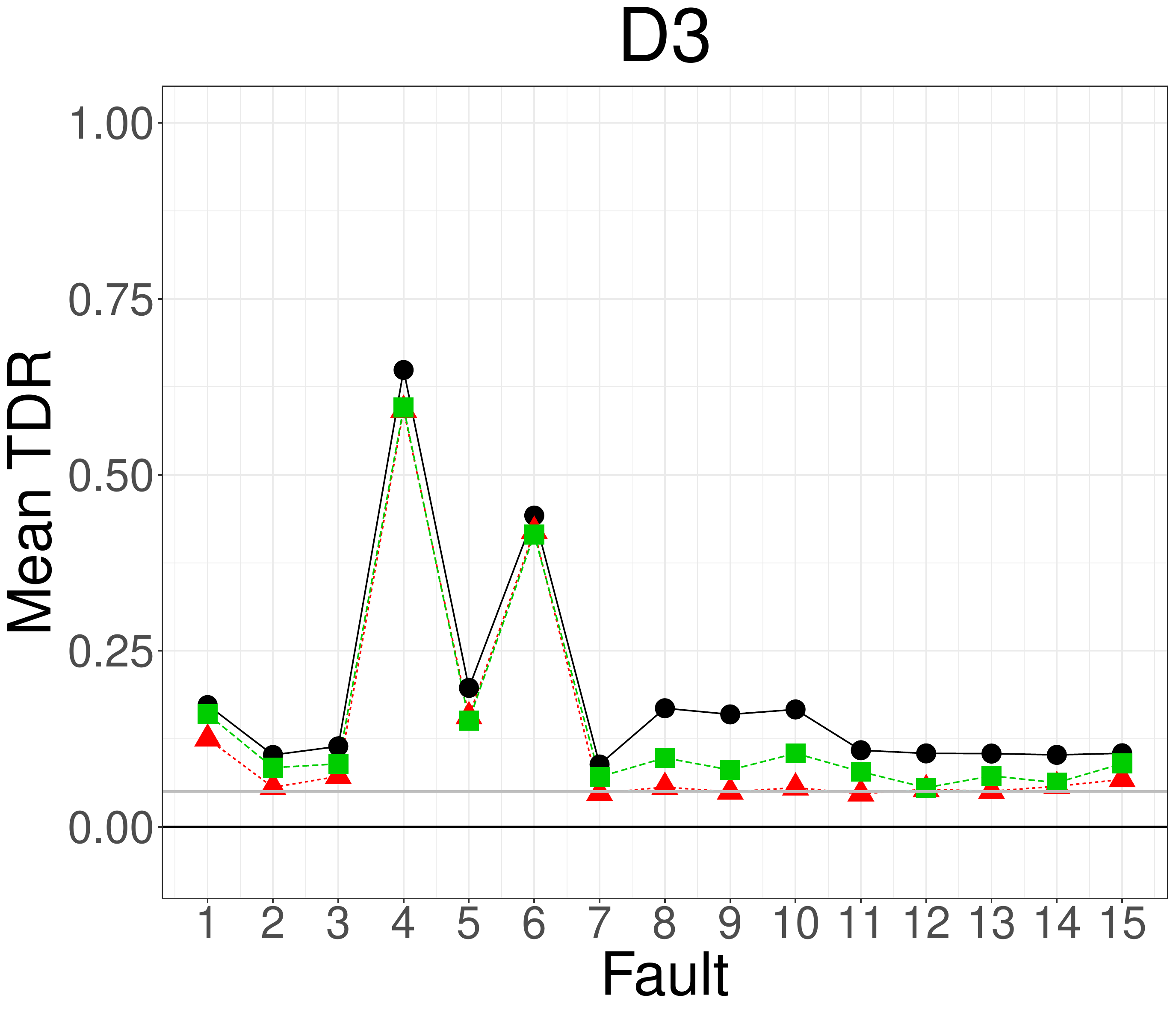}	&
		\includegraphics[width=.280\textwidth]{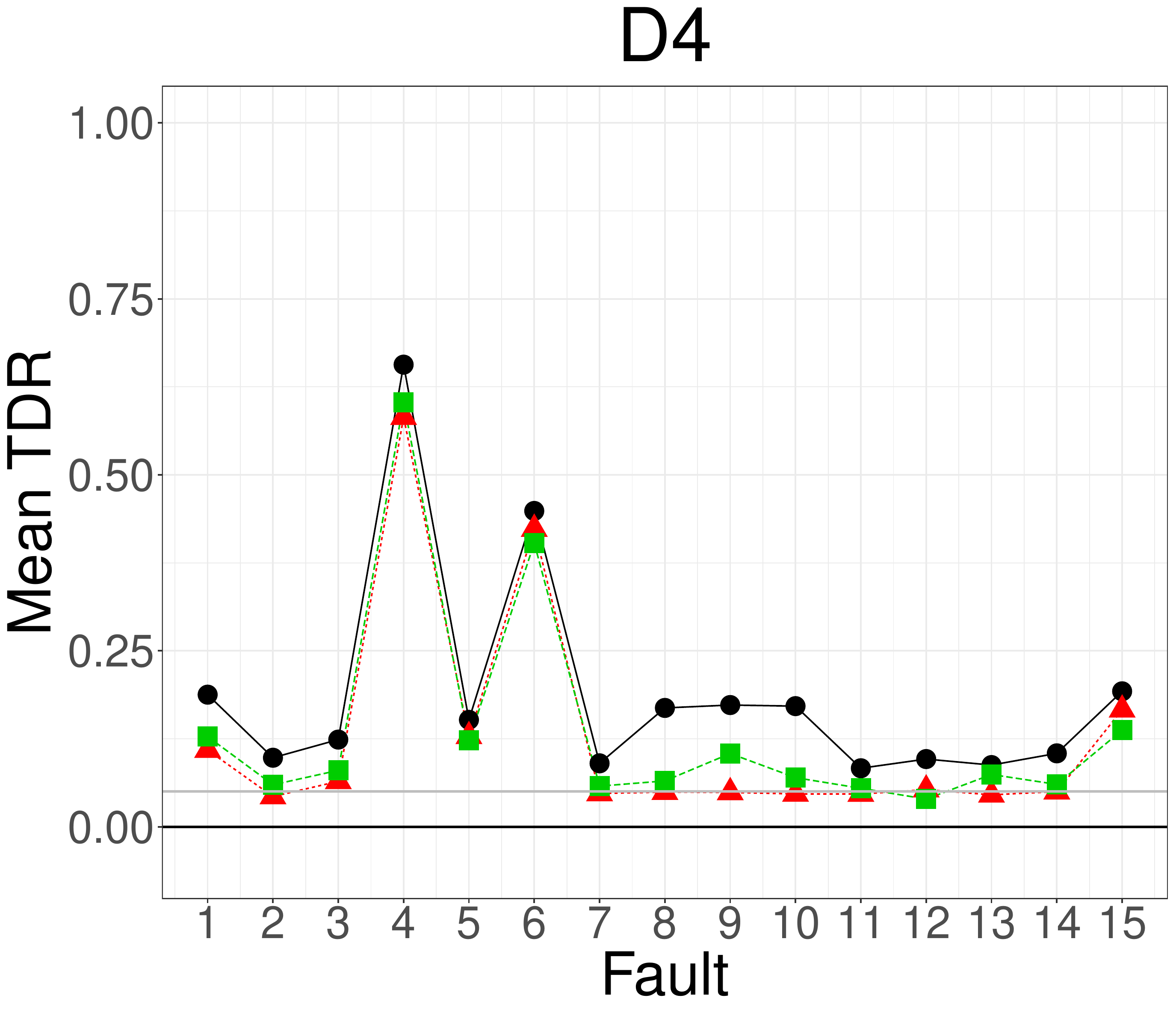}\\
		\textbf{BFR}& \includegraphics[width=.280\textwidth]{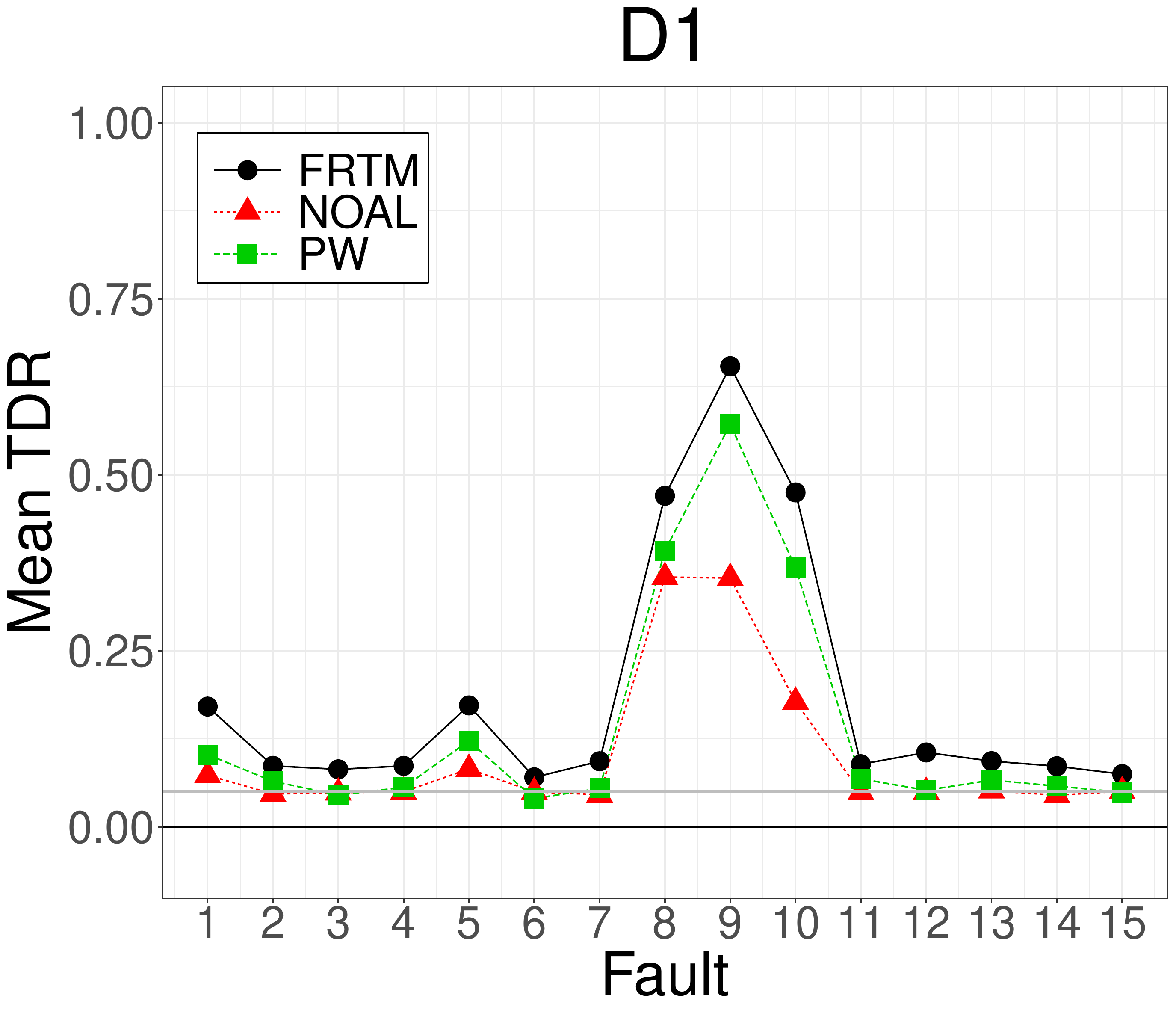}
		&\includegraphics[width=.280\textwidth]{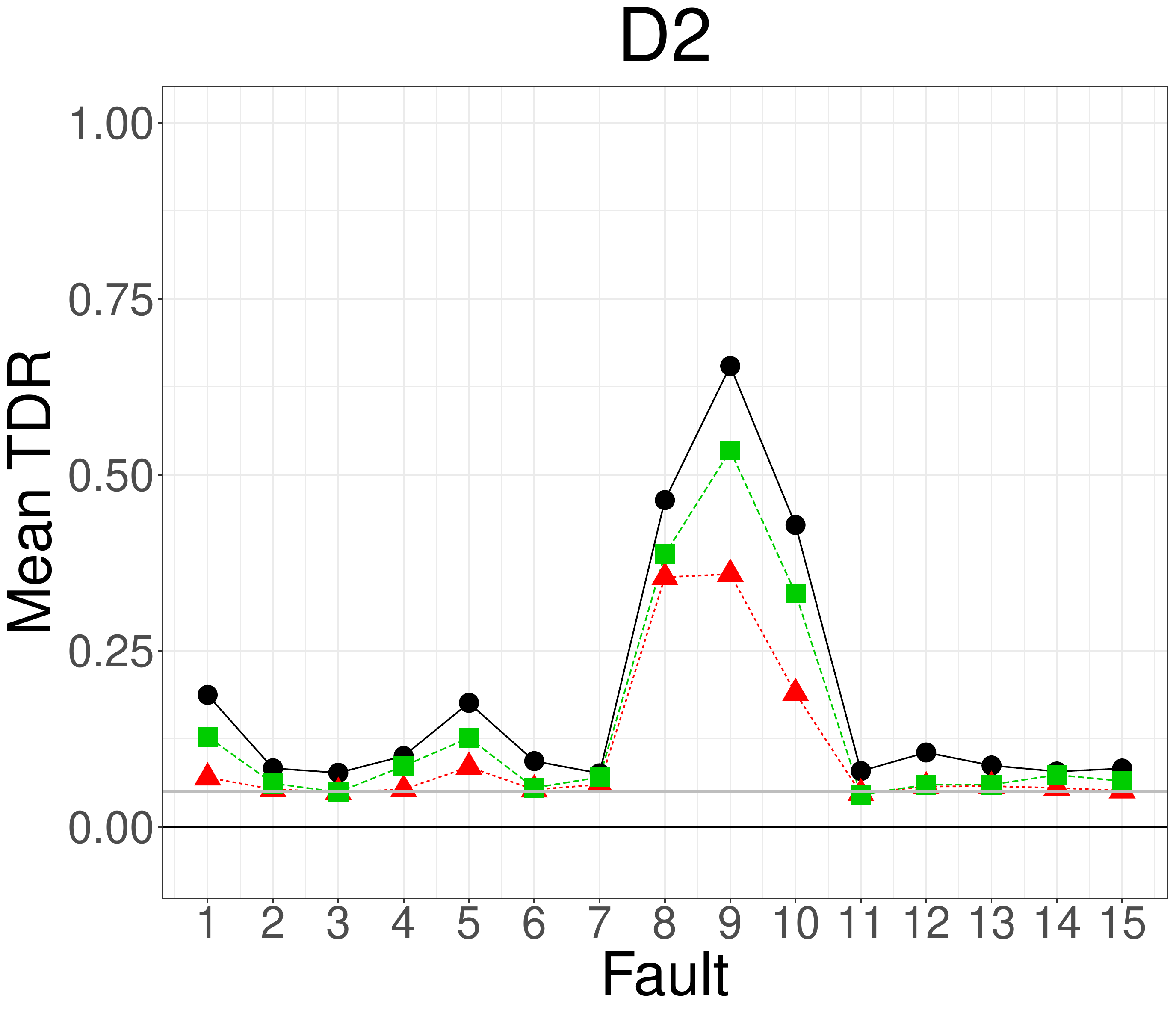}& \includegraphics[width=.280\textwidth]{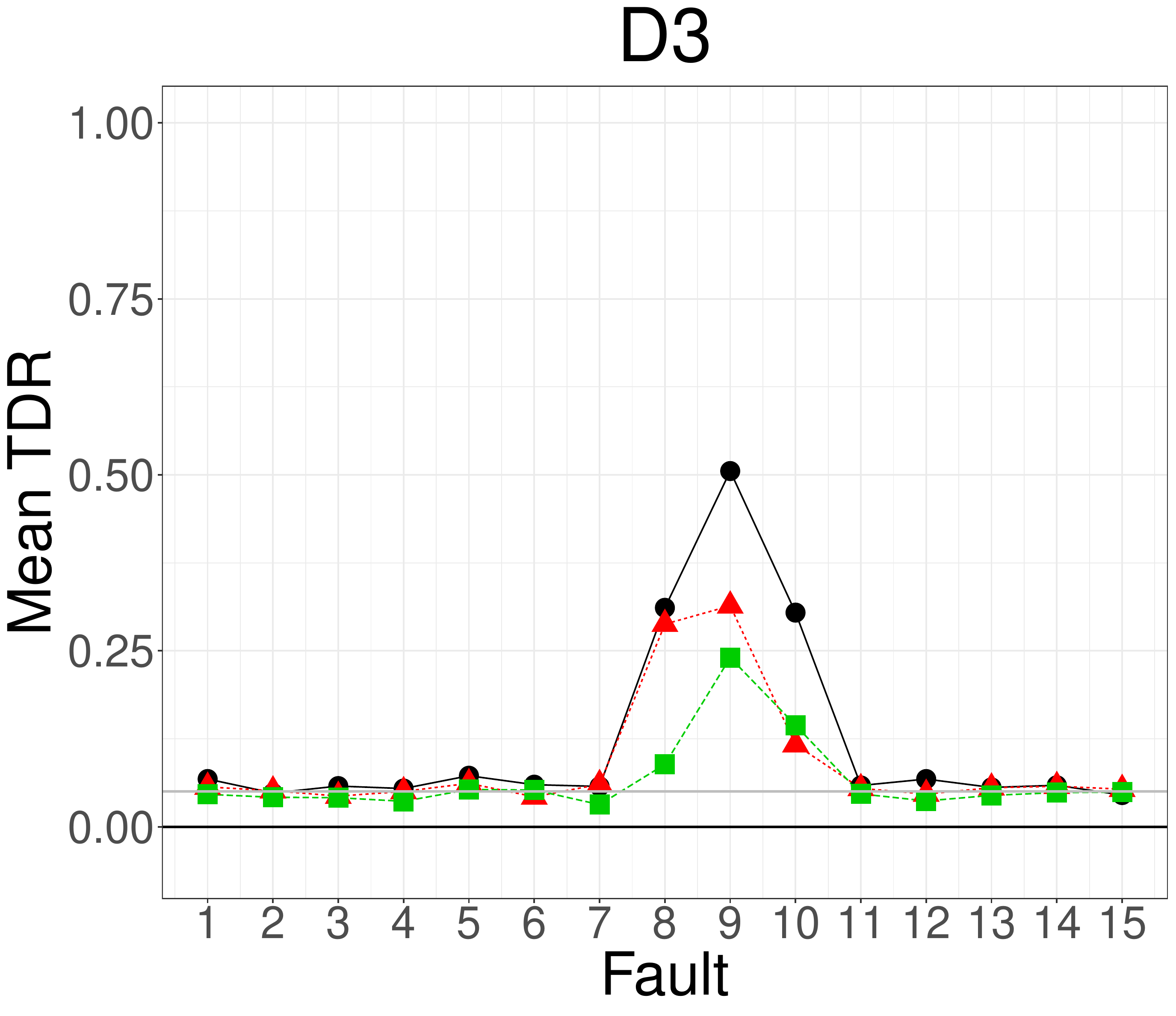}
		&\includegraphics[width=.280\textwidth]{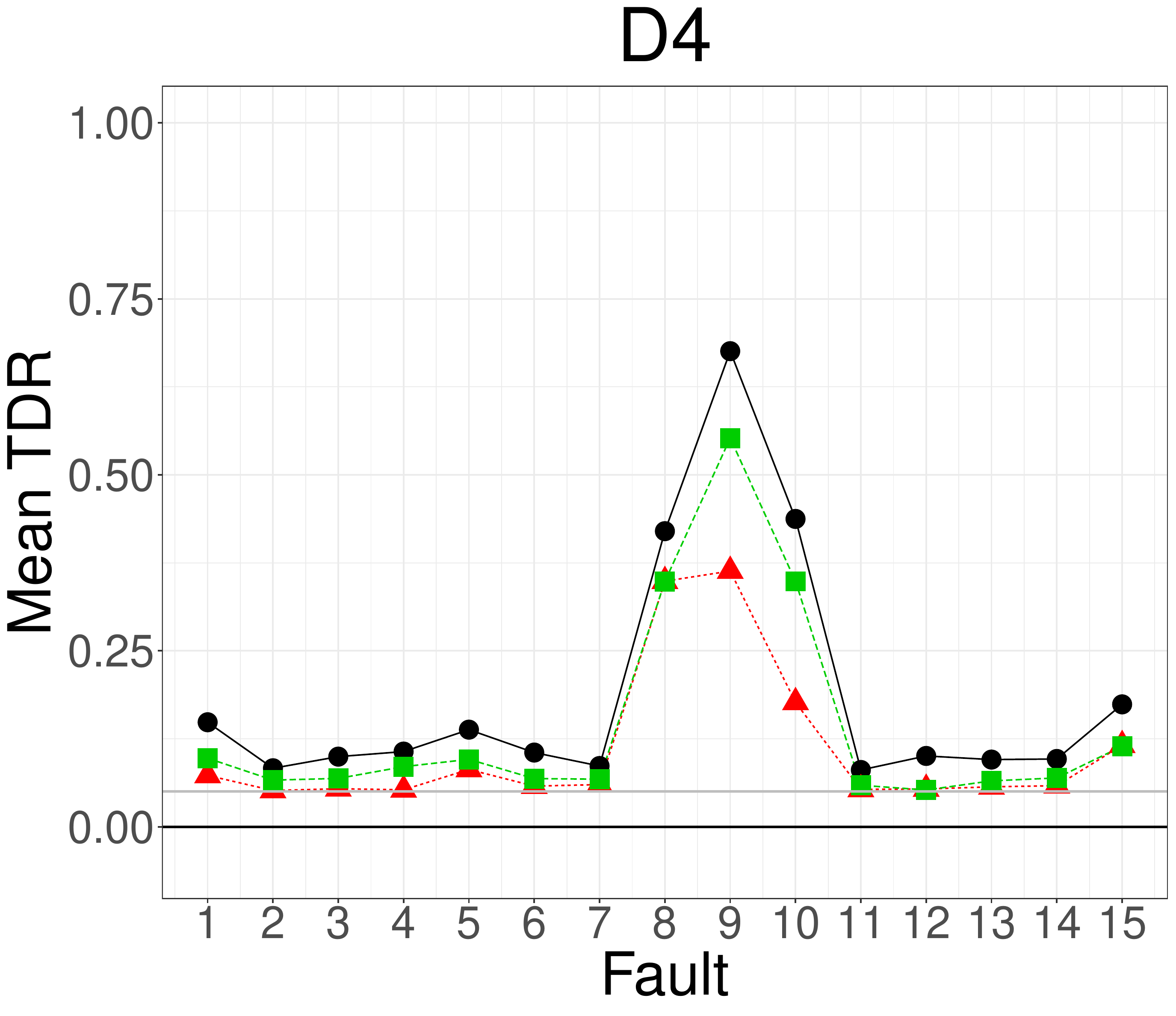}\\
	\end{tabular}
	\vspace{-.5cm}
\end{figure}

In  Figure \ref{fi_results_cs}, the mean TDR of FRTM and the competing methods applied to the O\textsubscript{2} and the BFR (row-wise) are plotted against the 15 fault types for each dataset (column-wise). Note that the mean TDR at each fault type is obtained by averaging across all fault magnitudes.
%,  disaggregated values are presented in  Supplementary Materials.
 Although the FDR is not explicitly shown here, we confirm that all  methods achieve values that are close to nominal $ \alpha $. Thus, they can be fairly compared in terms of TDR.
FRTM achieves the largest mean TDR for almost all the considered settings.
Specifically, in the monitoring of O\textsubscript{2}, FRTM outperforms  competitors in detecting faults 1-3 and 7-15 that  produce  moderate values of TDR. This behaviour reflects the fact that  FRTM  is more inclined to detect small OC conditions  than the PW method, due to its functional nature able to account for the whole curve pattern.  The NOAL method, which should also have benefited from its functional nature, in this setting  is badly affected by the presence of phase variation and, thus, shows  worse performance than FRTM.
%, which, on the contrary, is able to successfully deal with the presence of phase variation.
For faults 4-6, the performance of FRTM and the competing methods are more similar even though the former  is always the best one.
The performance of the three methods slightly changes  from D1 to D4. 
In the monitoring of the BFR,  the considered methods show the largest mean TDR for faults 8-10. Also in this case, the NOAL method is particularly affected by the presence of phase variation.
For faults 1-7 and 11-15, which represent OC conditions of moderate intensity, FRTM achieves the best performance.

\begin{figure}
	\caption{$T_{2}$ and $SPE$  control charts based on FRTM and NOAL methods,  and the PW control chart corresponding to five batches  randomly sampled from the quality characteristic BFR of the dataset D4. The thick solid  lines correspond to 	 control limits.}
	\label{fi_cs_exa}
	
	\resizebox{\textwidth}{!}{
		\begin{tabular}{M{0.3\textwidth}M{0.3\textwidth}M{0.3\textwidth}}
			\footnotesize{FRTM}&\footnotesize{NOAL}&\footnotesize{PW}\\
			\includegraphics[width=0.3\textwidth]{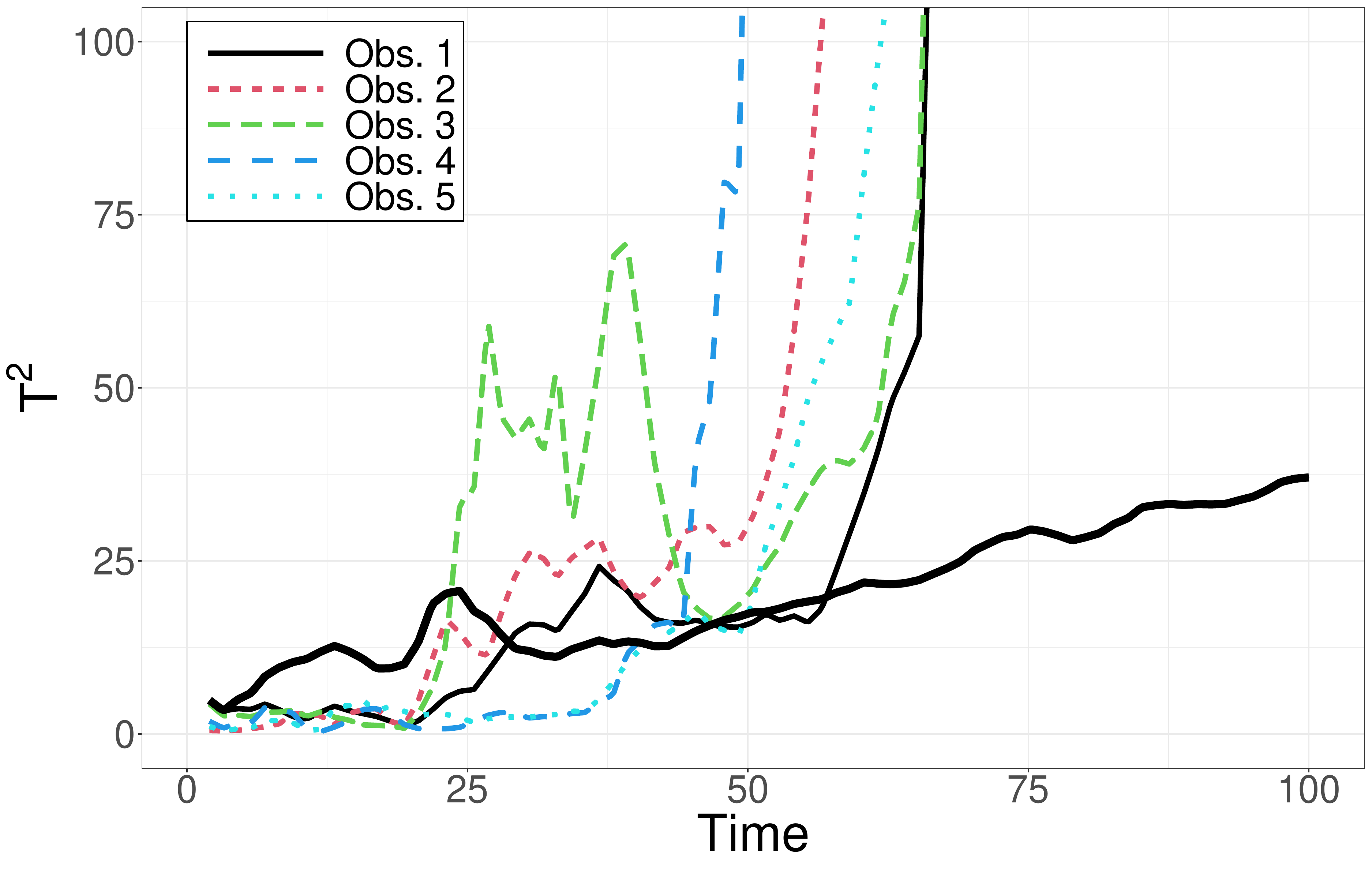}&\includegraphics[width=0.3\textwidth]{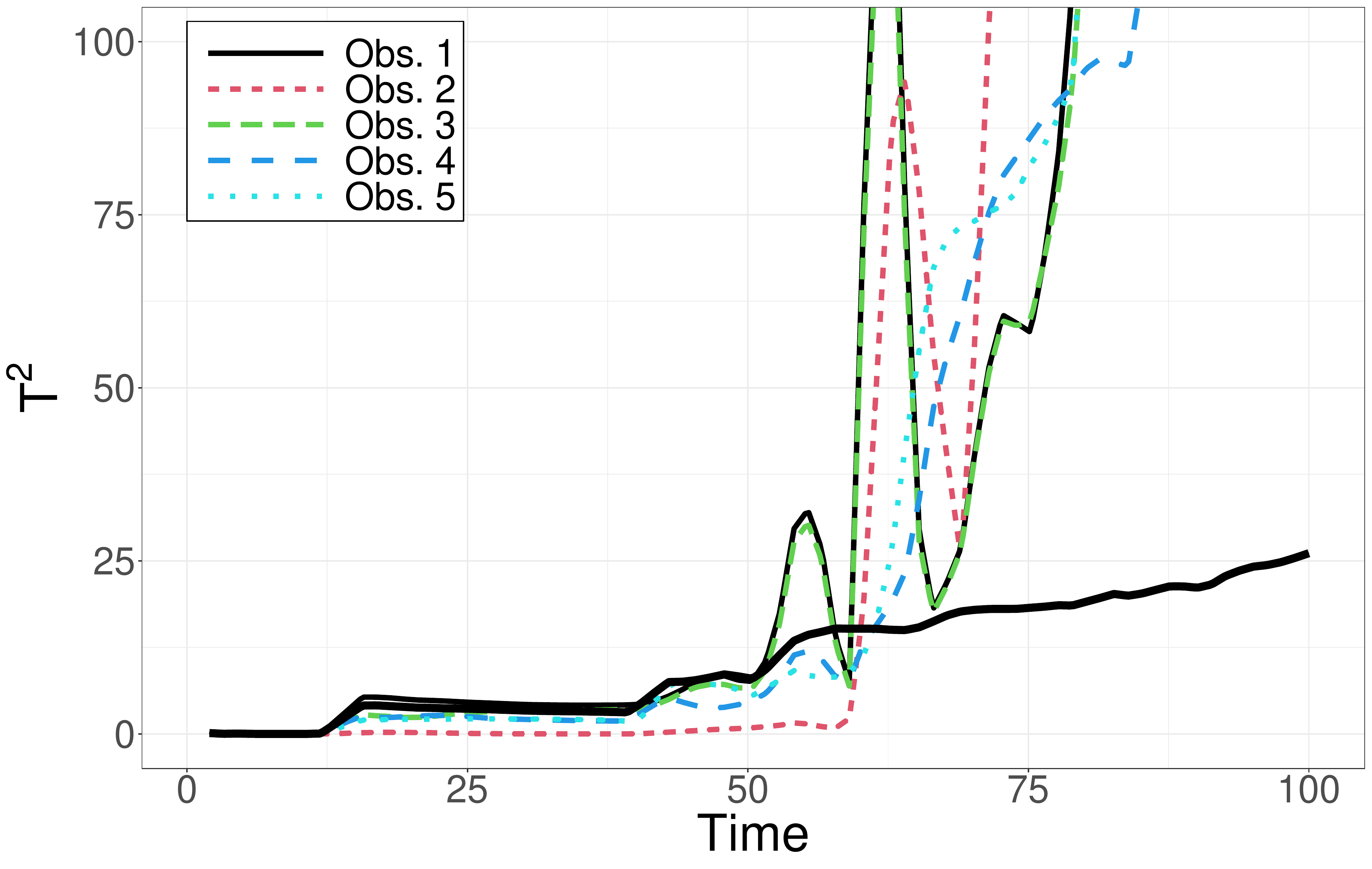}&\multirow{2}{*}{\includegraphics[width=0.3\textwidth]{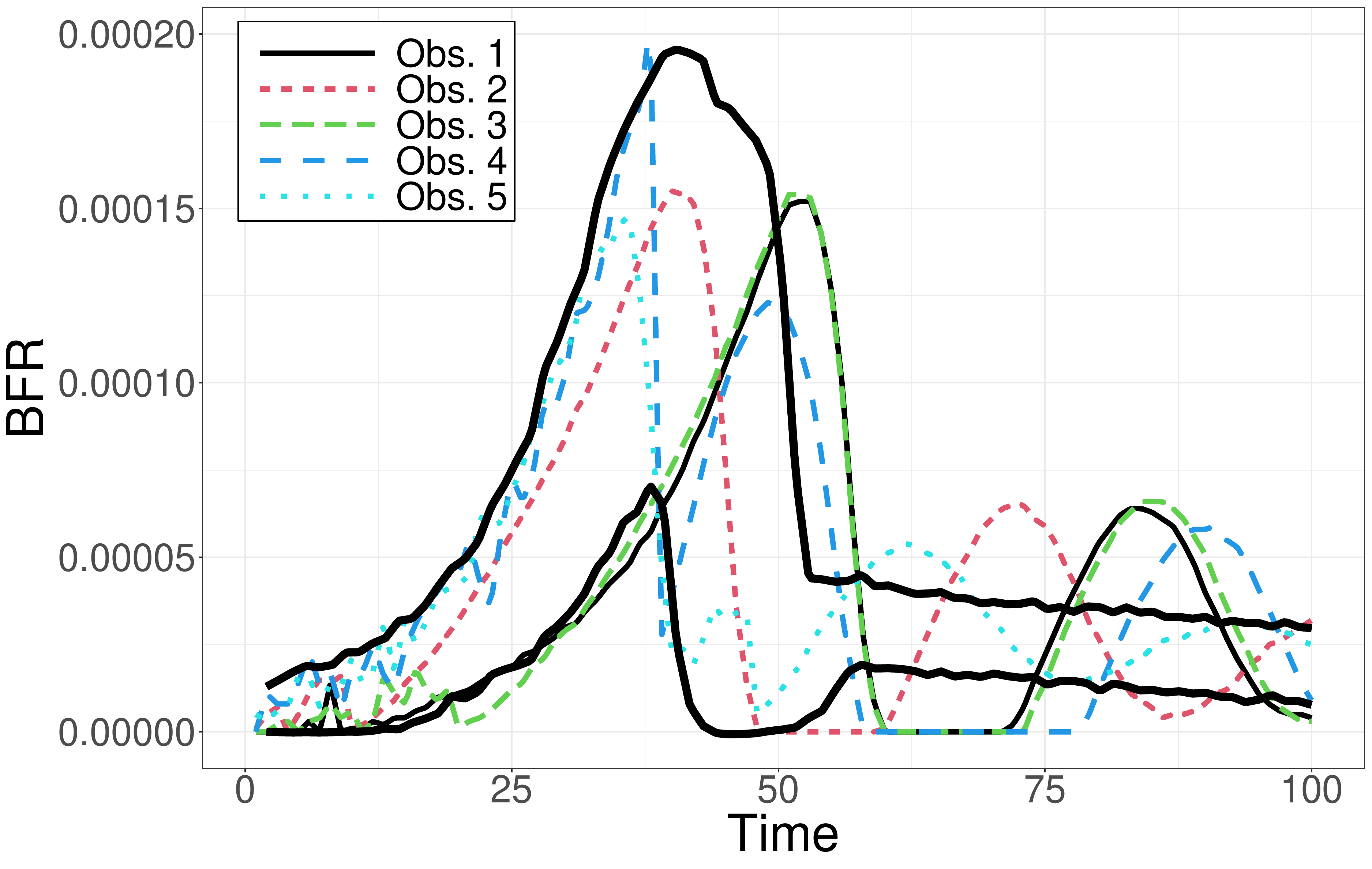}}\\
			\includegraphics[width=0.3\textwidth]{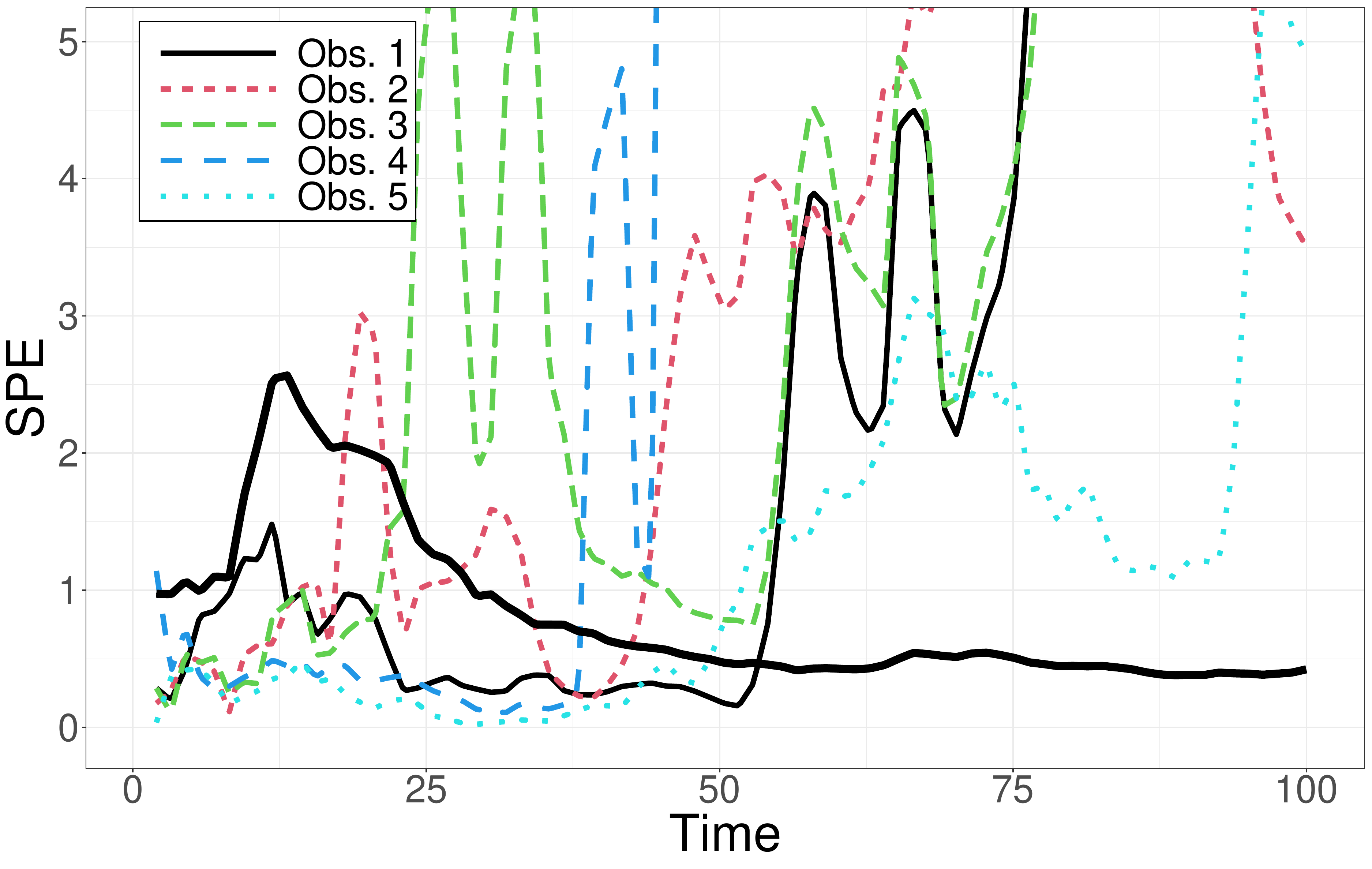}&\includegraphics[width=0.3\textwidth]{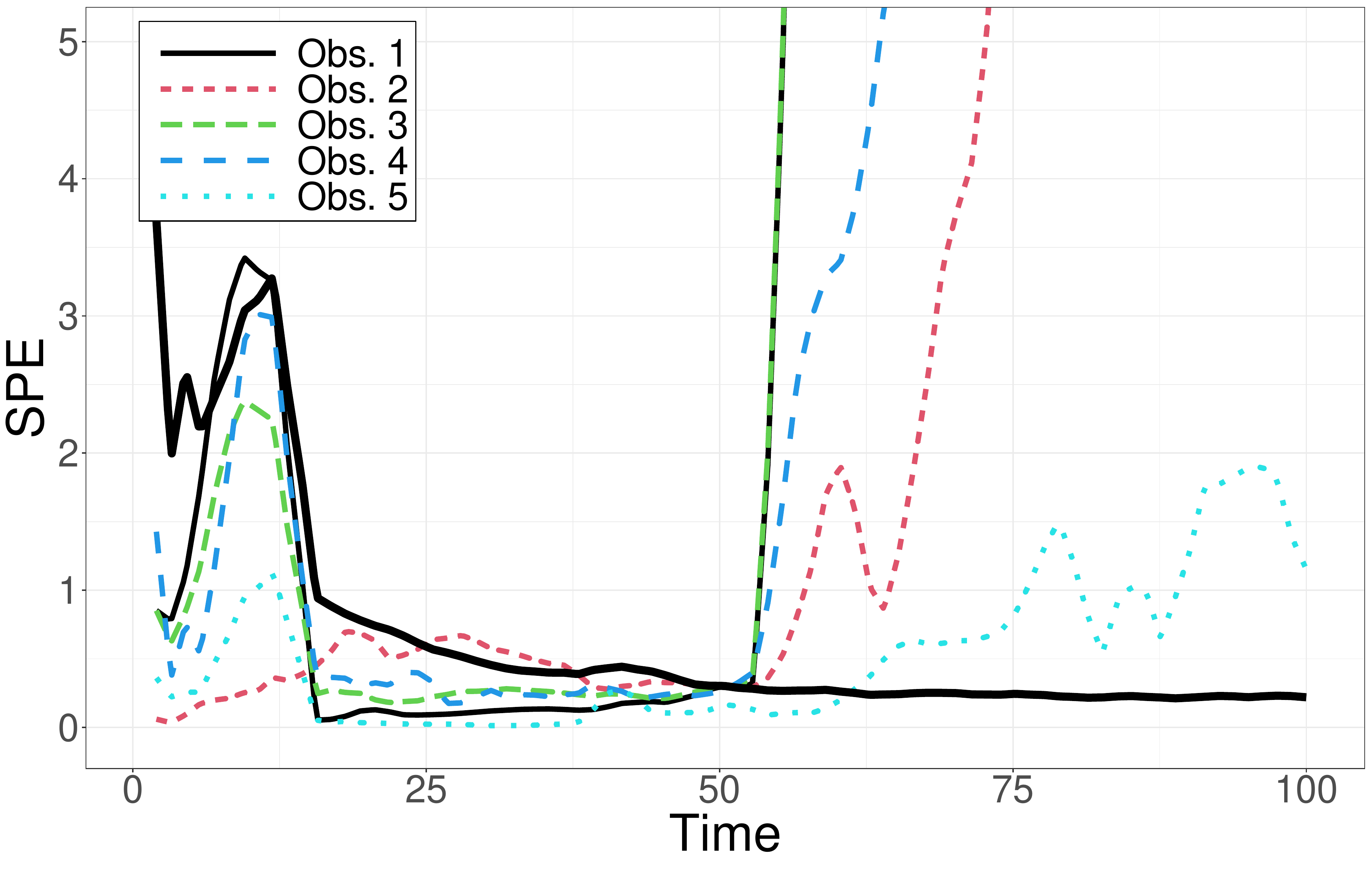}&\\
		\end{tabular}}
	
	\vspace{-1cm}
\end{figure}

For illustrative purposes, Figure \ref{fi_cs_exa} reports the application of the considered methods to monitor  five batches  randomly sampled from the quality characteristic BFR of the dataset D4. 
Also in this case, the mean TDR of FRTM is definitely larger than that of the NOAL and  PW methods and confirms the results of Figure \ref{fi_results_cs}.
As an example, by looking at the third observation, we clearly see that   FRTM signals the OC condition  much earlier  than the competing methods.
%\begin{figure}
%	\caption{$T_{2}$ and $SPE$  control charts for  FRTM and NOAL method,  and the PW control chart corresponding to five batches  randomly sampled from D4 of the BFR. The thick solid  lines corresponds to the control limits.}
%	\label{fi_cs_exa}
%
%	\resizebox{\textwidth}{!}{
%		\begin{tabular}{cM{0.45\textwidth}M{0.45\textwidth}}
%			\footnotesize{FRTM}&\includegraphics[width=0.45\textwidth]{Figures/cs_exa_FRTM_T2}
%			&\includegraphics[width=0.45\textwidth]{Figures/cs_exa_FRTM_SPE}\\
%			\footnotesize{NOAL}&\includegraphics[width=0.45\textwidth]{Figures/cs_exa_NOAL_T2}
%			&\includegraphics[width=0.45\textwidth]{Figures/cs_exa_NOAL_SPE}\\
%			\footnotesize{PW}&\multicolumn{2}{M{\textwidth}}{\includegraphics[width=0.45\textwidth]{Figures/cs_exa_PW.pdf}}
%	\end{tabular}}
%	
%	\vspace{-1cm}
%\end{figure}
\section{Conclusions}
\label{se_conclusions}
In this paper, we propose a new method for  monitoring a functional quality characteristic in real-time, referred to as \textit{functional real-time monitoring} (FRTM).
FRTM is designed to assess the  presence of assignable causes of variation before the process completion by taking into account both the phase and amplitude sources of variation. 
For instance, this could be of interest for long and expensive processes.
% where the real-time knowledge of the process state may be promptly corrected and yield cost reduction. 
Specifically, FRTM is based on three main elements: a curve registration  based on an open-end/open-begin version of the functional dynamic time warping, a dimensionality reduction through a new version of the mixed functional principal component analysis, and a monitoring strategy based on  Hotelling's $ T^2 $ and $ SPE $ control charts. 
%Starting from a reference sample, which is needed to characterize  the IC state and set up the control charting scheme (Phase I),	these elements are combined  in a Phase II  monitoring method,  where the partially observed  functional quality characteristic is firstly registered to the template function, then,  its amplitude and phase components are projected on an appropriate finite dimensional space, and, finally, the $ T^2 $ and $ SPE $ statistics are computed and compared with the corresponding control limits.
To the best of the authors' knowledge, FRTM is the first monitoring scheme that is able to  monitor a functional quality characteristic in real-time by successfully combining the information in both the amplitude and phase components.
Indeed,  methods already present in the literature  either skip the registration step overlooking the phase component or  consider the phase variability as a nuisance effect to be removed. Both approaches have drawbacks, with the former risking  producing low power procedures because phase variation could mask the true data structure, while the latter may not be able to  account for OC conditions that arise in the phase component.

The performance of FRTM is assessed through an extensive Monte Carlo simulation study where it is compared with two competing real-time monitoring methods, named NOAL and PW, which overlook  the presence of phase variation and the functional nature of the quality characteristics, respectively. Conversely, the ability of the proposed method to firstly separate and then combine  amplitude and phase variations bring FRTM to  outperform  competitors in all  considered scenarios, which covered a variety of types of phase components  and OC conditions. 
Lastly, the practical applicability of the proposed method is illustrated through a data example,  which addresses the issue of monitoring batches from a  penicillin production process. Also in this case, FRTM shows better performance than the competitors  in the identification of OC condition of the \textit{dissolved O\textsubscript{2} concentration}, and the \textit{base flow rate}.

Future research could be performed to extend the proposed framework to the case of multivariate functional data, which poses non trivial challenges from both  computational and methodological points of view. To further improve the detection power of FRTM, the uncertainty in the estimation of the warping components could be included in the dimensionality reduction phase and an optimality criterion could be developed to better choose the weights in the mFPCA.

\section*{Supplementary Materials}
Supplementary Materials contain additional details about the proposed method (A.1-A.3), details about the data generation process in the simulation study (B), additional simulation results (C), and additional plots for the data example (D), as well as the R code  to reproduce graphics and results over competing methods in the simulation study.

\bibliographystyle{chicago}
{\small
\bibliography{References}}

\end{document}

\def\spacingset#1{\renewcommand{\baselinestretch}%
{#1}\small\normalsize} \spacingset{1}

%%%%%%%%%%%%%%%%%%%%%%%%%%%%%%%%%%%%%%%%%%%%%%%%%%%%%%%%%%%%%%%%%%%%%%%%%%%%%%

\if0\blind
{
  
\title{Supplementary Materials to ``Real-time Monitoring of Functional Data''}
\author[1]{Fabio Centofanti}
\author[1]{Antonio Lepore \thanks{Corresponding author. e-mail: \texttt{antonio.lepore@unina.it}}}
\author[2,3]{ Murat Kulahci }
\author[2]{Max Peter Spooner}

\affil[1]{Department of Industrial Engineering, University of Naples Federico II, Piazzale Tecchio 80, 80125, Naples, Italy}
\affil[2]{Technical University of Denmark, Department of Applied Mathematics and Computer Science, Kongens Lyngby, Denmark}
\affil[3]{Luleå University of Technology, Department of Business Administration, Technology and Social Sciences, Luleå, Sweden}
\setcounter{Maxaffil}{0}
\renewcommand\Affilfont{\itshape\small}
\date{}
\maketitle
} \fi

\if1\blind
{
  \bigskip
  \bigskip
  \bigskip
  \begin{center}
    {\LARGE\bf Supplementary Materials to ``Real-time Monitoring of Functional Data''}
\end{center}
  \medskip
} \fi

\vfill
\hfill {\tiny technometrics tex template (do not remove)}

\newpage
\spacingset{1.45} % DON'T change the spacing!
\appendix
\numberwithin{equation}{section}
\section{Additional Details about  FRTM}
\subsection{The Numerical Solution of the OEB-FDTW}
Given the   template curve $ Y \in L^2\left(\mathcal{D}_{Y}\right)$, the OEB-FDTW estimates the warping function $ h_i $ corresponding to the functional observation $ X_i $  by solving the  variational problem in Equation (2).
Let  $\mathcal{D}_{h_i}=\left[a_i,b_i\right]$,  $\mathcal{D}_{Y}=\left[a_y,b_y\right]$ and $ \mathcal{D}_{X_{i}}=\left[a_{x,i},b_{x,i}\right] $. Then, the numerical solution is obtained by considering a discretized version of the objective function in Equation (2), that is 
\begin{multline}\label{eq_fdtwdis}
 \hat{a}_i,\hat{b}_i,\dots,\hat{h}_i(t_j),\dots ,\hat{\alpha}=\\	\arginf_{\tilde{a}_i,\tilde{b}_i\in S_t,h_i(t_j): j\in I_{t,i},\alpha\in[0,1]}\frac{1}{\tilde{b}_i-\tilde{a}_i}\sum_{j\in I_{t,i}\backslash\lbrace \tilde{a}_i \rbrace}(t_{j}-t_{j-1})	\left[F_a(Y,X_i,Y',X'_i,h_i,\alpha,t_j\right)\\+\lambda R\left(\frac{h_i(t_j)-h_i(t_{j-1})}{t_j-t_{j-1}}-T_i/T_0\right)\bigg],
\end{multline}
with 
\begin{multline*}
	F_a(Y,X_i,Y',X'_i,h_i,\alpha,t_j)=\alpha^2\left(\frac{Y\left(t_j\right)}{||Y||}-\frac{X_i\left(h_i\left(t_j\right)\right)}{||X_i||}\right)^2\\+(1-\alpha)^2\left(\frac{Y'\left(t_j\right)}{||Y'||}-\frac{(h_i(t_j)-h_i(t_{j-1}))X'_i\left(h_i\left(t_j\right)\right)}{(t_j-t_{j-1})||X'_i||}\right)^2,
\end{multline*}
under the constraints $\tilde{a}_i<\tilde{b}_i $ and $h_i(t_j)<h_i(t_{l})$, for $ j<l $, where
 $ I_{t,i}=\lbrace j=0,1,\dots,N_t:\tilde{a}_i\leq t_j \leq \tilde{b}_i, t_j\in S_t\rbrace $,   $ S_t=\lbrace t_j\rbrace_{j=0,1,\dots,N_t}$, with $ a_y=t_0\leq t_1\leq\dots\leq t_{N_t-1}\leq t_{N_t}=b_y $, is a finite grid in  $ \mathcal{D}_{Y} $.
 The function $ R(u) $ in Equation \eqref{eq_fdtwdis} is
 \begin{equation*}
 	R(u)=\begin{cases}
 	u^2&s^{min}\leq u+T_i/T_0\leq s^{max}\\
 		\infty &otherwise,
 	\end{cases}   
 \end{equation*}
where $ s^{min} $ and $ s^{max} $  are the given minimum and maximum allowed values of the first derivative of $ h_i $. Note that  the discretized version in Equation \eqref{eq_fdtwdis} is the simply the Riemman approximation of the integral in Equation (2), which implicitly assumes that both $  Y $ and $ h_i $ are well-approximated by piecewise constant functions with knot points at $ t_j $ with values $  Y(t_j) $ and $ h_i(t_j) $, respectively, and the $ h'_i $ values   in $ t_j $ approximated by $ \frac{h_i(t_j)-h_i(t_{j-1})}{t_j-t_{j-1}} $. Therefore,  the solution of  the optimization problem in Equation (2) is found with respect to the warping function domain boundaries $ \tilde{a}_i$ and $\tilde{b}_i$,  the strictly increasing values  $h_i(t_j) $, and $ \alpha $, by avoiding a minimization  with respect to an infinite dimensional function $ h_i $ and a scalar $ \alpha $.
As already introduced in Section 2.1, the problem in Equation \eqref{eq_fdtwdis} cannot be solved by directly applying  dynamic programming because the  solutions of each iteration in the algorithm now depend on $ \tilde{a}_i$ and $\tilde{b}_i$.  To deal with this issue, we approximate the optimal solution by following the idea of \cite{shallom1989dynamic}. 
We  assume, for each $ j=0,\dots,N_t$, $ h_i(t_j) $ to take values $ \tau_{j,k} $ from a discrete set in $ \mathcal{D}_{X_i} $, $ a_{x,i}\leq l_j=\tau_{j,1}\leq\dots\leq  \tau_{j,M_x}=u_j\leq b_{x,i} $.
To be specific, we assume also a linear spacing between  $ l_j $ and $ u_j $  defined  as
\begin{align}\label{key}
	l_j=\max(s^{min}t_j+a_{x,i}-s^{min}(a_y+\delta_{t,s}),t_js^{max}+b_{x,i}-\delta_{x,e}-b_ys^{max},a_{x,i}),\nonumber\\
	u_j=\min(s^{max}t_j+a_{x,i}+\delta_{x,s}-a_ys^{max},t_js^{min}+b_{x,i}-(a_y-\delta_{t,e})s^{min},b_{x,i}),
\end{align}
where $ \delta_{t,s}, \delta_{x,s} $ and $ \delta_{t,e}, \delta_{x,e} $ are the maximum allowed ranges of mismatch  at the begin and the end of the domain $ \mathcal{D}_{Y} $. Specifically, we have that $ a_y\leq \tilde{a}_i\leq a_y+ \delta_{t,s}$ and $b_y- \delta_{t,e} \leq \tilde{b}_i\leq b_y $, and $ a_{x,i}\leq h_i(\tilde{a}_i)\leq a_{x,i}+ \delta_{x,s}$ and $b_{x,i}- \delta_{x,e} \leq h_i(\tilde{b}_i)\leq b_{x,i} $.
An approximated solution of   the optimization problem in Equation \eqref{eq_fdtwdis} can be found  as weighted shortest path through the graphs with $ M_x(N_t+1) $ nodes associated to the values $ \tau_{j,k} $, $ j=0,1,\dots,N_t $, $ k=1,\dots,M_x $ \citep{deriso2019general}.
From each node $ \tau_{j,k} $,  the only possible transitions are  to $ \tau_{j+1,k} $, $ k=1,\dots,M_x $, with  costs equal to
\begin{multline}\label{key}
	\frac{1}{\Delta_{j,k}+(t_{j+1}-t_{j})}(t_{j+1}-t_{j})	\left[F_a(Y,X_i,Y',X'_i,h_i,\alpha,t_{j+1}\right)\\+\lambda R\left(\frac{h_i(t_{j+1})-h_i(t_{j})}{t_{j+1}-t_{j}}-T_i/T_0\right)\bigg],
\end{multline}
where $\Delta_{j,k}$ is the current domain size corresponding to the optimal path that enters   $ \tau_{j,k} $, if it exists. It is zero if no such path exists. 
Note that, by considering  $ \Delta_{j,k} $, the cost associated with a given path depends on its current length, and thus, shorter paths are not preferred in each  algorithm iteration.
Moreover, the starting costs associated to $ \tau_{0,k} $ for $ k:\tau_{0,k}\leq a_{x,i}+\delta_{x,s} $, and $ \tau_{j,1} $, for $ j:t_j\leq a_y+\delta_{t,s} $ are set equal to zero, or infinite otherwise. The shortest path can be found through dynamic programming with computational cost  order $ M_x^2N_t $.
The error associated to the discretization of the values of $ h_i(t_j) $ by $ \tau_{j,k} $ could be reduced by updating $ l_j $ and $ u_j $ as described by \cite{deriso2019general}.

Practically, $ \delta_{t,s}, \delta_{t,e} $ and $ \delta_{x,s}, \delta_{x,e} $ could be fixed by considering a given fraction of $ \mathcal{D}_{Y} $ and $  \mathcal{D}_{m}  $, respectively. In the Monte Carlo simulation study of Section 3 and the data example of Section 4, they are set equal to the $ 10\% $  of $ \mathcal{D}_{Y} $ and $  \mathcal{D}_{m}  $, and then, they are refined according to the band constraint.  Moreover, the evenly spaced values  $  t_j $, $ j=0,1,\dots,N_t $ with  $ N_t=100 $, $ M_x=50 $,   $ l_j $ and $ u_j $ are updated 3 times through the procedure described by \cite{deriso2019general}, and $ \alpha $ is selected among $0,0.5$, and $1$.
\subsection{Missing Data Imputation of the Functional Observations}
The output of the OEB-FDTW in Section 2.1   and the warping functions $ (X^*_i, h_i) $ are allowed to take values on different domains $ \mathcal{D}_{h_i}\subseteq \mathcal{D}_{Y} $, because of the open-begin/open-end feature of the registration method. This causes difficulty in the estimation of the mFPCA models, which requires functional observations  to be defined on the same domain.
To overcome this issue, we consider a missing data imputation problem. Specifically, we set $ \mathcal{D}_{Y} $ as the reference domain of definition  and the missing parts of the functional observations  $ (X^*_i, h_i) $ are imputed by considering as reference the template curve as reference opportunely shifted to ensure continuity. That is, let us set $\mathcal{D}_{h_i}=\left[a_i,b_i\right]$,  $\mathcal{D}_{Y}=\left[a_y,b_y\right]$ and $ \mathcal{D}_{X_{i}}=\left[a_{x,i},b_{x,i}\right] $, then the observations  $ (X^*_i, h_i) $ are substituted by the corresponding completed observations $ (X^*_{i,c}, h_{i,c}) $ defined for $ t\in \left[a_y,b_y\right] $ as
\begin{equation}\label{key}
	X^*_{i,c}(t)=\begin{cases}
			Y(t)+X^*_{i}(a_i)-Y(a_i)&t\in\left[a_y,a_i\right)\\
		X^*_{i}(t) &t\in\left[a_i,b_i\right]\\
			Y(t)+X^*_{i}(b_i)-Y(b_i)&t\in\left(b_i,a_y\right]
	\end{cases}   
\end{equation}
and
\begin{equation}\label{key}
	h_{i,c}(t)=\begin{cases}
			Y(t)+ct+d_1&t\in\left[a_y,a_i\right)\\
		h_i(t) &t\in\left[a_i,b_i\right]\\
		Y(t)+ct+d_2&t\in\left(b_i,a_y\right],
	\end{cases}   
\end{equation}
with $ c=\frac{b_{x,i}-a_{x,i}}{b_i-a_i} $ and $ d_1=-c(b_i-a_i) $ and $ d_2=b_{x,i}-cb_i $.
More complex data imputation methods could be used as well. However,  for moderate level of open-end/open-begin behaviours, such procedures are not able to significantly provide overall better results than the ones obtained through the proposed procedure.
Note that, mFPCA models should be estimated   for each $ t\in\mathcal{D}_{Y} $. In this case, the missing data imputation problem is employed by  using as reference domain   $ \mathcal{D}_{Y,t} $  (Section 2.4.1)   in place of $ \mathcal{D}_{Y} $.

\subsection{Details on the Adaptive Band Constraint Calculation}
Let us consider the functional observation $ X_{i,x^*}:\mathcal{D}_{X_{i,x^*}}\subseteq\mathcal{D}_{X_{i}} $, which is the partial functional observation $ X_i $ observed until $ x^* $, and  $ h_{i,x^*} $, the partial warping function corresponding to $ X_{i,x^*} $. Moreover, let us consider a point $ x^*_{d}=x^*+dx $, with $ dx $  a small positive value.
To relax the band constraint,  we compare the values of $ h_{i,x^*} $ and $ h_{i,x^*_{d}} $ and their first derivatives on their domain right boundaries. If both quantities are similar, then we allow  the band constraint to be relaxed. This is based on the idea that, if the real-time registration step produces similar warping functions for subsequent time points, then the band constraint could be adapted based on the available information, which is thus, sufficient for a reliable registration.
Specifically, let us consider the conditions
\begin{equation}\label{eq_condband}
	|h_{i,x^*}(t_{i}^*) -h_{i,x^*_d}(t_{i,d}^*)|\leq \delta_v \quad
	\frac{|h'_{i,x^*}(t_{i}^*) -h'_{i,x^*_d}(t_{i,d}^*)|}{|h'_{i,x^*}(t_{i}^*)|}\leq \delta_d,
\end{equation}
where $ \delta_v,\delta_d>=0 $,  $t_{i}^*: h_{i,x^*}(t_{i}^*)=x^*$, $t_{i,d}^*: h_{i,x^*_{d}}(t_{i,d}^*)=x^*_{d}$, and, $h'_{i,x^*} $ and $ h'_{i,x^*_{d}} $ are the first derivatives of $ h_{i,x^*} $ and $ h_{i,x^*_{d}} $, respectively.
When the conditions in Equation \eqref{eq_condband} are simultaneously satisfied for all  $ x \in\left[x^*,x^*+\Delta\right]$, with $ \Delta\geq 0 $, then, from $ x\geq x^*+\Delta+dt $, the band constrain is relaxed as $l(x)=t_{c,x}-\delta_{c}$ and $u(x)=t_{c,x}+\delta_{c}$, $ \delta_c\geq 0 $, with $ t_{c,x}=\frac{dt+h'_{i,x-dt}(t_{i,x-dt})t_{i,x-dt}}{h'_{i,x-dt}(t_{i,x-dt})} $, $t_{i,x-dt}: h_{i,x^*}(t_{i,x-dt})=x-dt$. Note that, $ t_{c,x} $ is the value of the line passing from $ (t_{i,x-dt},x-dt) $ with slope equal to $ h'_{i,x-dt}(t_{i,x-dt}) $  corresponding to $ x $.
As soon as the conditions in Equation \eqref{eq_condband} are violated, the original band constrain is employed again.

In the practical implementation of the proposed method, the monitoring domain $ \mathcal{D}_{m} $ is discretized over a regular grid thus, $ dt $ is set  as the discretization step, whereas $ \Delta $ is considered as $ \delta_{\Delta}|\mathcal{D}_{m}| $.
In the Monte Carlo simulation study of Section 3 and the data example of Section 4, we  set $ \delta_{\Delta}=0.1, \delta_v=0.03,\delta_d=0.05 $, and $ \delta_c=0.04  $. Moreover, in Equation \eqref{eq_condband} and  the calculation of $ t_{c,x} $, the pointwise first derivatives are substituted by the median first derivatives calculated over a left  neighbourhood of the domain right boundaries to mitigate  the effects of possible estimation errors.

\section{Details on Data Generation in the Simulation Study}
\label{sec_appB}
The data generation process is inspired by the works of \cite{wang1997alignment,gasser1999synchronizing}. An IC functional observation $ X_i $  is generated by considering the following model
\begin{equation}\label{eq_gd}
	X_i\left(x\right)=g_i\left(h_i^{-1}\left(x\right)\right)+\varepsilon_i\left(x\right) \quad x \in \mathcal{D}_{X_i},
\end{equation}
with $ \varepsilon_i $  a white noise function, where for each $ x \in \mathcal{D}_{X_i} $, $ \varepsilon_i\left(x\right) $ is a normal random function with zero mean and standard deviation $ \sigma_e $.
Let us assume that $ \mathcal{D}_{X_i}=\left[0,T_i\right]  $, with $ T_i $  normal distributed with mean $ \mu_{end} $ and standard deviation $ \sigma_{end} $, and that the amplitude function $ g_i $ is defined as
\begin{multline}\label{eq_g}
	g_i(t)=A_i(t)=a_i [15\exp(-20(t-0.7)^2))-5(\exp(-50(t-0.45)^2))+6(\exp(-100(t-0.3)^2))\\-6(\exp(-150(t-0.2)^2))+5(\exp(-200(t-0.15)^2))] \quad t \in \left[0,1\right]
\end{multline}
where $ a_i $ is a normal random variable with mean $ \mu_a $ and standard deviation $\sigma_a $.
Then, Scenario 1 and Scenario 2 are defined by the form of the inverse warping function $ h_i^{-1} $.

In Scenario 1, we consider the following inverse warping function
	\begin{equation}\label{eq_h1}
		h_i^{-1}\left(x\right)=P_i^{S1}\left(x\right)=\tau_i(x)+b_i\tau_i(x)\left(\tau_i(x)-1\right)\quad x\in \left[0,T_i\right],
	\end{equation}
where $\tau_i(x)= \frac{x(t_{ei}-t_{bi})}{T_i}+t_{bi} $, with $ t_{bi},t_{ei}\in\left[0,1\right] $, and $ t_{bi}<t_{ei} $.
The condition $ |b_i|<1 $ should be satisfied to make $ 	h_i^{-1} $ strictly increasing.  Whereas, in  Scenario 2, we consider the following inverse warping function 
		 \begin{equation}\label{eq_h2}
			h_i^{-1}\left(x\right)=P_i^{S2}\left(x\right)=\tau_i(x)+b_i\left[\tau_i(x)\left(\tau_i(x)-1\right)-0.2\sin\left(3\pi\tau_i(x)\right)^2\right]\quad x\in \left[0,T_i\right].
	\end{equation}
In this case, $ h_i^{-1} $ is strictly increasing for $ |b_i|<0.367 $.
In both scenarios, we assume that $ b_i $ is a normal random variable with zero mean and standard deviation $ \sigma_b $. A reasonable value of $ \sigma_b $ is needed to produce strictly increasing  warping functions. However, if an  inadequate $ b_i $ is generated, it is iteratively reduced by $ 5\% $  until the corresponding condition is met.

The scalars $ t_{bi}$ and $t_{ei} $ are needed to simulate settings suitable for  the open-end/open-begin features of the OEB-FDTW.
Specifically, values of $ t_{bi}$ and $t_{ei} $ different from $ 0 $ and $ 1 $ allow the resulting functional observation $ X_i $ to be generated from a fraction of the amplitude function $ g_i $. That is, $ t_{bi}$ and $t_{ei} $ are chosen such that  $ h_i^{-1}\left(0\right)=s_i $ and $ h_i^{-1}\left(T_i\right)=e_i $, where $ s_i $ and $ e_i $ are normal random variables with  means equal to $ 0.05 $ and $ 0.95 $, respectively, and common  standard deviation  equal to $ \sigma_{be} $. It means  that, on average $ X_i $ is generated from  the values of $ g_i $  on the interval $ \left[0.05,0.95\right] $. If the values of $ t_{bi}$ and $t_{ei} $ are lower than 0 or larger than 1, they are truncated to 0 and 1, respectively.
The  parameters used to generate IC samples in Scenario 1 and Scenario 2 are listed in Table \ref{ta_1}, for the three  models, each with a different  level of misalignment indicated by M1,M2, and M3.
%For illustrative purposes, a sample of $100$ randomly generated realizations of IC observations for models M1,M2, and M3 are shown in Figure \ref{fi_dataic}.
\begin{table}
	\caption{Parameters used to generate IC observations  for models M1,  M2 and M3 in Scenario 1 and Scenario 2 of the simulation study.}
	\label{ta_1}
	
	\centering
	\resizebox{0.3\textwidth}{!}{
		\begin{tabular}{cccc}
			\toprule
			&M1&M2&M3\\
			\midrule
$ \sigma_{a} $&0.15&0.15&0.15\\
$ \sigma_{b} $&0.2&0.05&0.025\\
$ \sigma_{e} $&0.2&0.2&0.2\\
$ \sigma_{end} $&0.25&0.0625&0.03125\\
$ \sigma_{be} $&0.01& 0.0025&0.00125\\
$ \mu_{end} $&10&10&10\\
$ \mu_{a} $&1&1&1\\

			\bottomrule
	\end{tabular}}
\end{table}

To simulate departures from IC patterns, we consider three different kinds of shift, namely Shift A, Shift B and Shift C, which affect  the phase component alone, only the amplitude component alone and both components, respectively.
Specifically,   OC observations are generated through the model in Equation \eqref{eq_gd} with amplitude function defined as follows
\begin{equation}
	g_i(t)=\begin{cases}
		A_i(t) &t\in\left[\right.0,t_{i,out})\\
		A_i(t)+t\delta_{g}-\delta_{g}t_{i,out} & t\in\left[t_{i,out},1\right],
	\end{cases}   
\end{equation}
and inverse warping function defined as 
\begin{equation}
	h^{-1}_i(x)=\begin{cases}
		P_i(x) &x\in\left[\right.0,x_{i,out})\\
		P_i\left(\tau_{i,out}(x)\right)+\tau_{i}\left(\tau_{i,out}(x)\right)^2\delta_{h,1}-\tau_{i}\left(\tau_{i,out}(x)\right)\delta_{h,2}+\delta_{h,3}  & x\in\left[x_{i,out},T_{i,out}\right],
	\end{cases}   
\end{equation}
\sloppy where $  \delta_{h,1}=\frac{\delta_{h,2}-\delta_{h,2}\tau(x_{i,out})}{1-\tau(x_{i,out})^2}$, $  \delta_{h,3}=\delta_{h,2}\tau(x_{i,out})-\delta_{h,1}\tau(x_{i,out})^2$, $ \tau_{i,out}(x)=T_i\left[\frac{\left(x-x_{i,out}\right)}{T_{i,out}-x_{i,out}}\left(1-\frac{x_{i,out}}{T_i}\right)+\frac{x_{i,out}}{T_i}\right]$,
%\begin{equation}\label{key}
%	h^{-1}_i(x)=\begin{cases}
%		P_i(x) &x\in\left[\right.0,x_{i,out})\\
%		P_i\left(\tau_{i,out}(x)\right)+\tau_{i}\left(\tau_{i,out}(x)\right)\delta_{h}-\delta_{h}\tau_i\left(\tau_{i,out}(x_{i,out})\right) & x\in\left[x_{i,out},T_{i,out}\right],
%	\end{cases}   
%\end{equation}
%where $ \tau_{i,out}=T_i\left[\frac{\left(x-T_ix_{i,out}\right)}{T_{i,out}-T_ix_{i,out}}\left(1-x_{i,out}\right)+x_{i,out}\right]$, 
$ t_{i,out}= h^{-1}_i(x_{i,out})$.
 The term  $ A_i $ is defined in Equation  \eqref{eq_g}, and $ P_i $ is equal to $ P_i^{S1} $ and   $ P_i^{S2} $ reported in Equation  \eqref{eq_h1}  and  \eqref{eq_h2}, for Scenario 1 and Scenario 2, respectively. Moreover, $ T_{i,out} $ is equal to   $ T_i+\delta_{end} $, and $ x_{i,out}=\frac{x^{*}_{out}-t_{bi}}{t_{ei}-t_{bi}}T_i $, with $ x^{*}_{out}\in\left[t_{bi},t_{ei}\right] $. The OC behaviour arises  at the $ (x^{*}_{out}100)\%$ of the total duration of
the generating process  that corresponds to time instants $ x_{i,out} $ and  $ t_{i,out} $ in the observed and transformed time scale, respectively.
The performance of  FRTM is studied for $ x^{*}_{out} $ equal to $ 0.3  $ and $ 0.6 $, which are representative of settings where  OC conditions appear in the first and second parts of the monitored process, respectively. Given $ x^{*}_{out} $, the three different types of shift are defined by the choice of $ \delta_{g}, \delta_{h,2},\delta_{end} $, as detailed in Table \ref{ta_2}, as a function of the severity level $ d\in\lbrace 0.25,0.50,0.75,1.00\rbrace $. Note that $ d=0 $ corresponds to the IC state. 
\begin{table}
	\caption{ The parameters $ \delta_{g}, \delta_{h,2},\delta_{end} $ as a function of the severity level $ d $ for Scenario 1 and Scenario 2 and  $ x^{*}_{out} $ equal to $ 0.3  $ and $ 0.6 $ in the simulation study.}
	\label{ta_2}
	
	\centering
	\resizebox{0.6\textwidth}{!}{
		\begin{tabular}{M{0.15\textwidth}cccccccc}
			\toprule
			&&\multicolumn{3}{c}{$ x^{*}_{out} = 0.3  $}&&\multicolumn{3}{c}{$ x^{*}_{out} = 0.6  $}\\
			\midrule
			&&$ \delta_{g} $&			$ \delta_{h,2} $&	$\delta_{end} $&&$ \delta_{g} $&			$ \delta_{h,2} $&	$\delta_{end}$\\\midrule
			\multirow{3}{*}{Scenario 1}&Shift A&$ 0 $&$ d $&$ 2d $&&$ 0 $&$ 1.5d $&$ 4d $\\
		&	Shift B&$ 20d $&$ 0 $&$ 0 $&&$ 80d $&$ 0 $&$ 0 $\\
			&Shift C&$ 10 $&$ 0.5d $&$ d $&&$ 40d $&$ 0.75d $&$ 2d $ \\\midrule
			\multirow{3}{*}{Scenario 2}&Shift A&$ 0 $&$ 1.5d $&$ 2d $&&$ 0 $&$ 2d $&$ 4d $\\
			&	Shift B&$ 20d $&$ 0 $&$ 0 $&&$ 80d $&$ 0 $&$ 0 $\\
			&Shift C&$ 10 $&$ 0.75d $&$ d $&&$ 40d $&$ 1d $&$ 2d $ \\

			\bottomrule
	\end{tabular}}
\end{table}
 Finally, $ X_i $ is assumed to be observed at 100 equally spaced time points   on either  $\left[0,T_i\right]$ for IC observations or  $\left[0,T_{i,out}\right]$ for OC observations.
 
\begin{figure}[h]
 	\caption{An example of $100$ randomly generated IC observations for each model M1,M2, and M3 in Scenario 1 and Scenario 2 of the simulation study.}
 	
 	\label{fi_dataic1}
 	
 	\centering
 	\begin{tabular}{cM{0.28\textwidth}M{0.28\textwidth}M{0.28\textwidth}}
 	Scenario 1&\includegraphics[width=.3\textwidth]{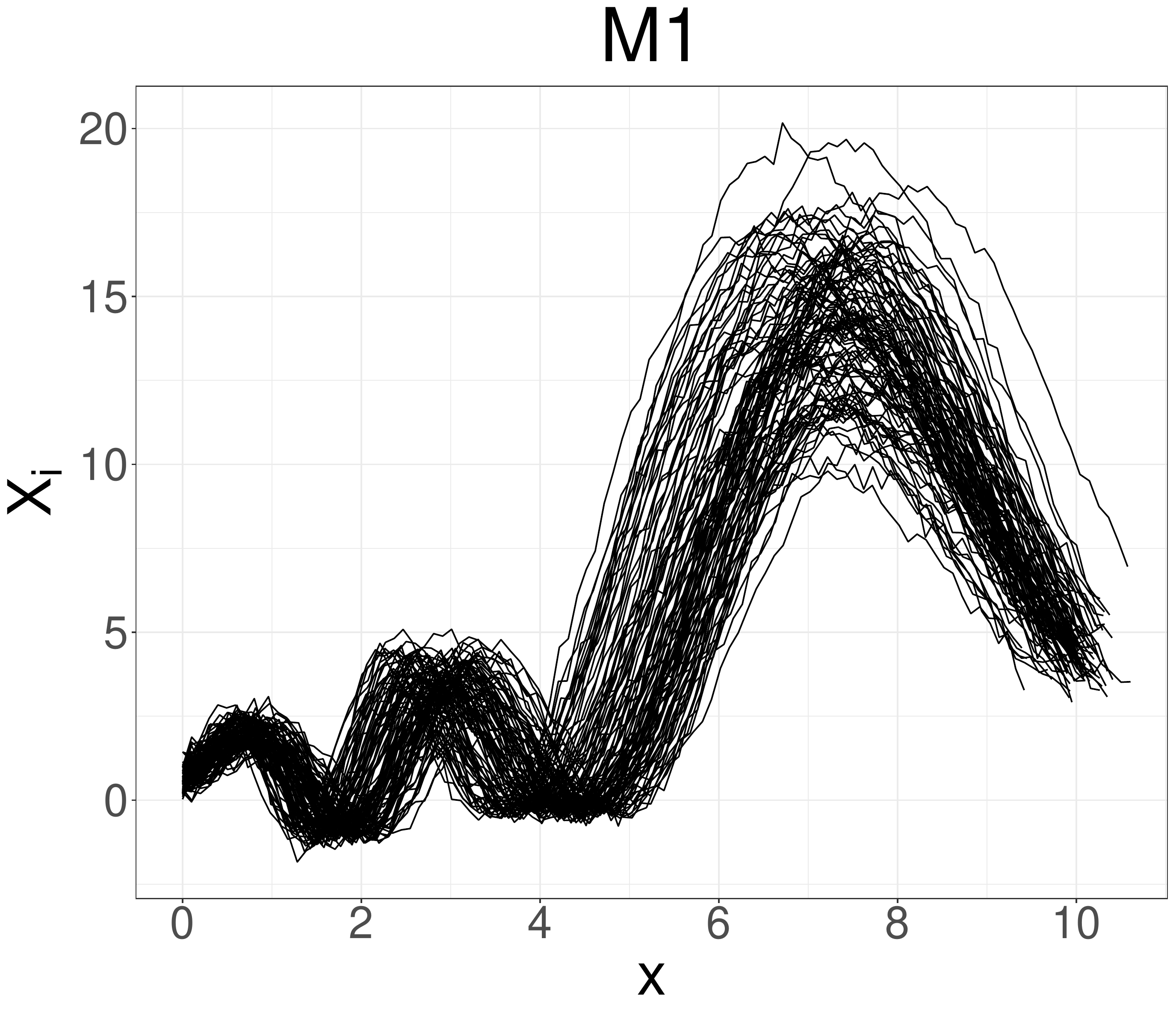}&
 	\includegraphics[width=.3\textwidth]{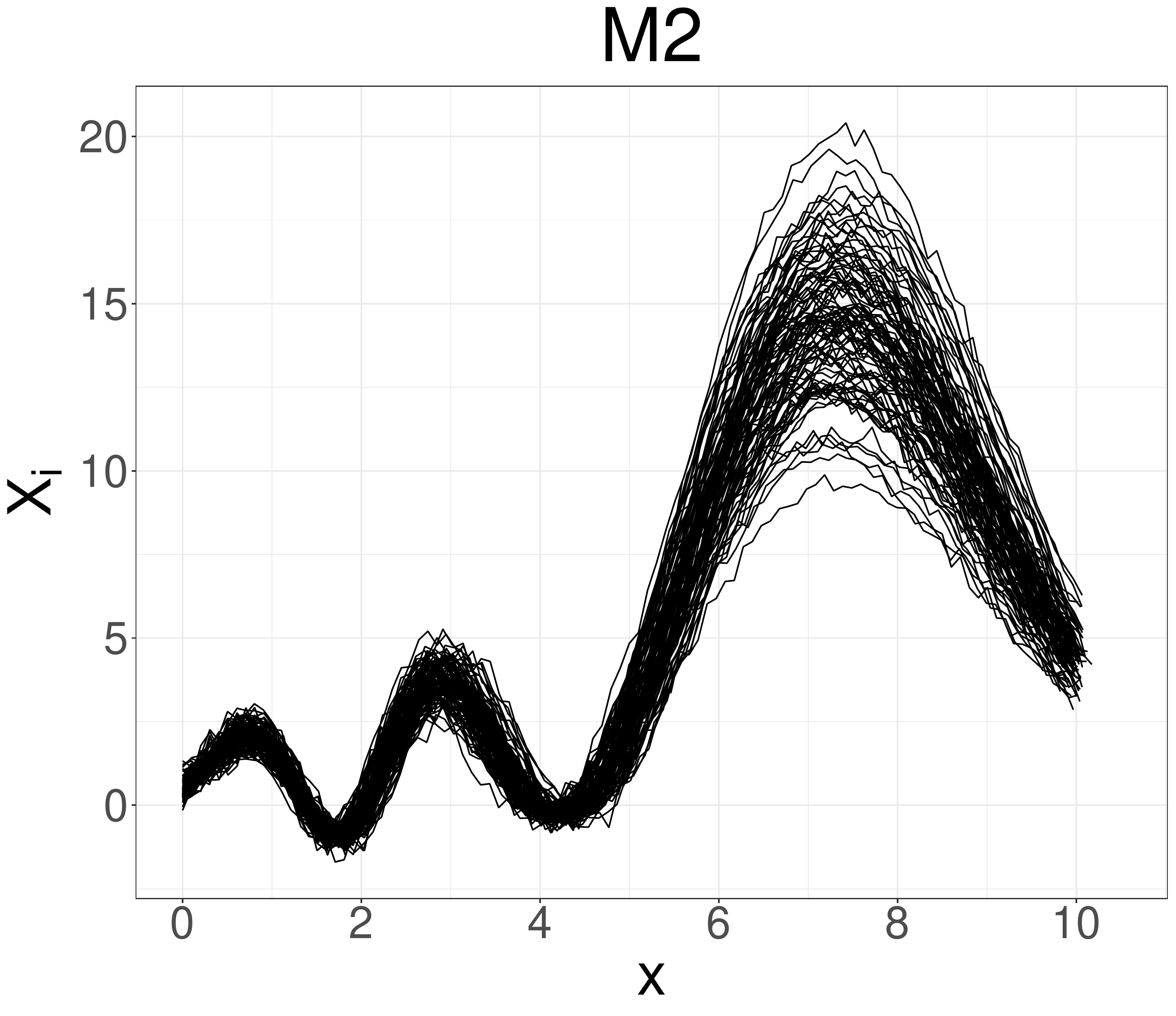}&
 	\includegraphics[width=.3\textwidth]{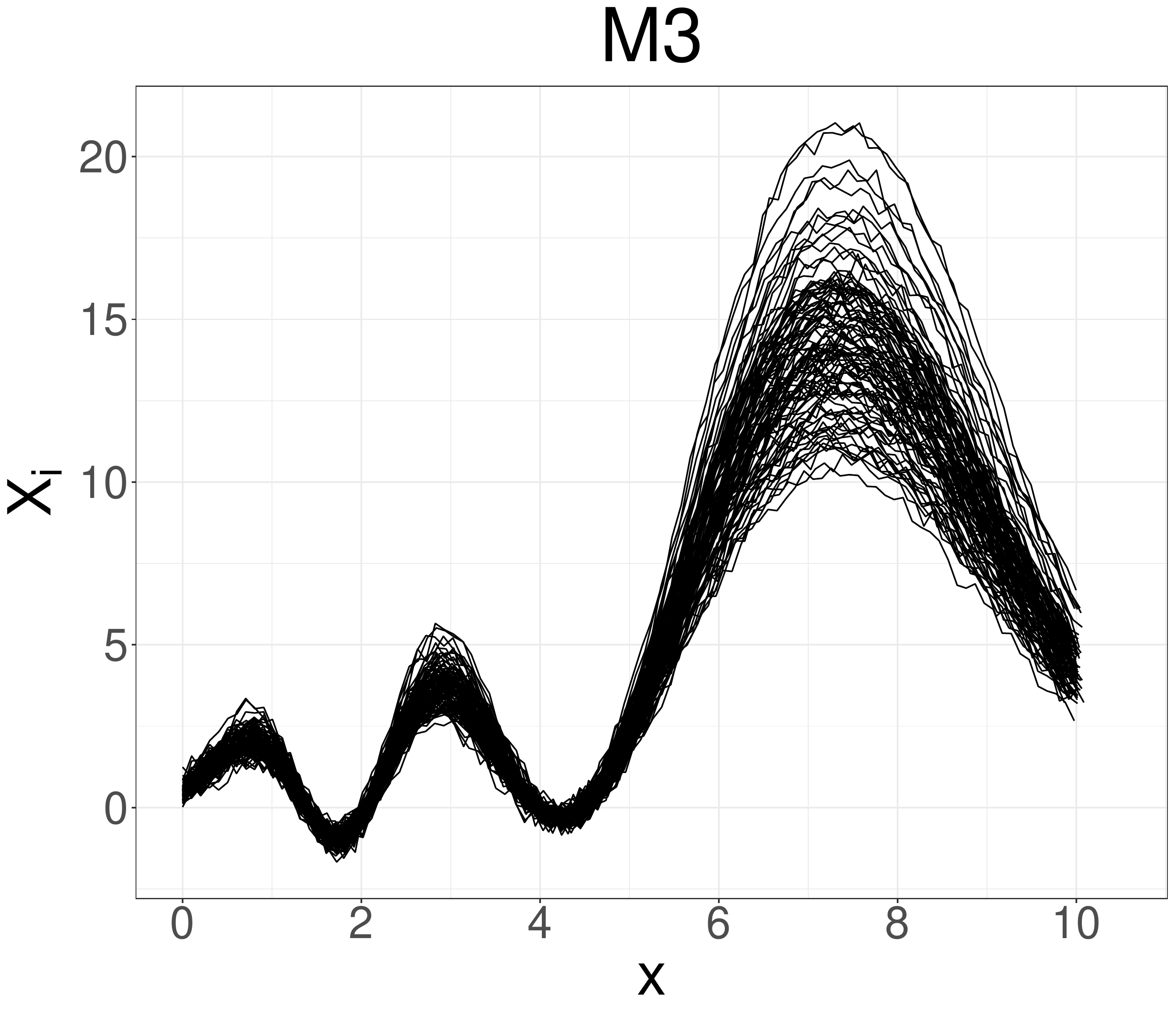}	\\
 	Scenario 2&\includegraphics[width=.3\textwidth]{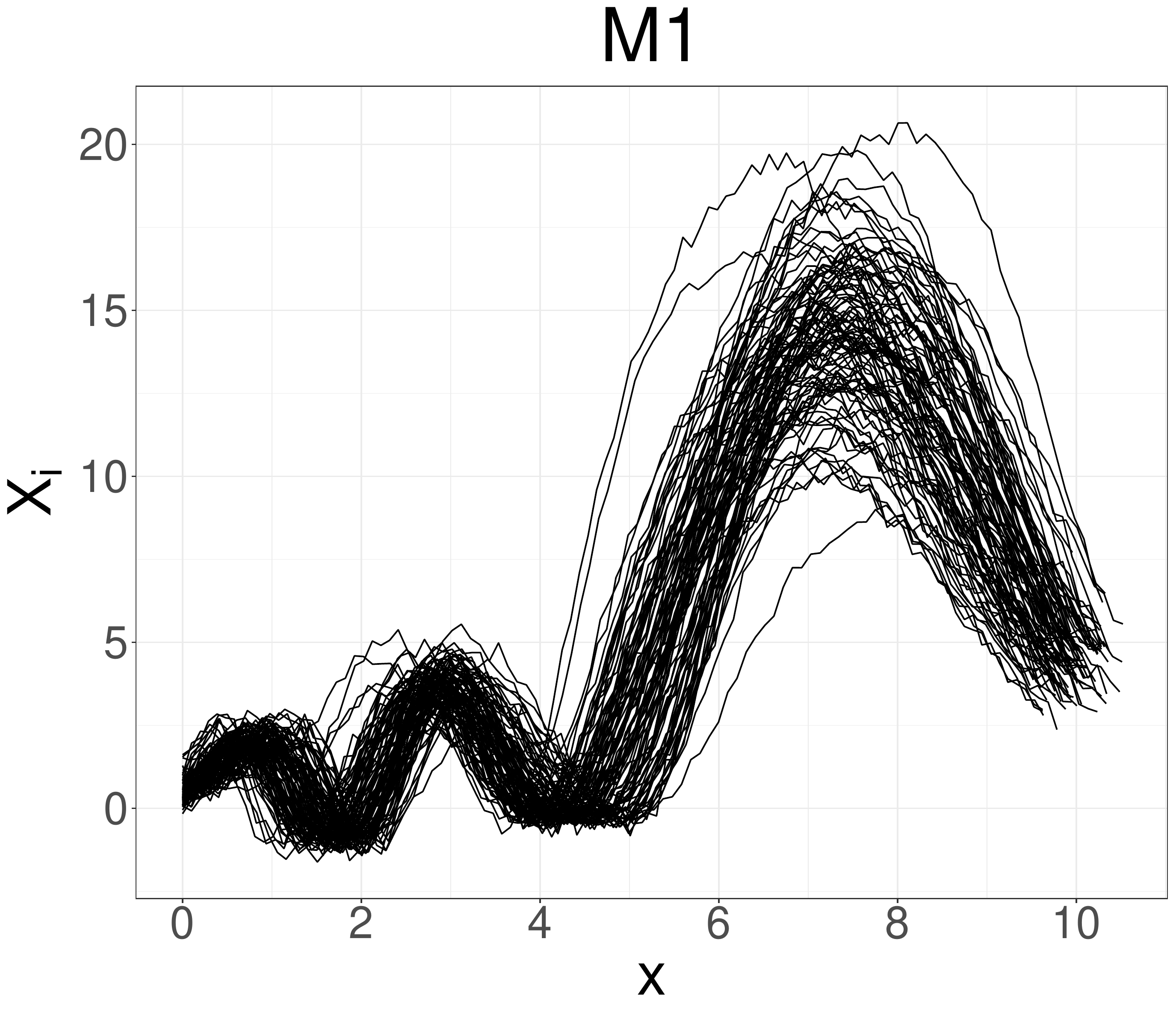}&
 	\includegraphics[width=.3\textwidth]{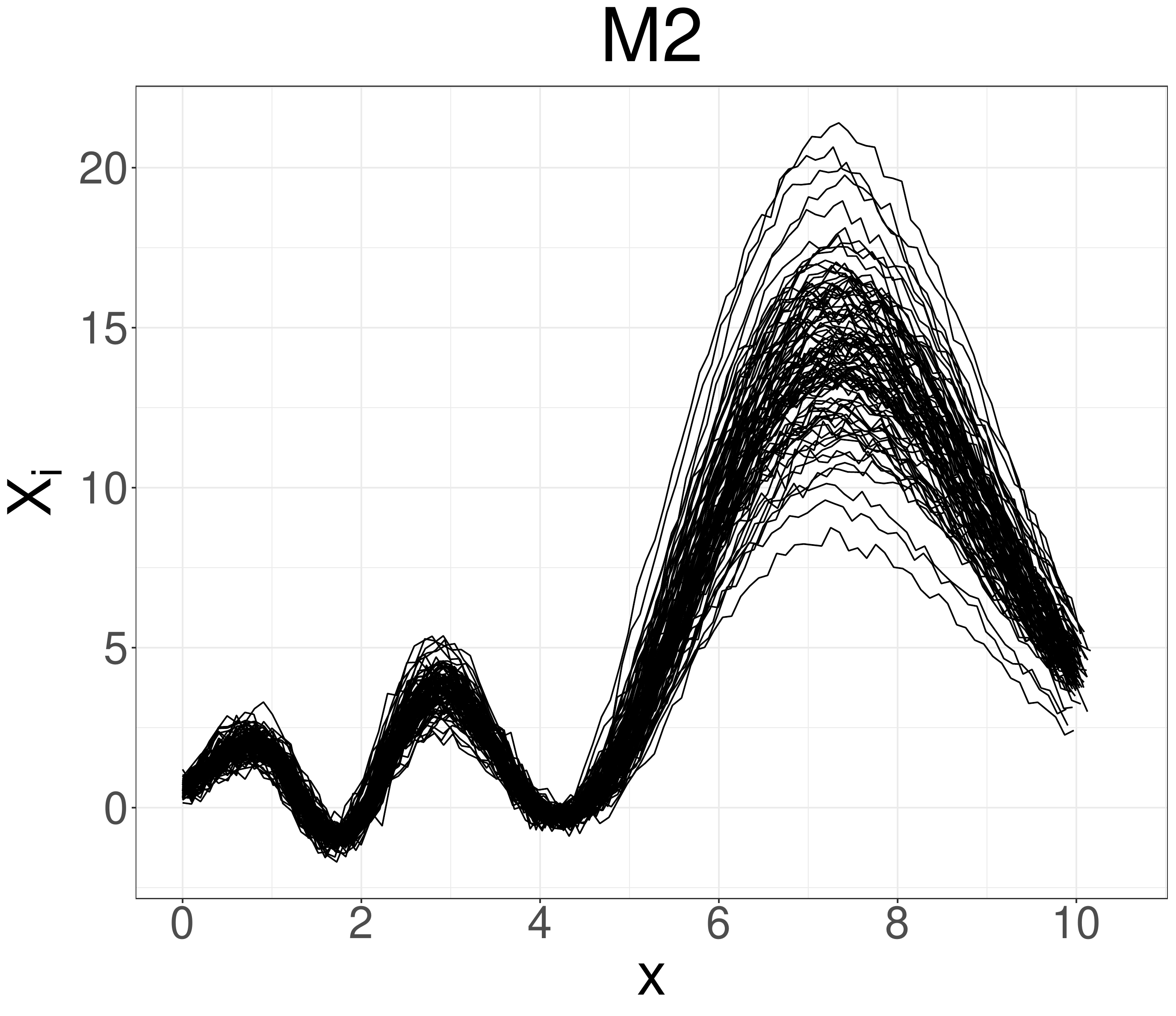}&
 	\includegraphics[width=.3\textwidth]{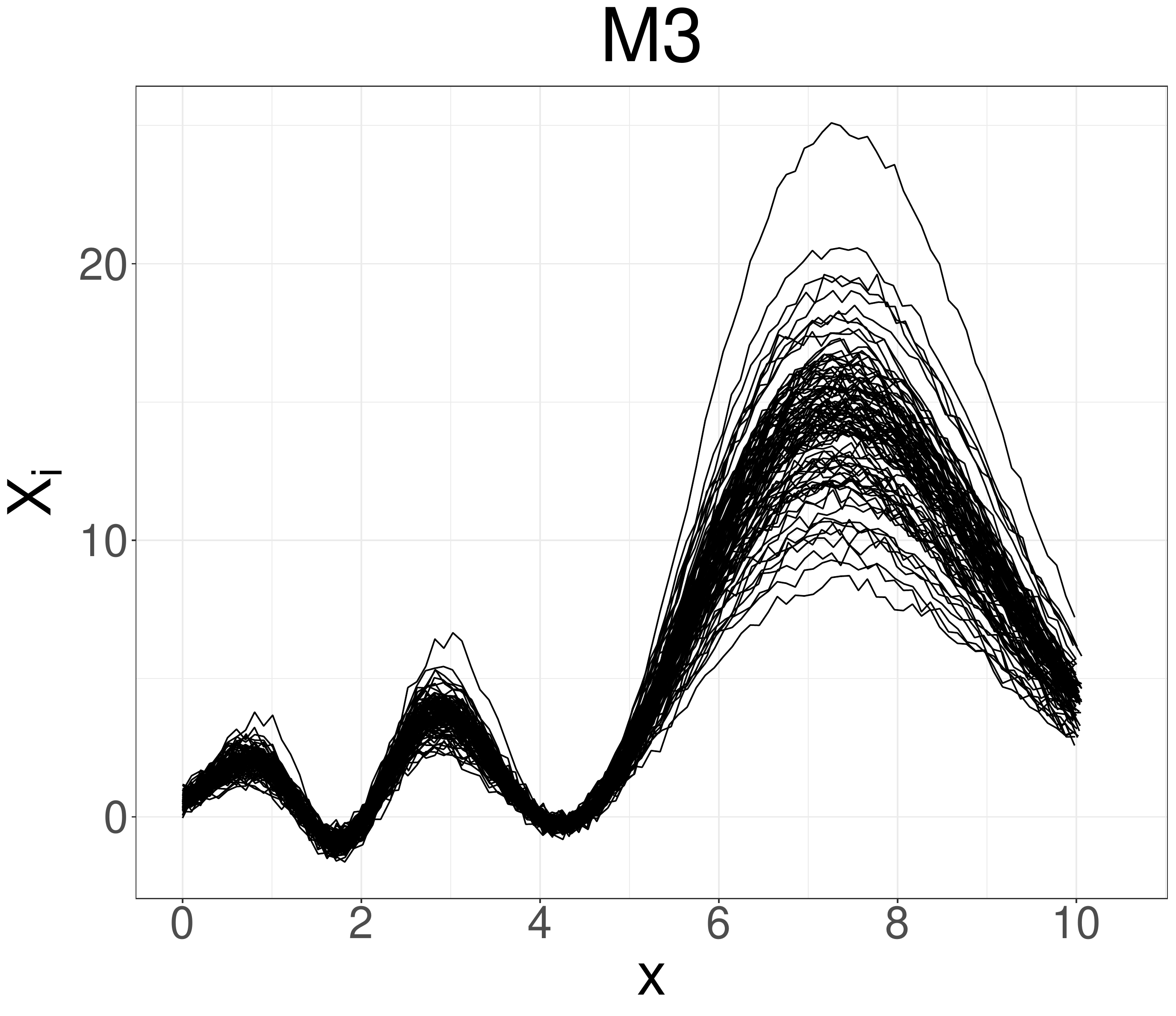}\\
 \end{tabular}
 	\vspace{-.5cm}
 \end{figure}

\begin{figure}[h]
 	\caption{A sample of $100$ randomly generated observations for models M1,M2, and M3, each type of shift and each severity level $ d $ in Scenario 1 of the simulation study with $ x^{*}_{out}=0.3	 $.}
 	
 	\label{fi_dataoc1}
 	
 	\centering
 	\begin{tabular}{cM{0.28\textwidth}M{0.28\textwidth}M{0.28\textwidth}}
 		\textbf{M1}&\includegraphics[width=.3\textwidth]{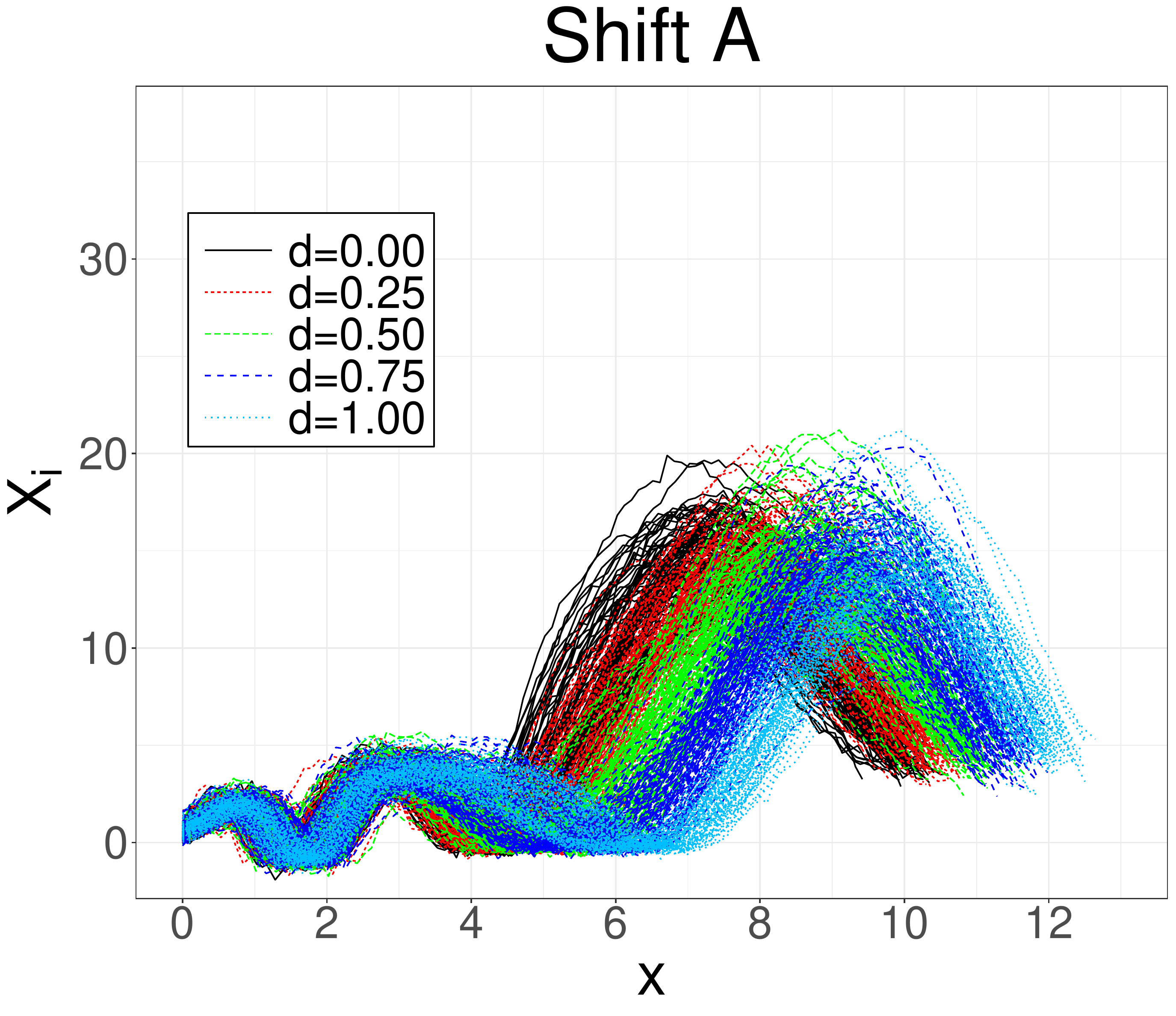}
 		&\includegraphics[width=.3\textwidth]{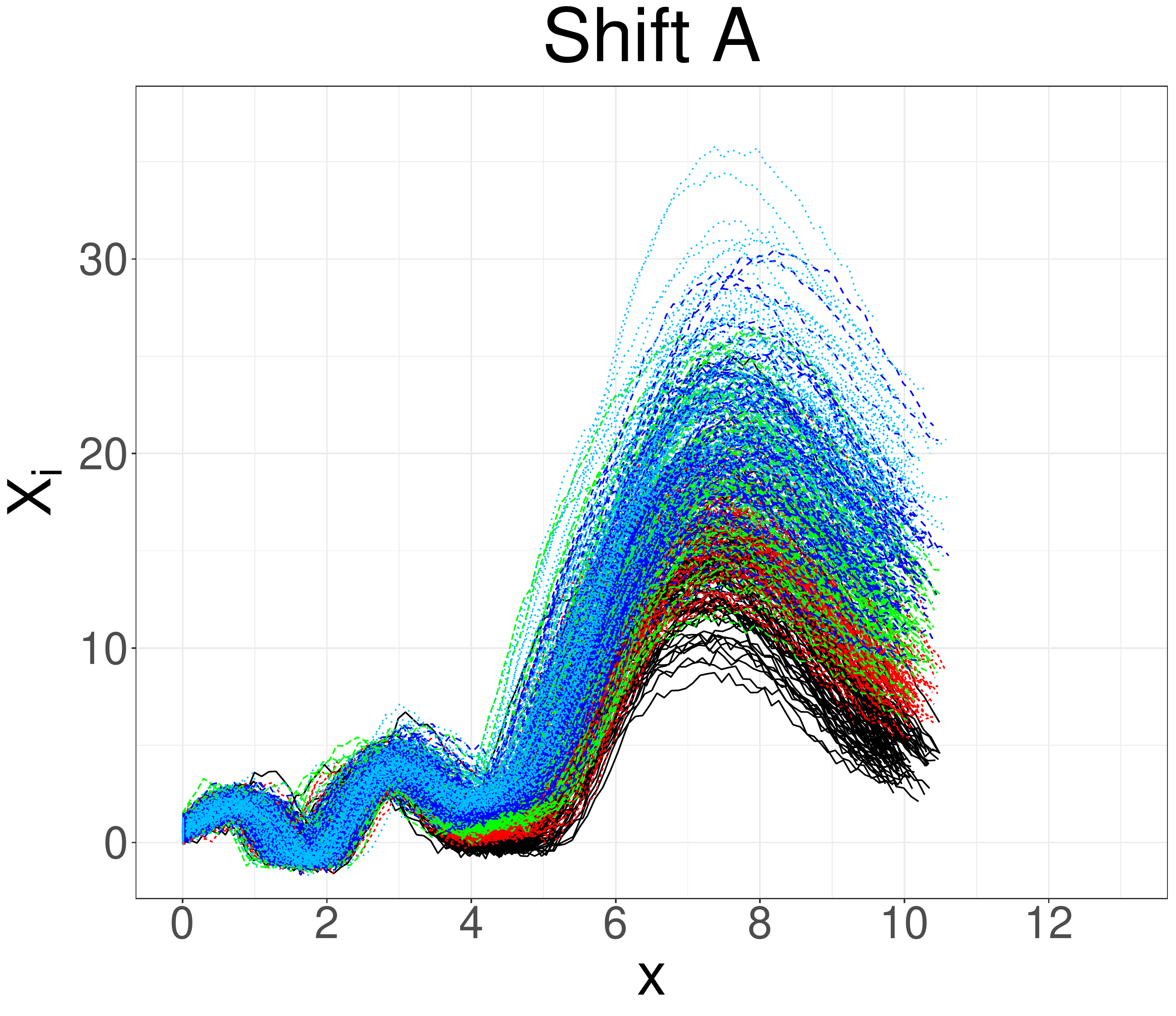}
 		&\includegraphics[width=.3\textwidth]{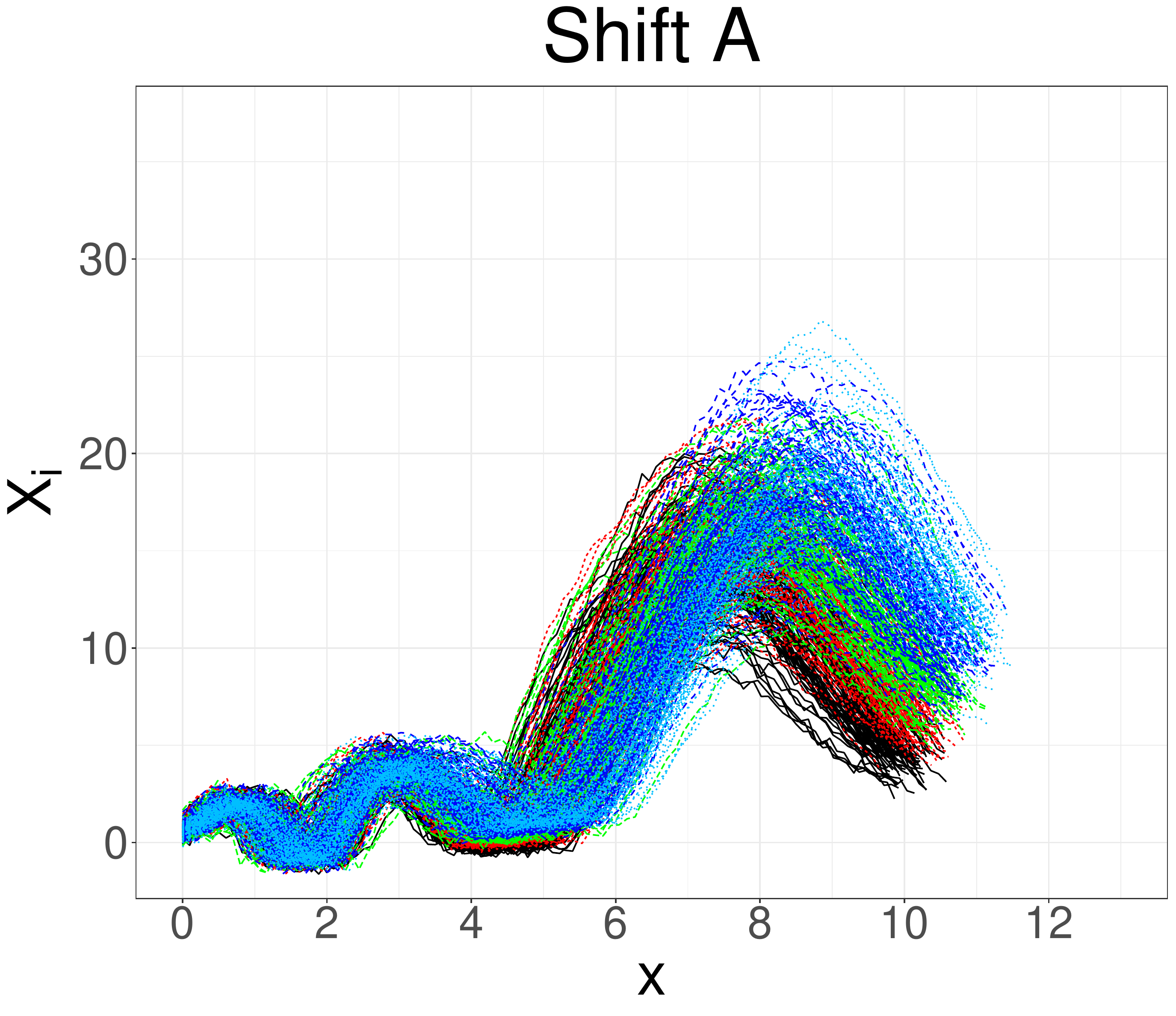}	\\
 		\textbf{M2}&\includegraphics[width=.3\textwidth]{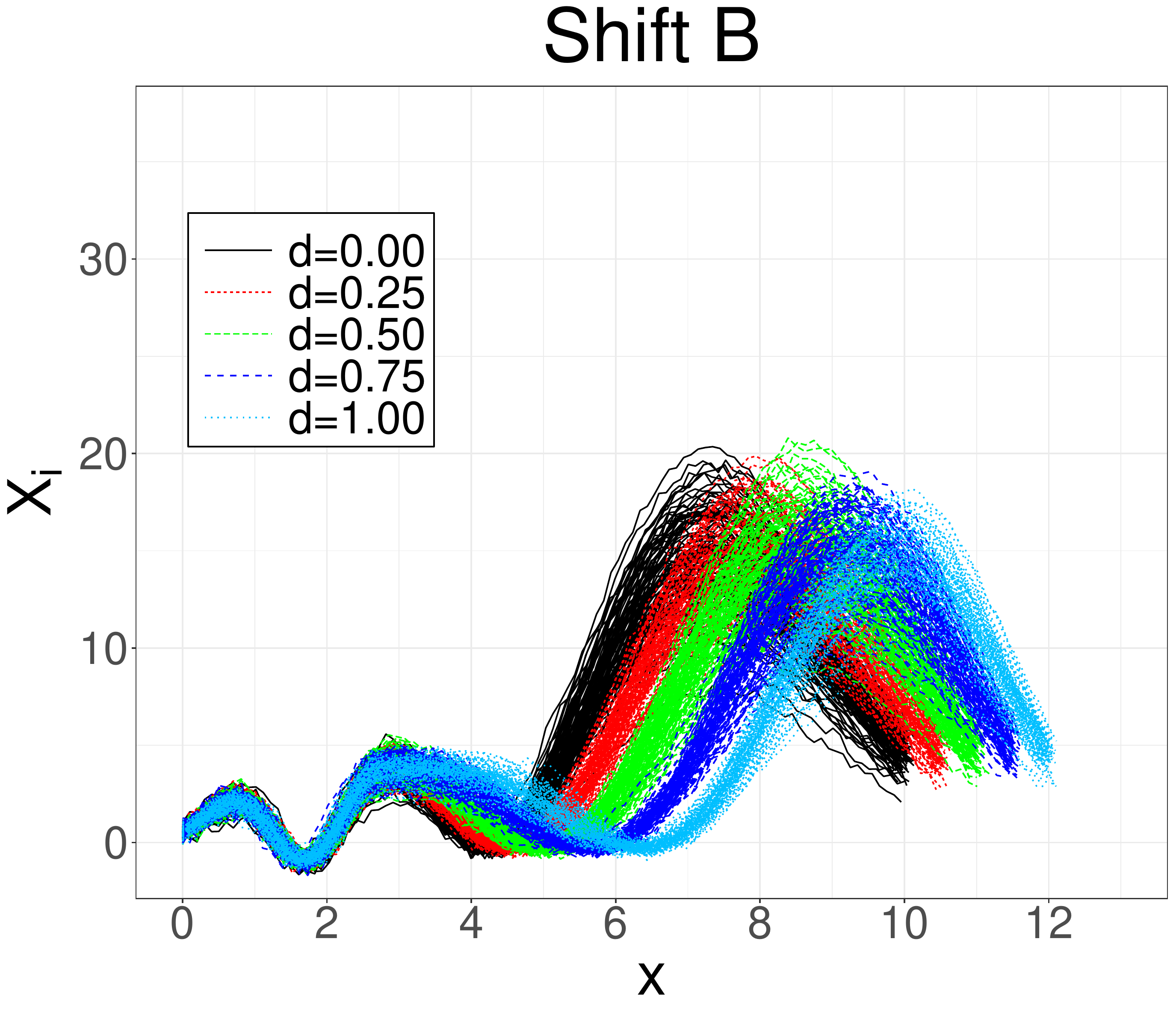}
 		&\includegraphics[width=.3\textwidth]{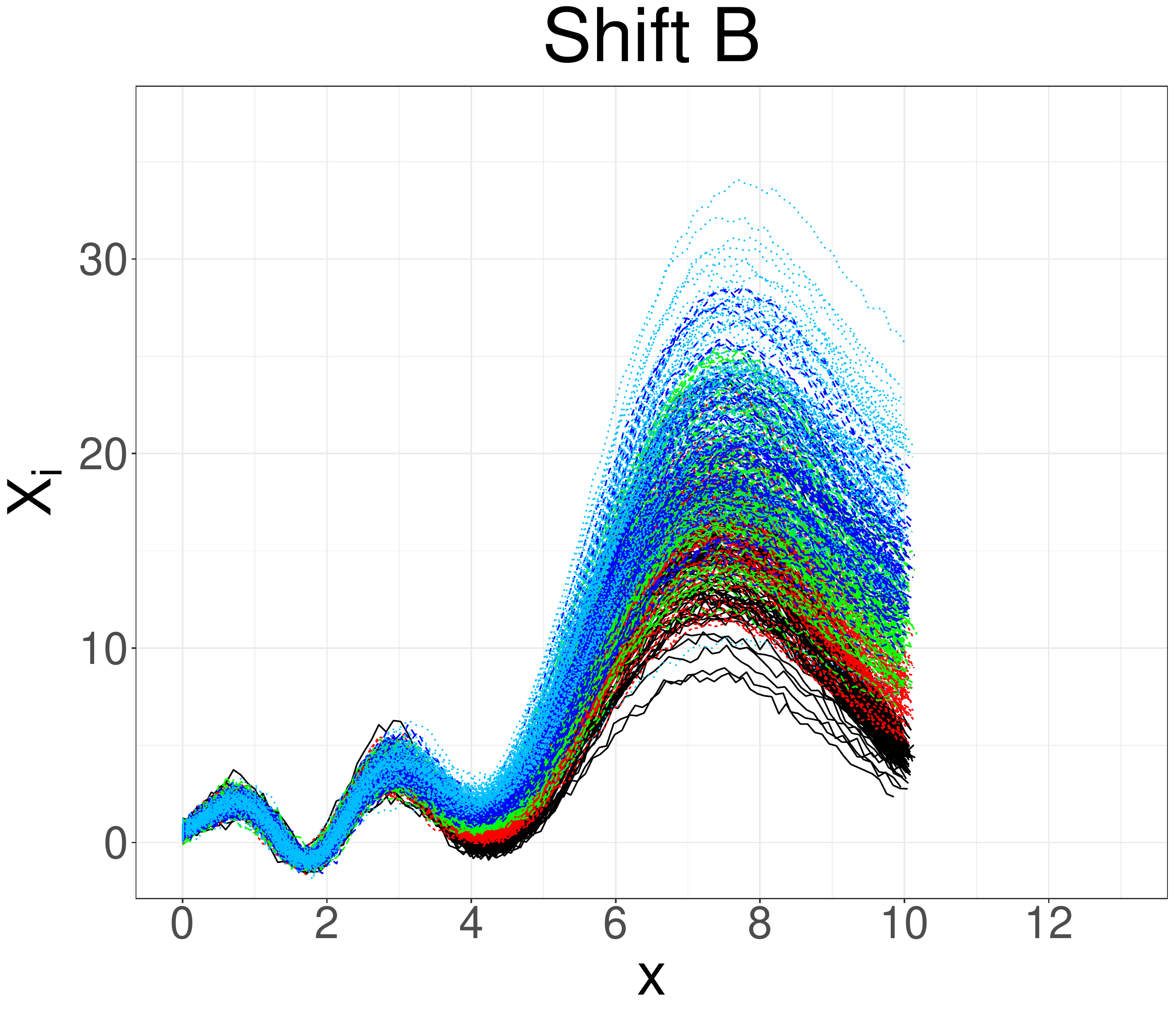}
 		&\includegraphics[width=.3\textwidth]{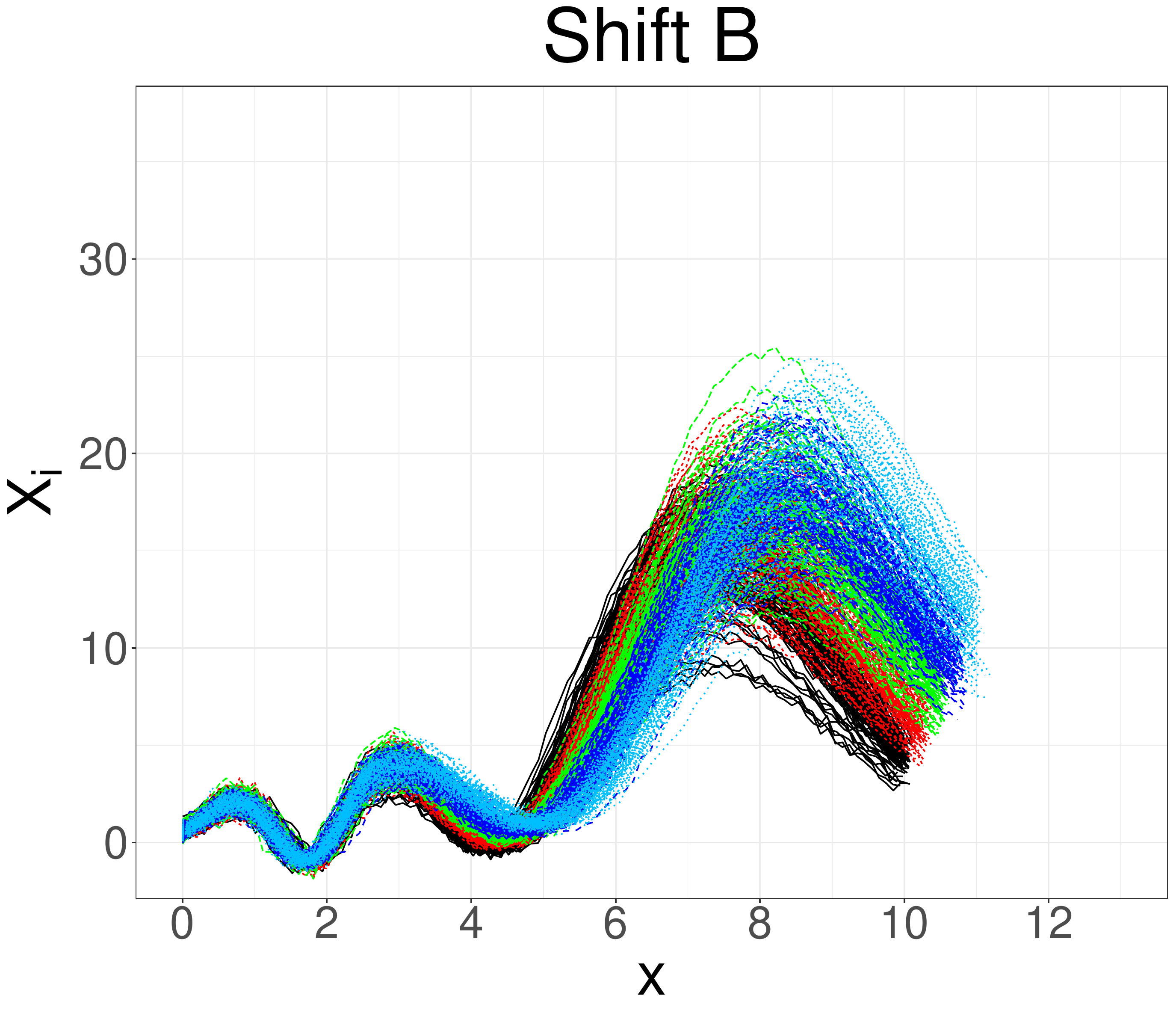}\\
 		\textbf{M3}&\includegraphics[width=.3\textwidth]{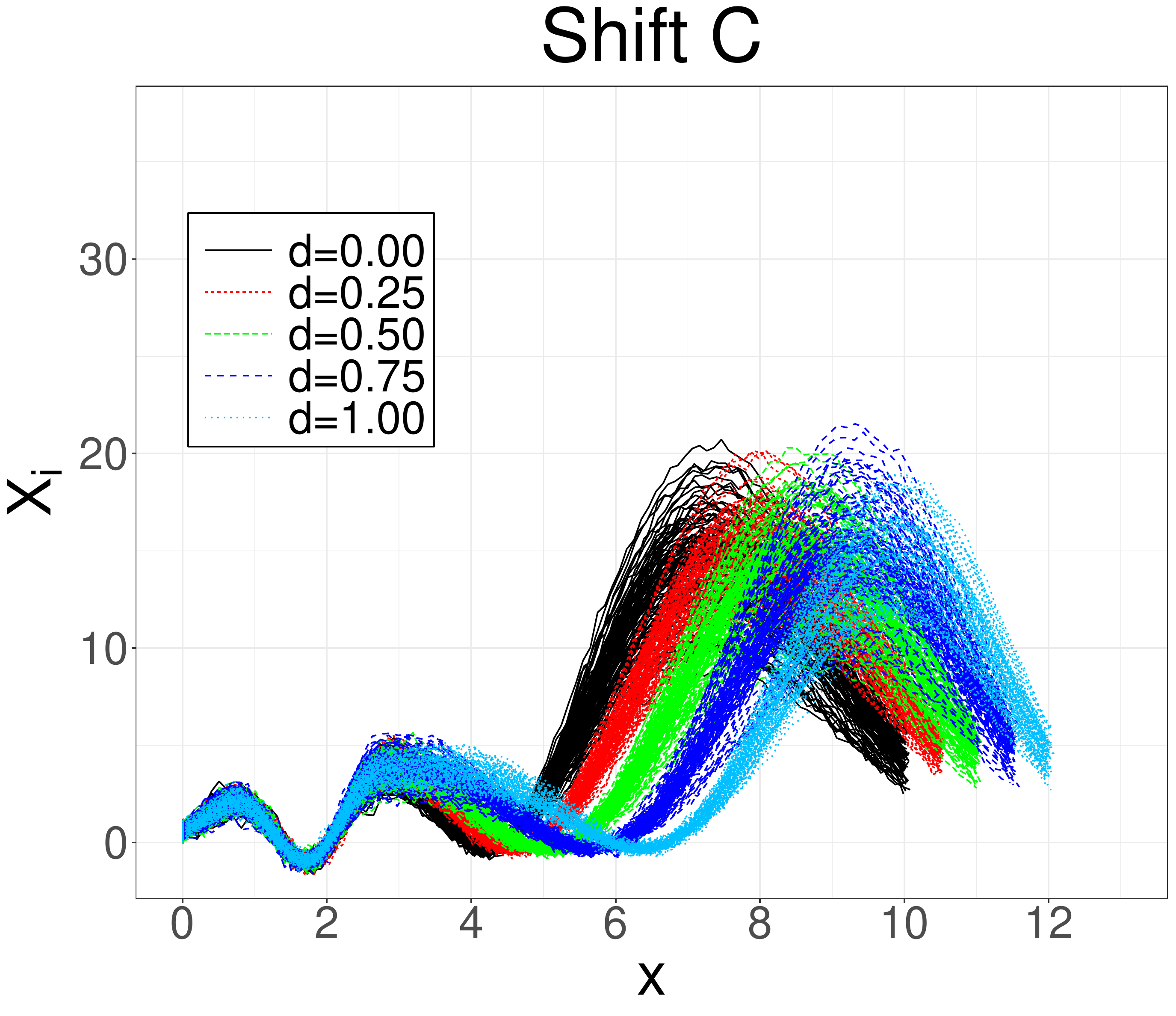}
 		&\includegraphics[width=.3\textwidth]{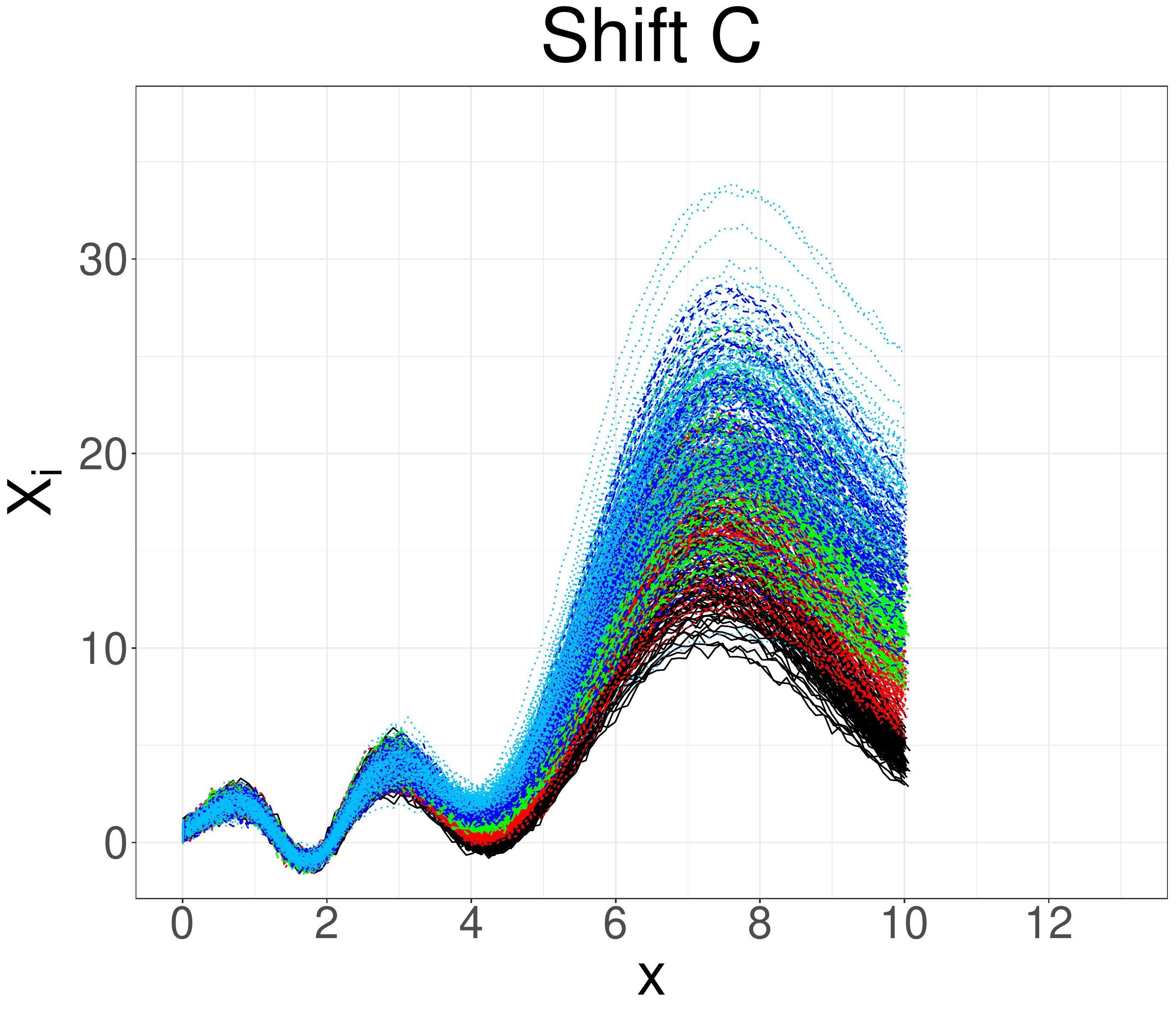}
 		&\includegraphics[width=.3\textwidth]{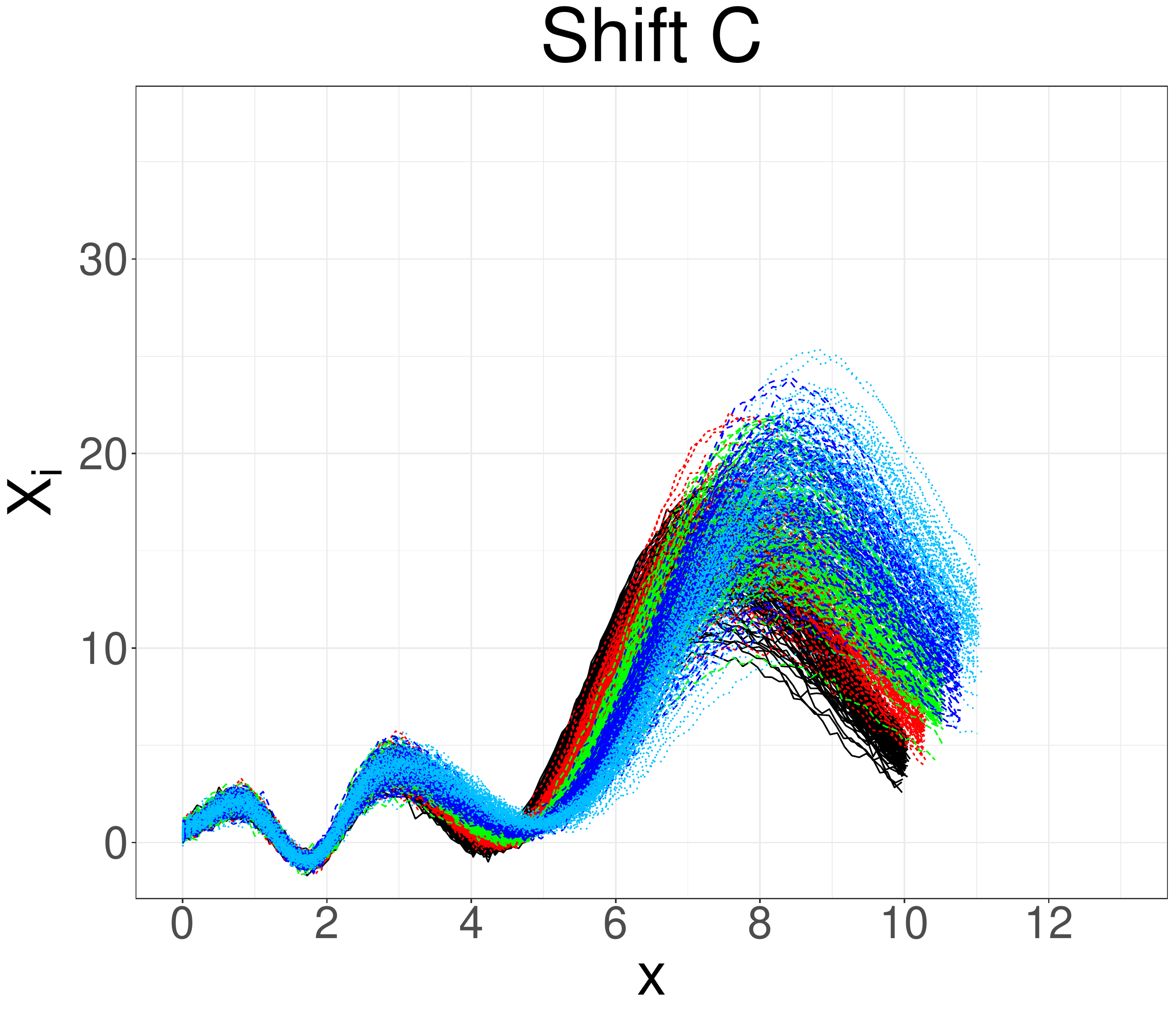}
 	\end{tabular}
 	\vspace{-.5cm}
 \end{figure}

 For illustrative purposes, a sample of $100$ randomly generated realizations of IC observations from models M1,M2, and M3 in Scenario 1 and Scenario 2 are shown in Figure \ref{fi_dataic1}. Whereas, Figure \ref{fi_dataoc1}  shows a sample of $100$ randomly generated observations in Scenario 1 for models M1,M2, and M3, each type of shift and  severity level $ d $,  with $ x^{*}_{out}=0.3	 $.
 
% \begin{figure}[h]
%	\caption{An example of $100$ randomly generated observations for models M1,M2, and M3 and each type of shift in Scenario 2 of the simulation study  as a function of the severity level $ d $.}
%	
%	\label{fi_dataoc2}
%	
%	\centering
%	\begin{tabular}{cM{0.28\textwidth}M{0.28\textwidth}M{0.28\textwidth}}
%		\textbf{M1}&\includegraphics[width=.3\textwidth]{Figures/sim_dataoc_Scenario_II_1_OC_h.pdf}
%		&\includegraphics[width=.3\textwidth]{Figures/sim_dataoc_Scenario_II_1_OC_x.pdf}
%		&\includegraphics[width=.3\textwidth]{Figures/sim_dataoc_Scenario_II_1_OC_xh.pdf}	\\
%		\textbf{M2}&\includegraphics[width=.3\textwidth]{Figures/sim_dataoc_Scenario_II_2_OC_h.pdf}
%		&\includegraphics[width=.3\textwidth]{Figures/sim_dataoc_Scenario_II_2_OC_x.pdf}
%		&\includegraphics[width=.3\textwidth]{Figures/sim_dataoc_Scenario_II_2_OC_xh.pdf}\\
%		\textbf{M3}&\includegraphics[width=.3\textwidth]{Figures/sim_dataoc_Scenario_II_3_OC_h.pdf}
%		&\includegraphics[width=.3\textwidth]{Figures/sim_dataoc_Scenario_II_3_OC_x.pdf}
%		&\includegraphics[width=.3\textwidth]{Figures/sim_dataoc_Scenario_II_3_OC_xh.pdf}
%	\end{tabular}
%	\vspace{-.5cm}
%\end{figure}
\newpage
%\section{The FDR and TDR over the Different Fault Magnitudes in the Case Study}
% Figure show the mean TDR for each dataset and fault over the different fault magnitudes of  the FRTM and the NOAL and PW methods. Details on the faults can be found in \cite{van2015extensive}.
% Figure that the detection power of the FRTM is overall higher than the competitors, except for few exceptions where the proposed method achieves performance comparable than the competitors. This confirms what already presented in Section 4. 
%\begin{figure}[h]
%	\caption{Mean TDR over the different fault magnitudes achieved by FRTM (solid line), NOAL (dotted line) and PW (dashed line) applied to the CO\textsubscript{2} and the BFR for each fault and dataset.}
%	
%	\label{fi_results_cs}
%	
%	
%	
%	\hspace{-2.2cm}
%	\begin{tabular}{cM{0.27\textwidth}M{0.27\textwidth}M{0.27\textwidth}M{0.27\textwidth}}
%		\textbf{CO\textsubscript{2}}&\includegraphics[width=.280\textwidth]{Figures/cs_TD_1_1.pdf}
%		&\includegraphics[width=.280\textwidth]{Figures/cs_TD_dis_2_1_1.pdf}&\includegraphics[width=.280\textwidth]{Figures/cs_TD_3_1.pdf}	&
%		\includegraphics[width=.280\textwidth]{Figures/cs_TD_4_1.pdf}\\
%		\textbf{BFR}& \includegraphics[width=.280\textwidth]{Figures/cs_TD_1_2.pdf}
%		&\includegraphics[width=.280\textwidth]{Figures/cs_TD_2_2.pdf}& \includegraphics[width=.280\textwidth]{Figures/cs_TD_3_2.pdf}
%		&\includegraphics[width=.280\textwidth]{Figures/cs_TD_4_2.pdf}\\
%	\end{tabular}
%	\vspace{-.5cm}
%\end{figure}
\section{Additional Simulation Results}
\begin{figure*}[h]
	\caption{Mean FAR ($ d=0 $) or TDR ($ d\neq 0 $) plus/minus one standard error achieved by FRTM (solid line), NOAL (dotted line) and PW (dashed line), for each shift type (Shift A, B and C) and increasing misalignment level (M1, M2 and M3)   as a function of the severity level $ d $ in Scenario 1, with $ x^{*}_{out}=0.6 $.}
	
	\label{fi_results_1_06}
	
	\centering
	\begin{tabular}{cM{0.28\textwidth}M{0.28\textwidth}M{0.28\textwidth}}
		\textbf{M1}&\includegraphics[width=.3\textwidth]{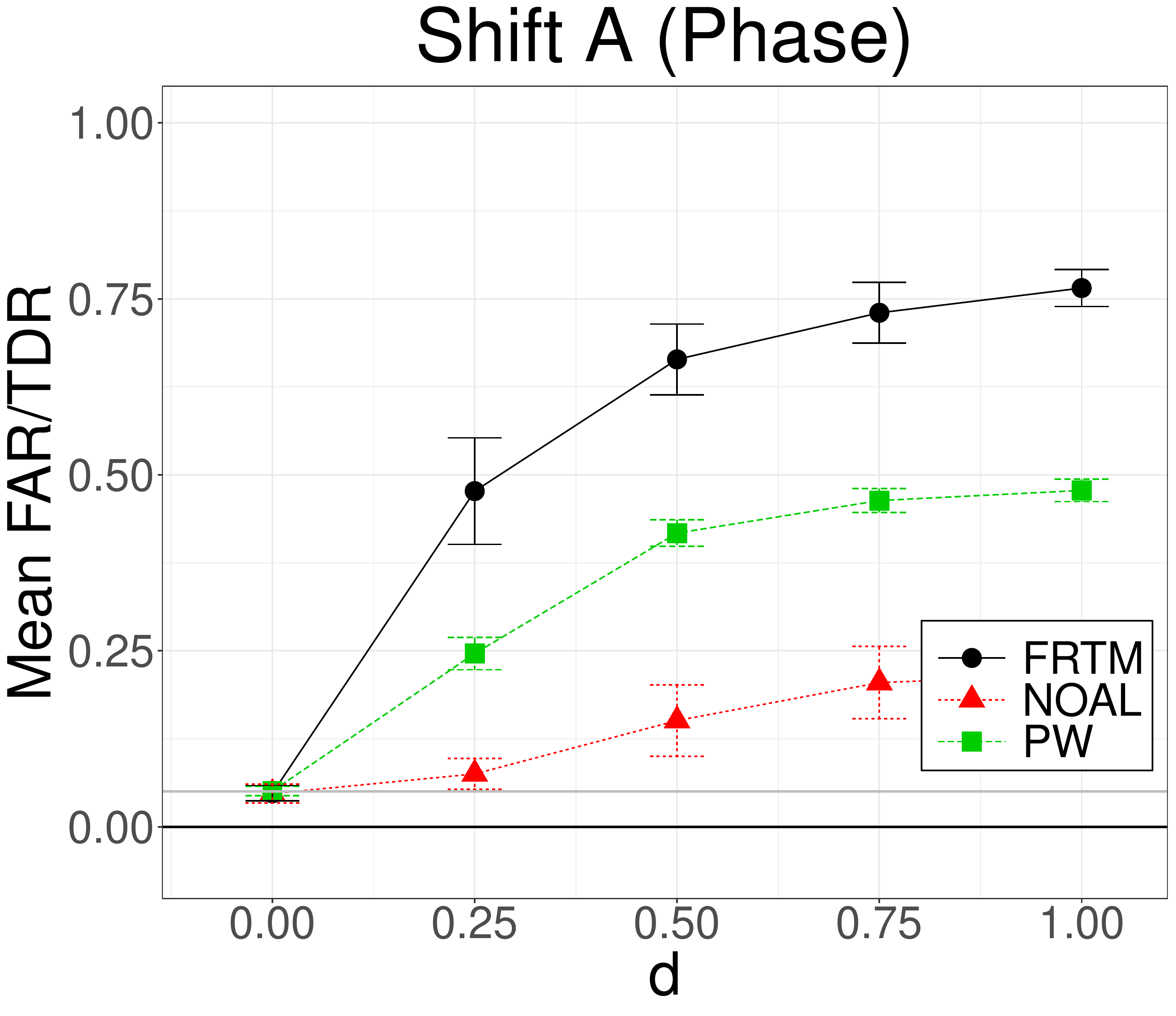}
		&\includegraphics[width=.3\textwidth]{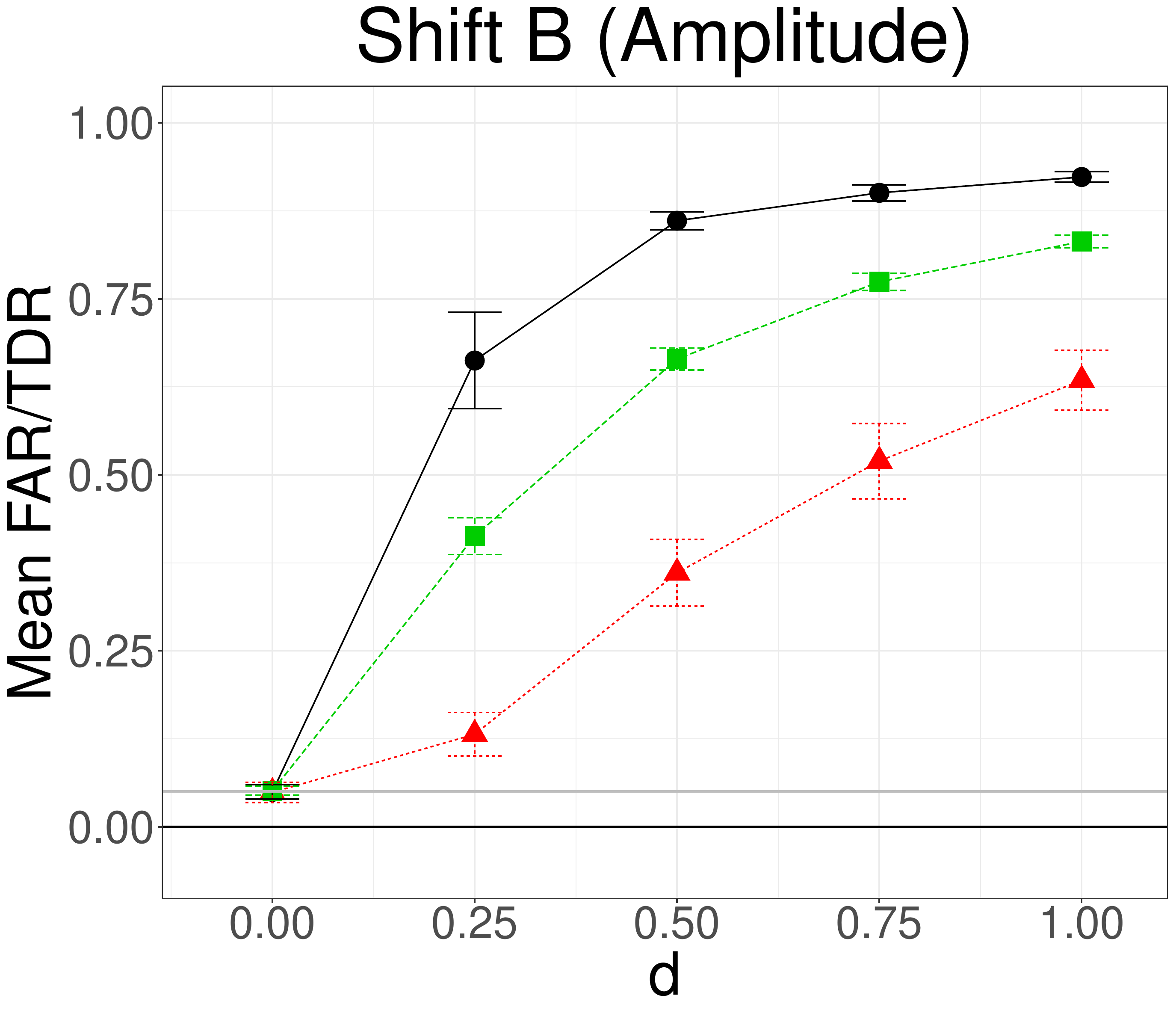}
		&\includegraphics[width=.3\textwidth]{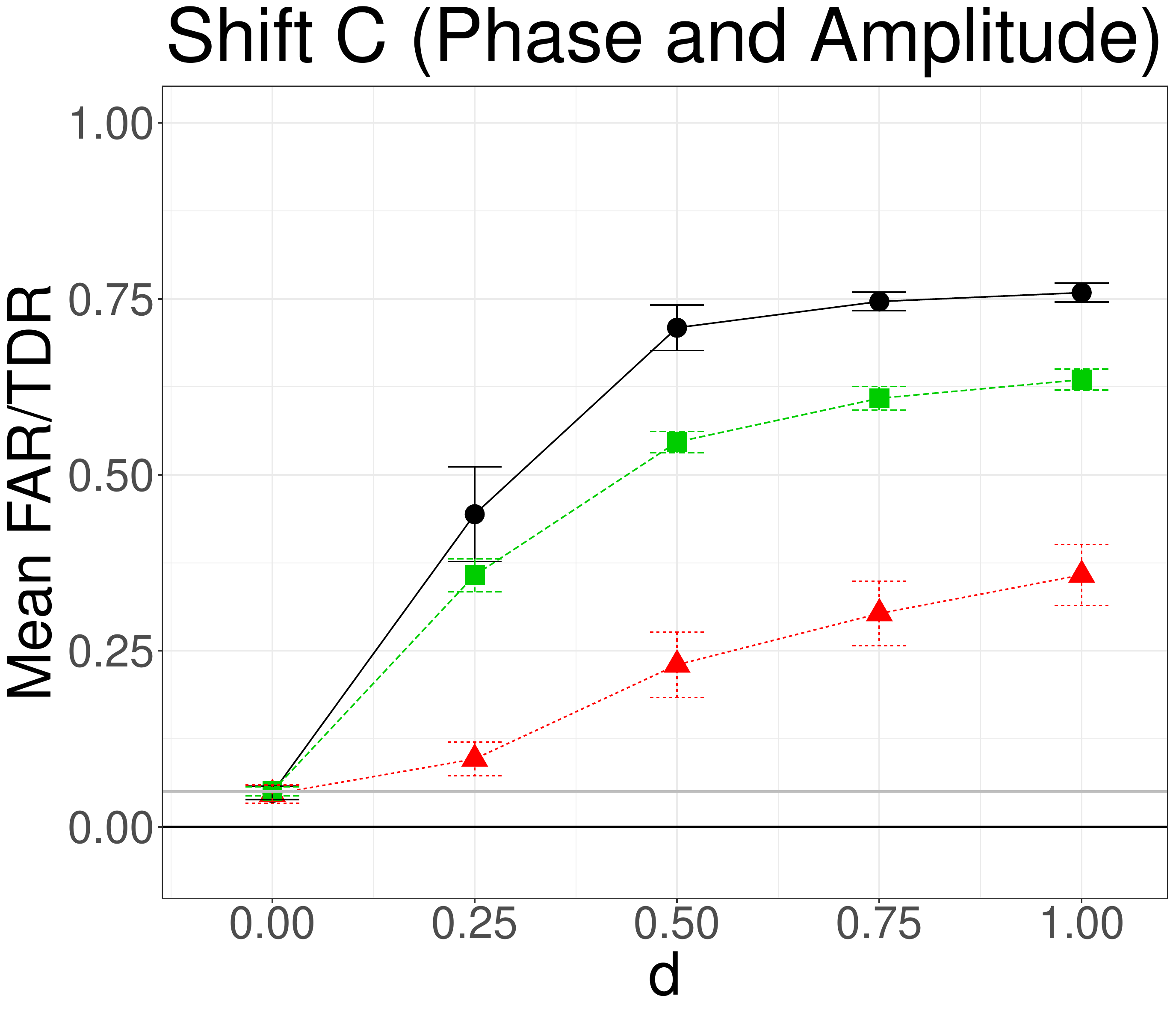}	\\
		\textbf{M2}&\includegraphics[width=.3\textwidth]{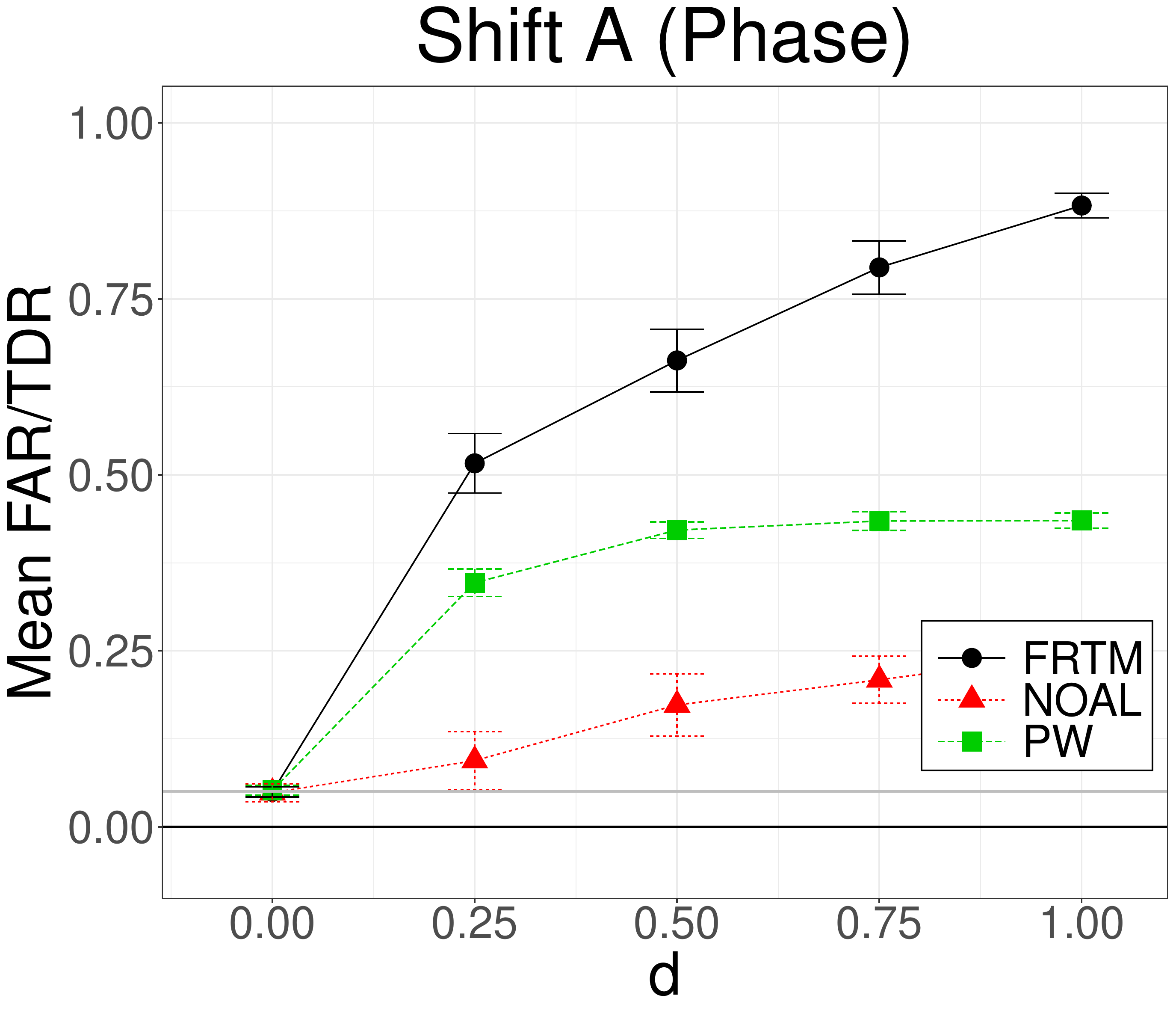}
		&\includegraphics[width=.3\textwidth]{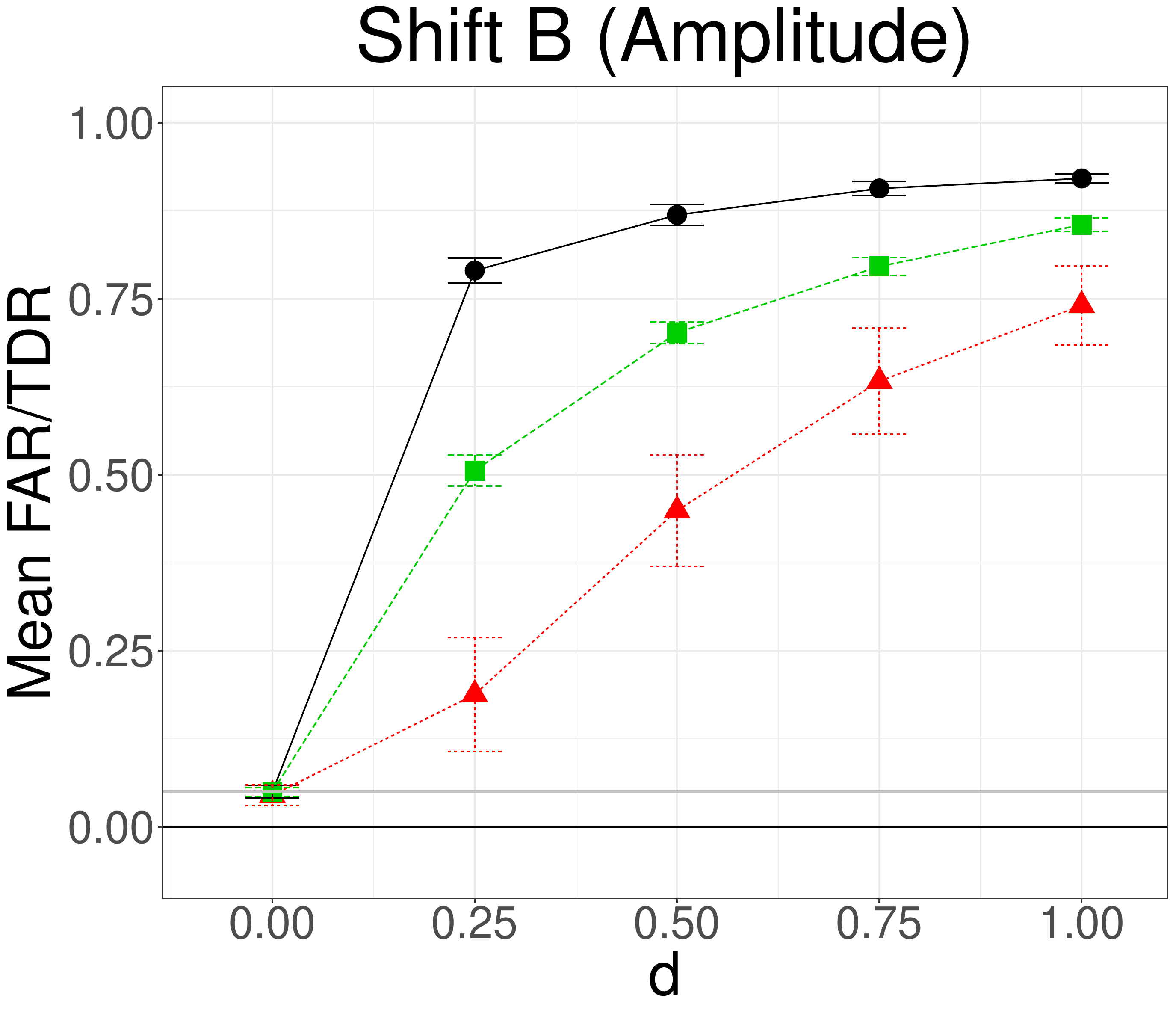}
		&\includegraphics[width=.3\textwidth]{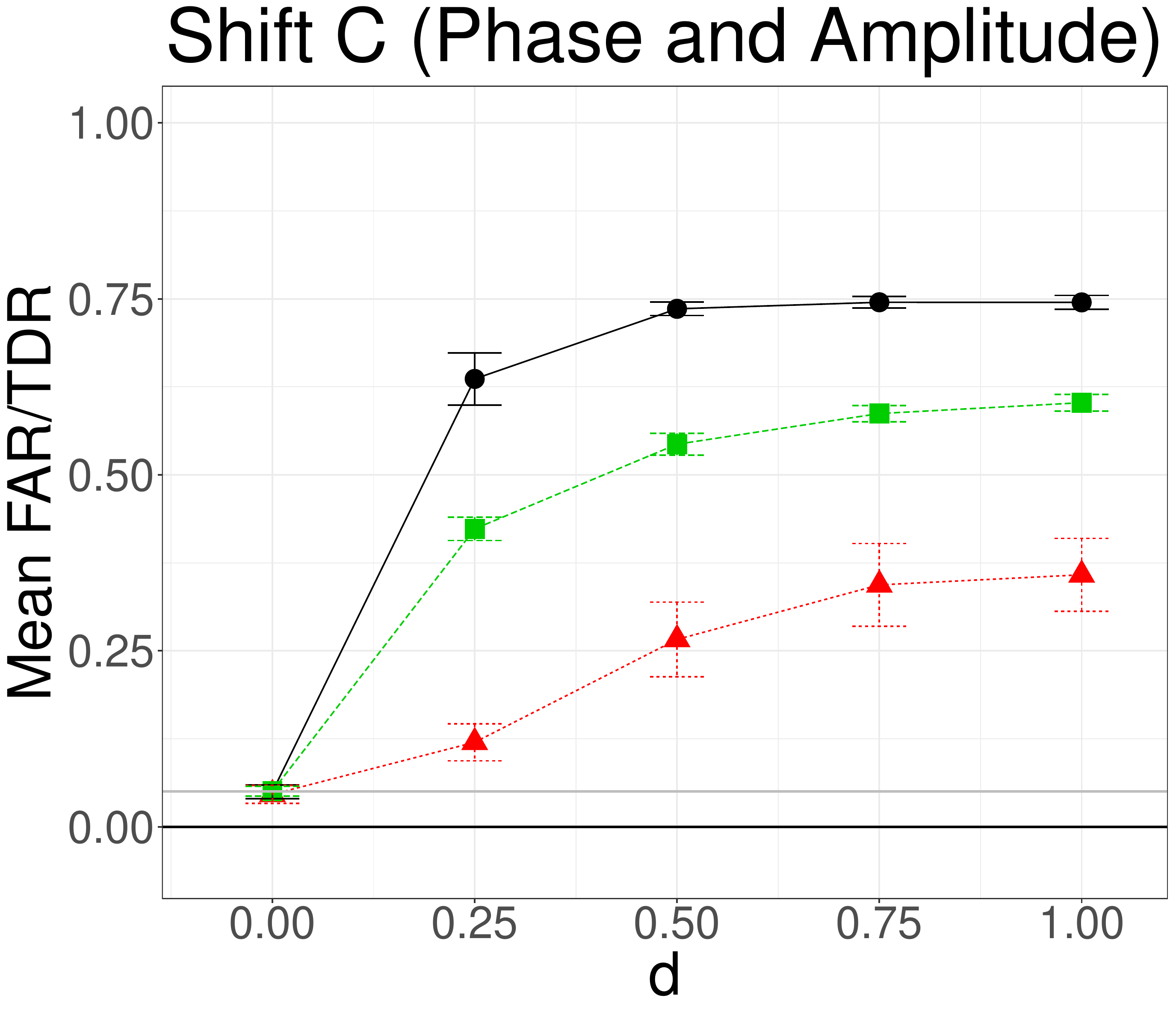}\\
		\textbf{M3}&\includegraphics[width=.3\textwidth]{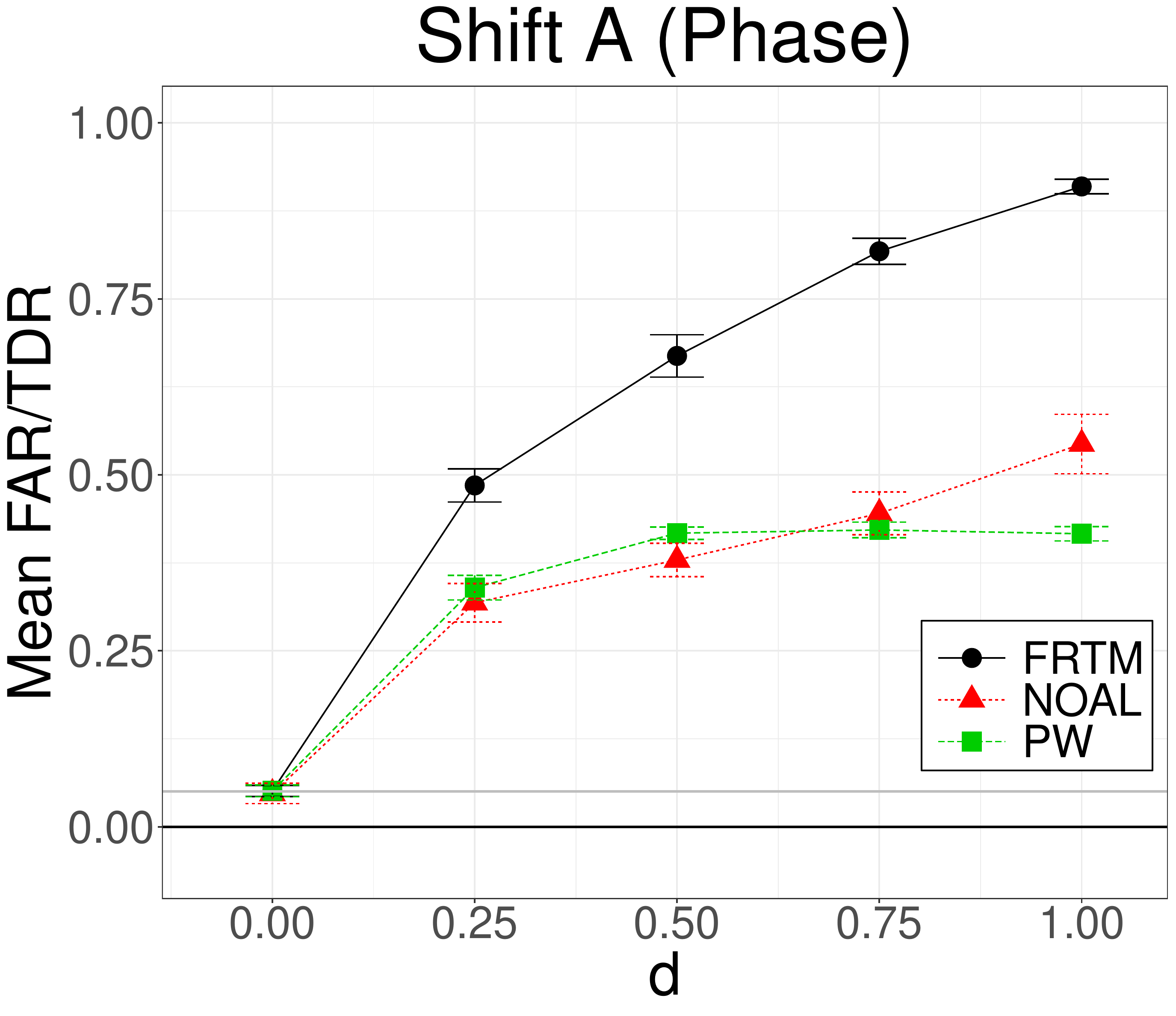}
		&\includegraphics[width=.3\textwidth]{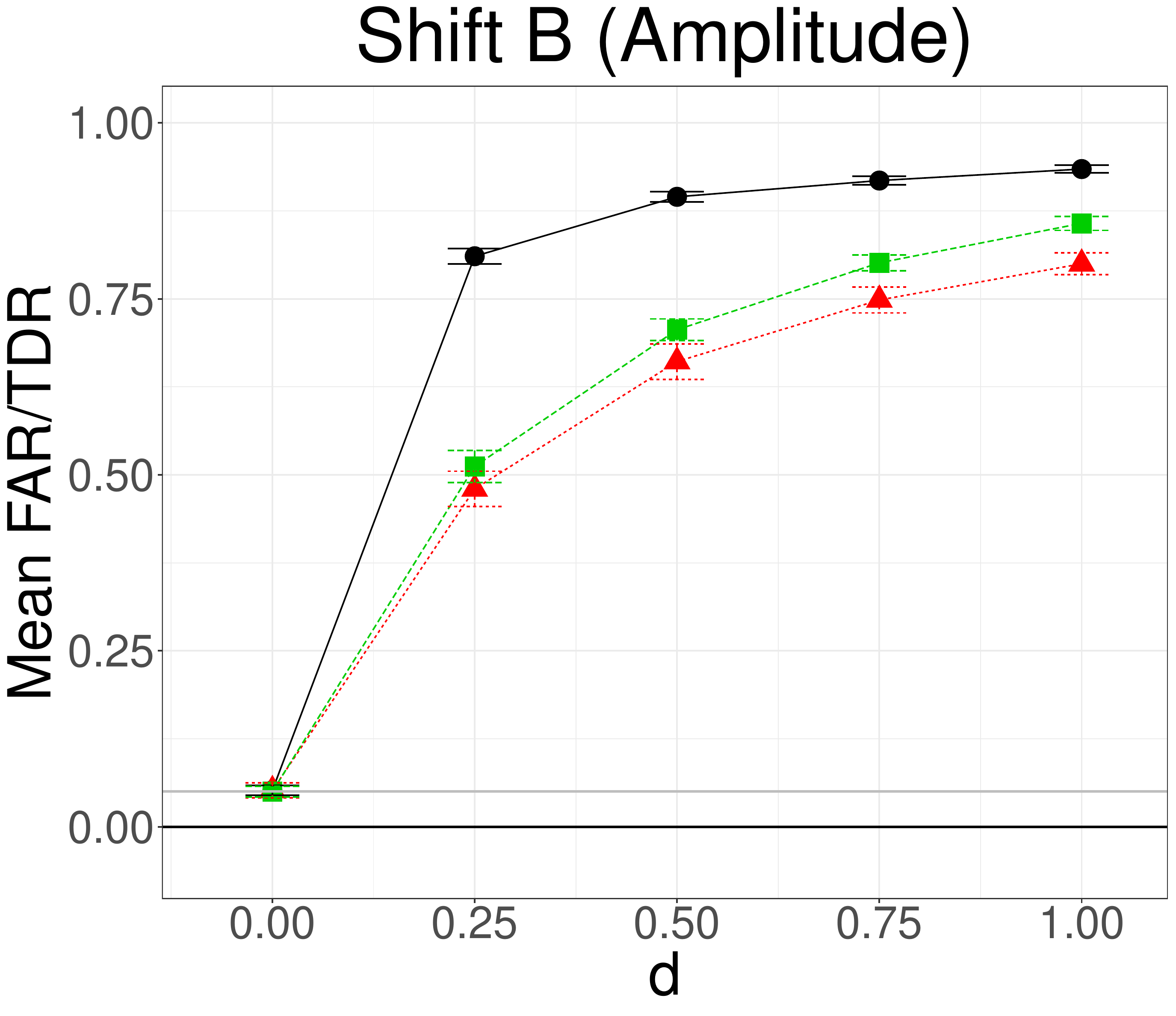}
		&\includegraphics[width=.3\textwidth]{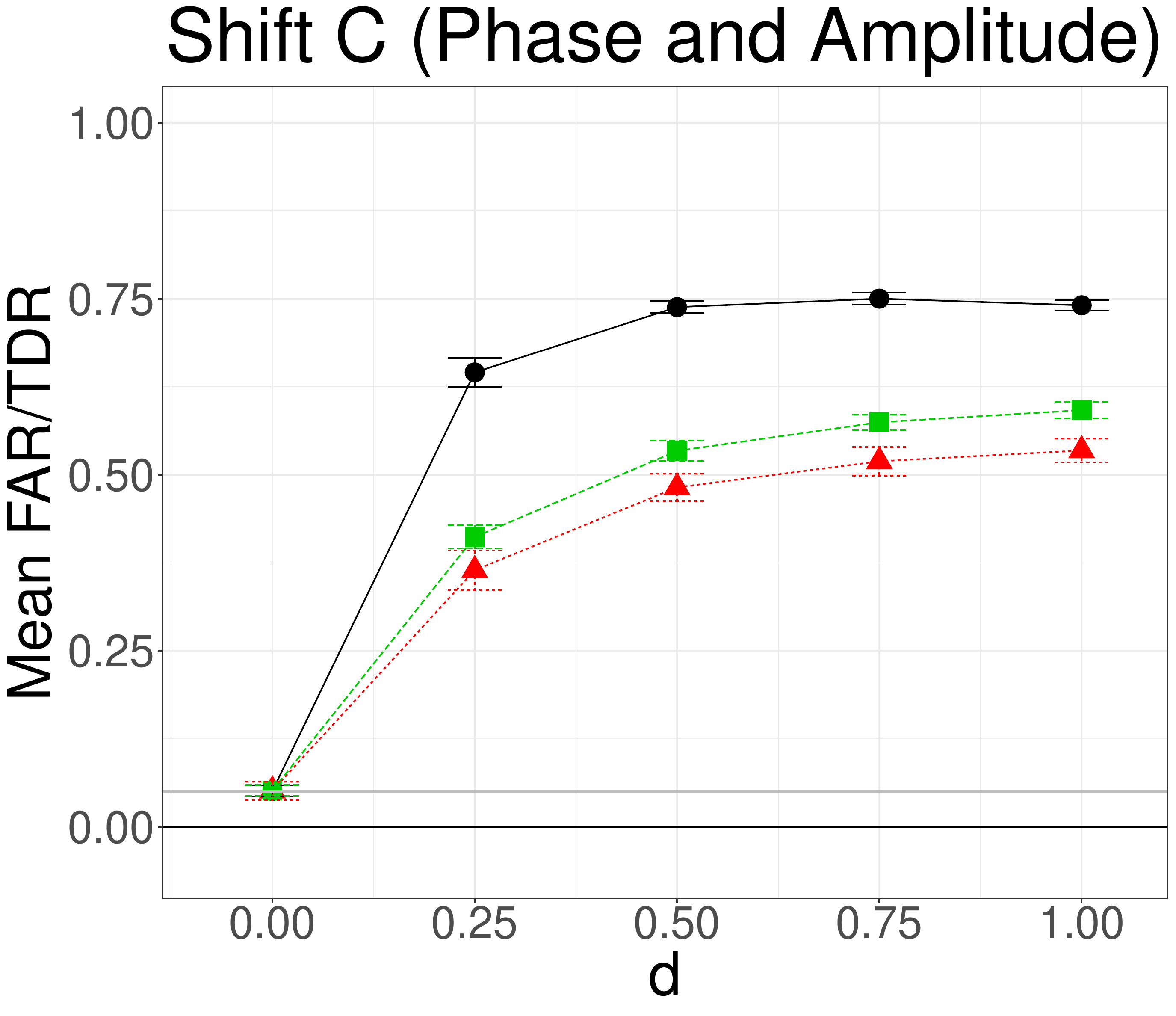}
	\end{tabular}
	\vspace{-.5cm}
\end{figure*}

\begin{figure}[h]
	\caption{Mean FAR ($ d=0 $) or TDR ($ d\neq 0 $) plus/minus one standard error achieved by FRTM (solid line), NOAL (dotted line) and PW (dashed line), for each shift type (Shift A, B and C) and increasing misalignment level (M1, M2 and M3)   as a function of the severity level $ d $ in Scenario 2, with $ x^{*}_{out}=0.6 $.}
	
	\label{fi_results_2_06}
	
	\centering
	\begin{tabular}{cM{0.28\textwidth}M{0.28\textwidth}M{0.28\textwidth}}
		\textbf{M1}&\includegraphics[width=.3\textwidth]{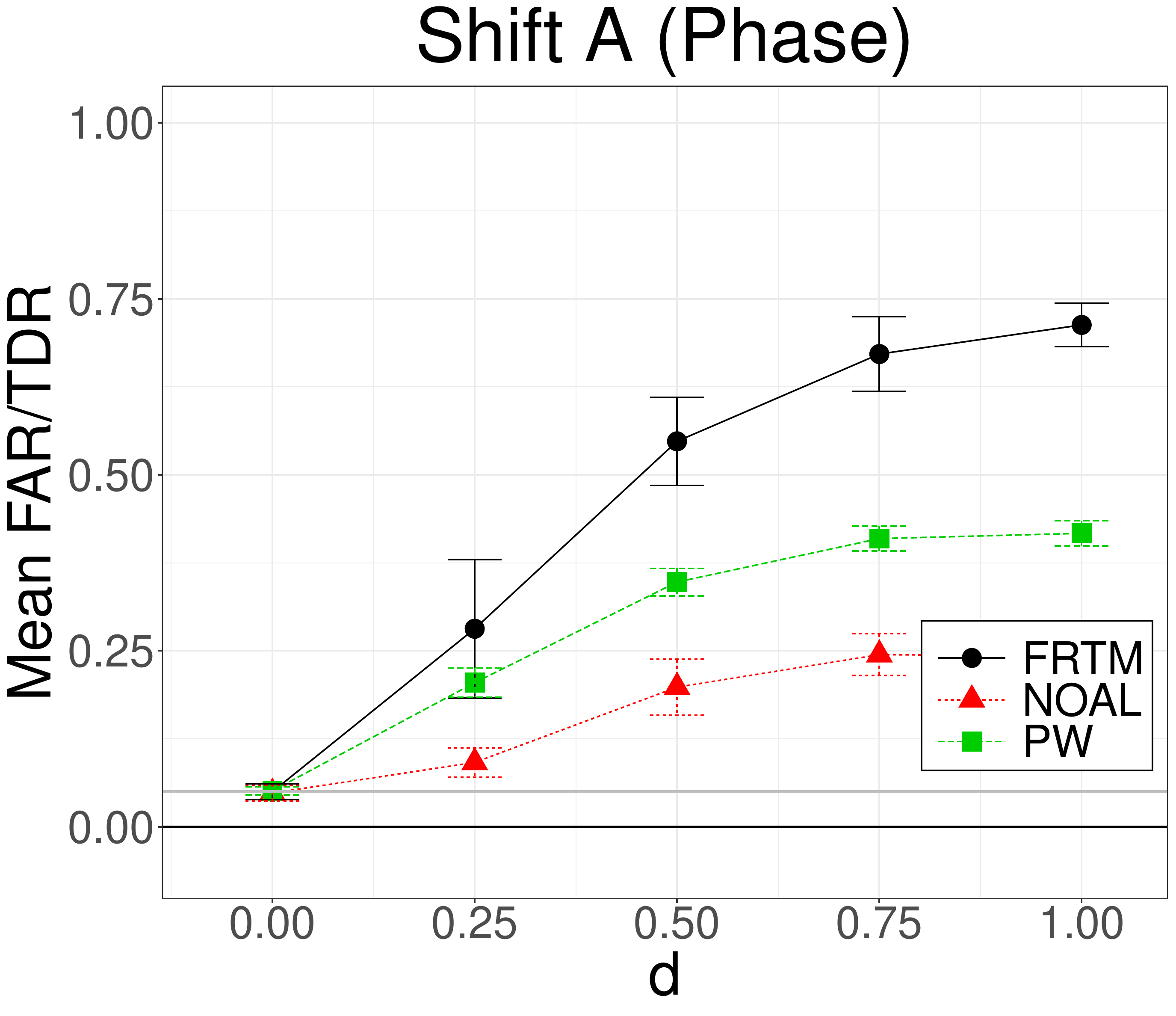}
		&\includegraphics[width=.3\textwidth]{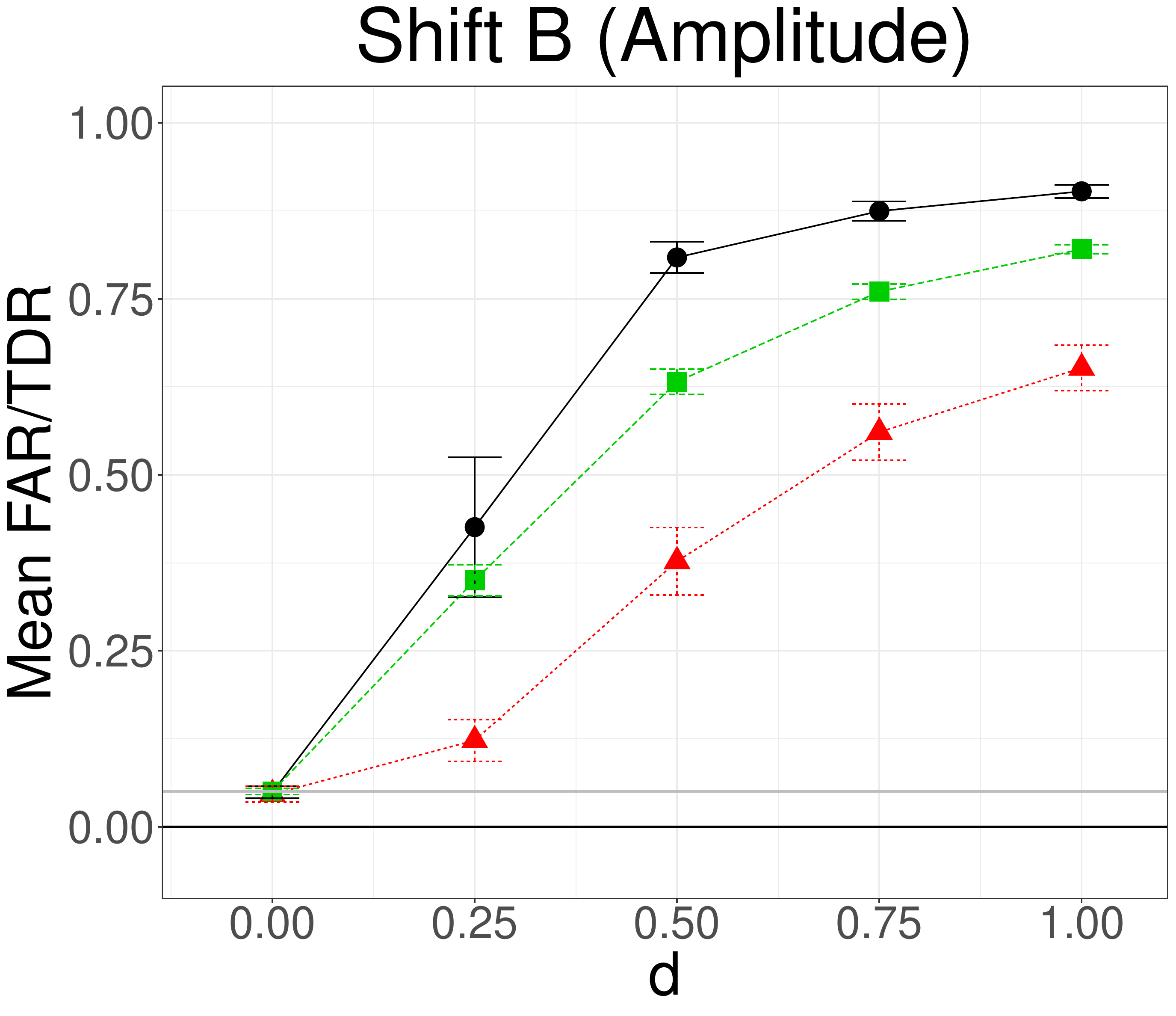}
		&\includegraphics[width=.3\textwidth]{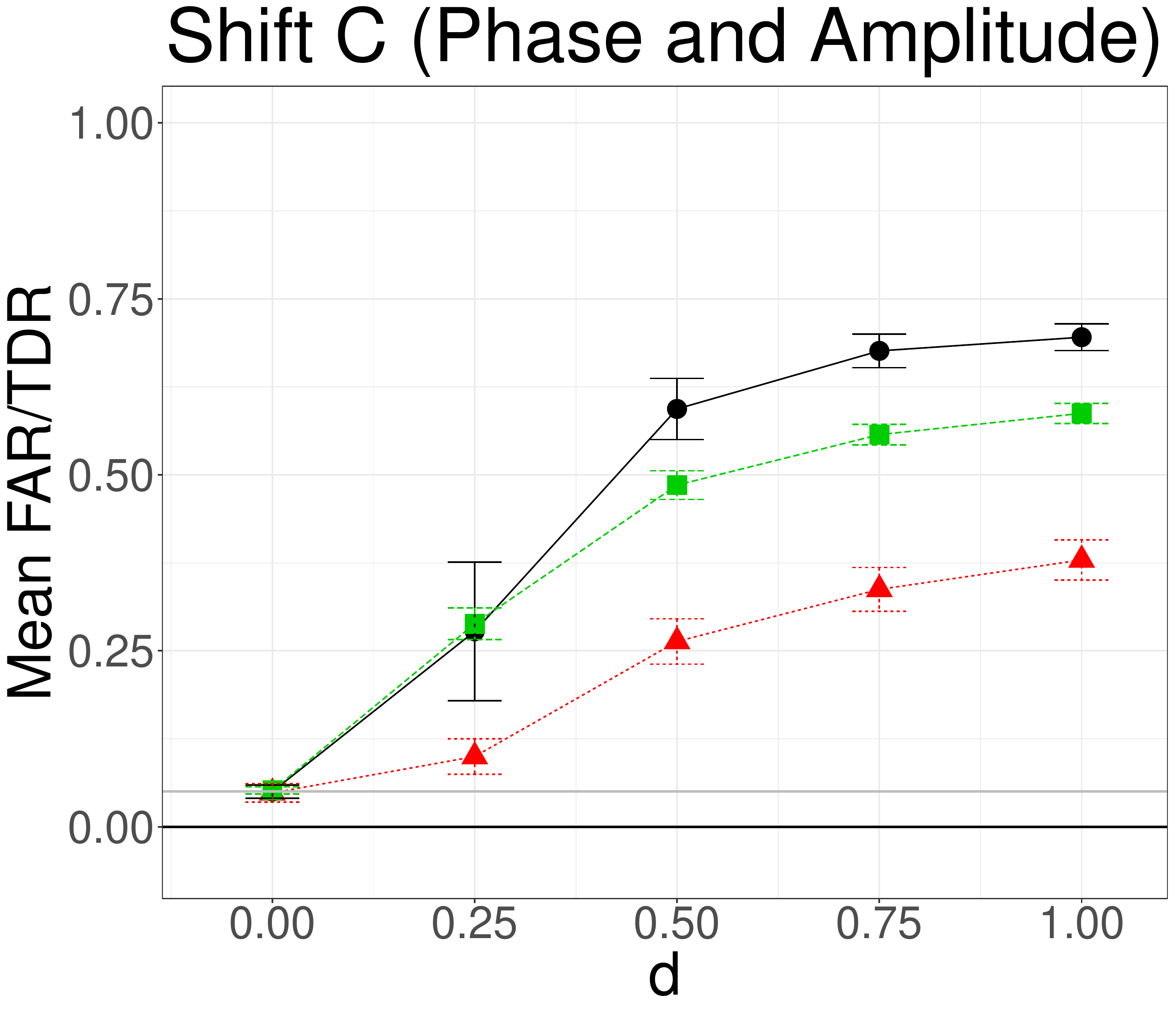}	\\
		\textbf{M2}&\includegraphics[width=.3\textwidth]{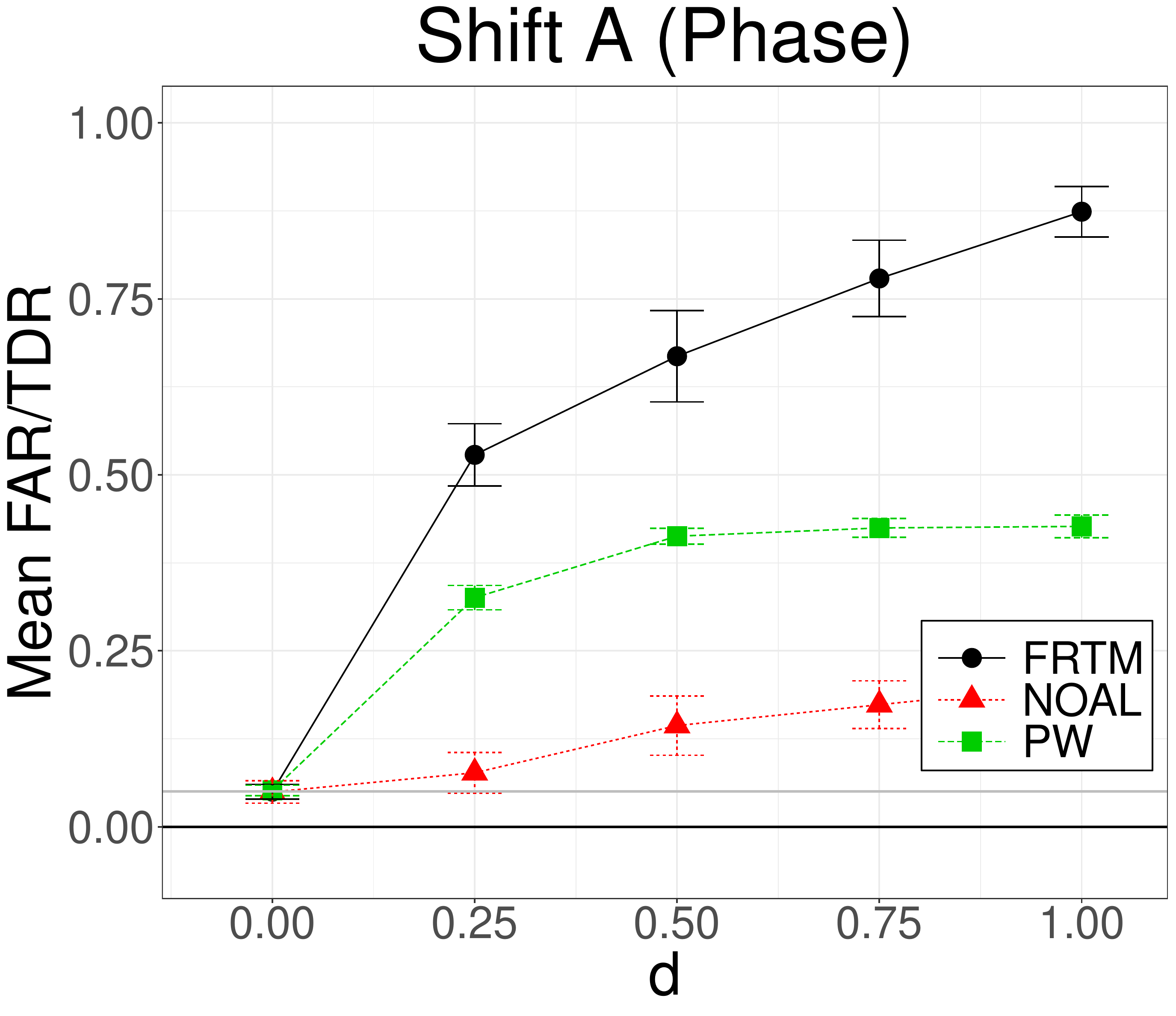}
		&\includegraphics[width=.3\textwidth]{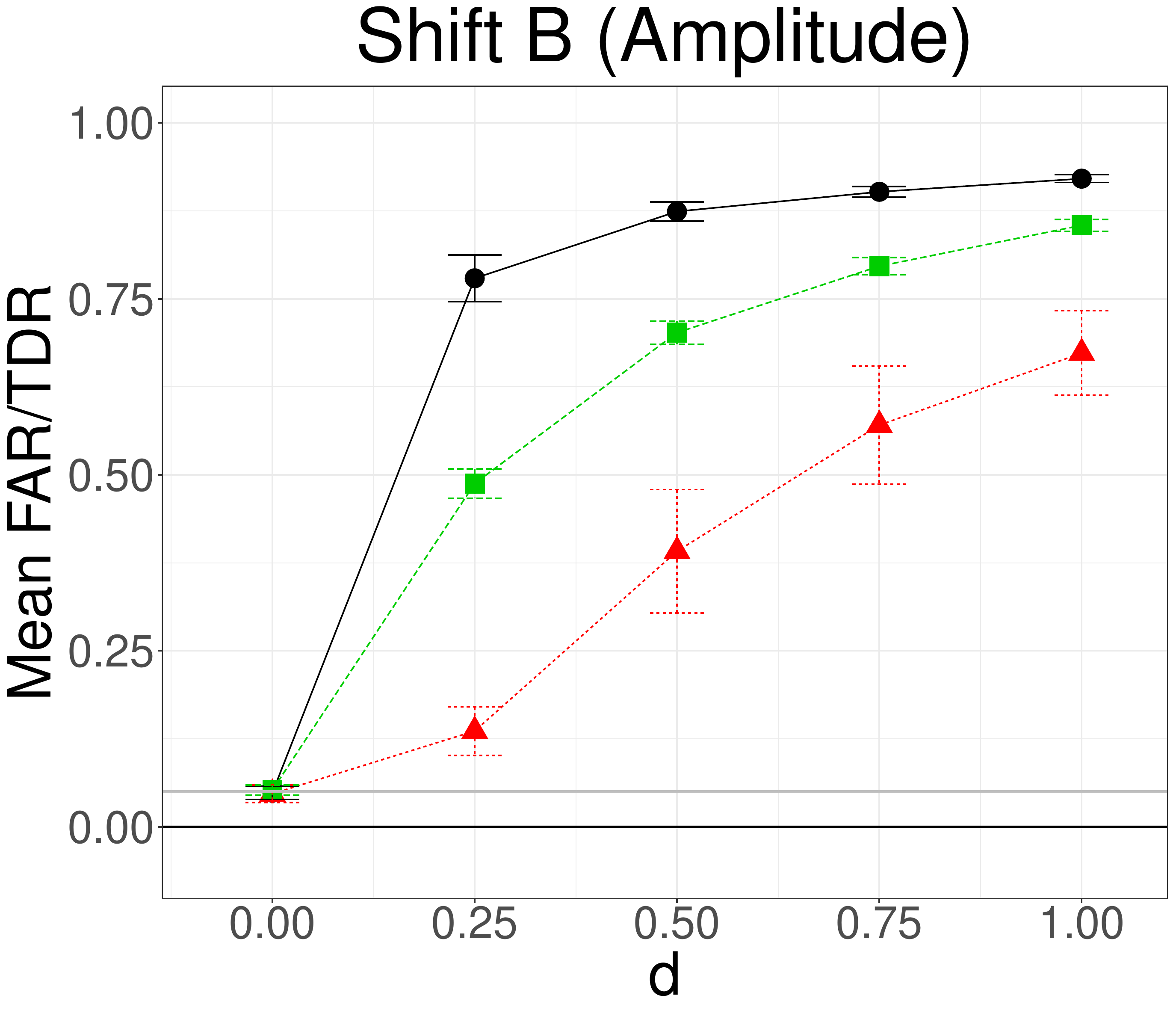}
		&\includegraphics[width=.3\textwidth]{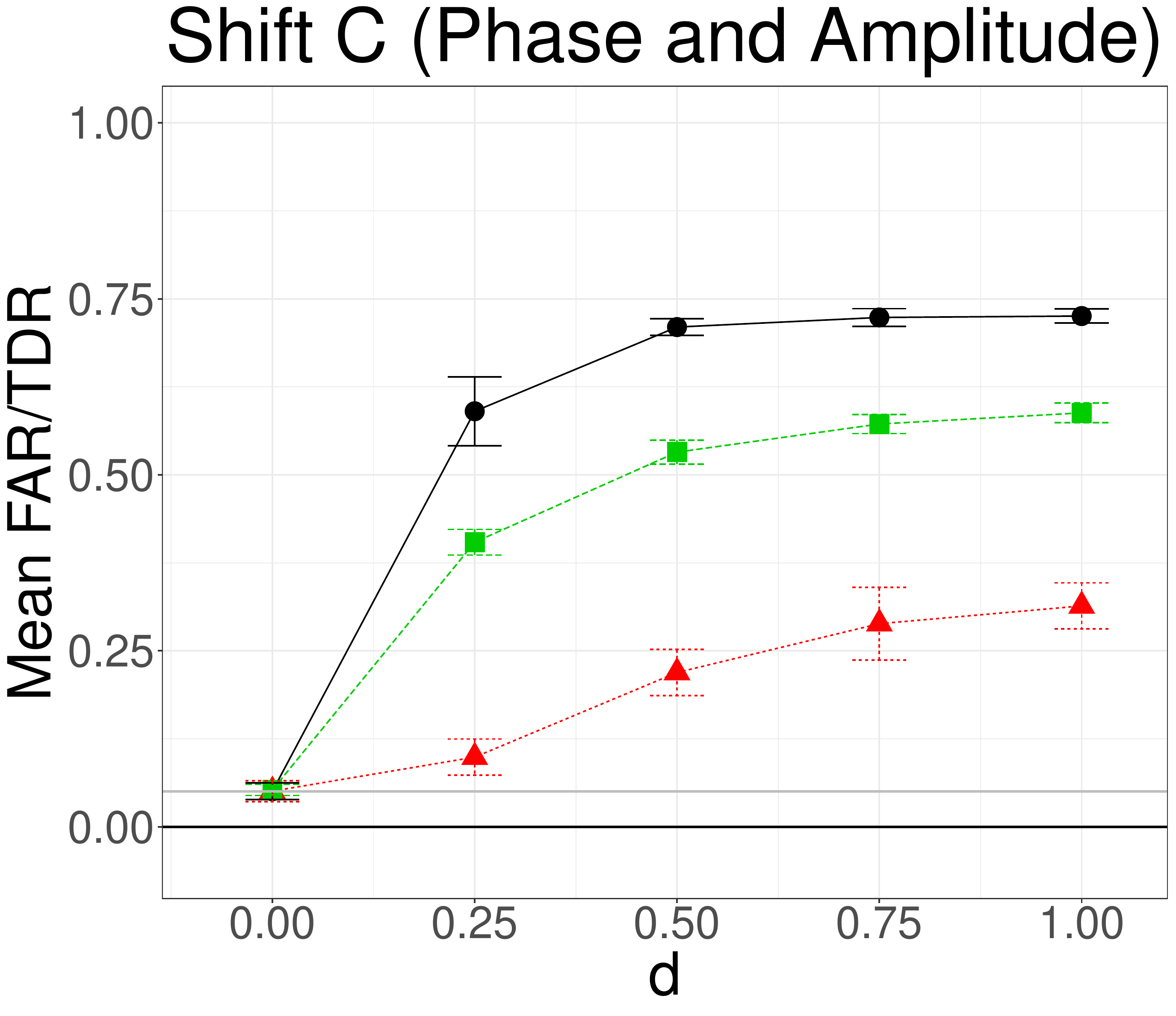}\\
		\textbf{M3}&\includegraphics[width=.3\textwidth]{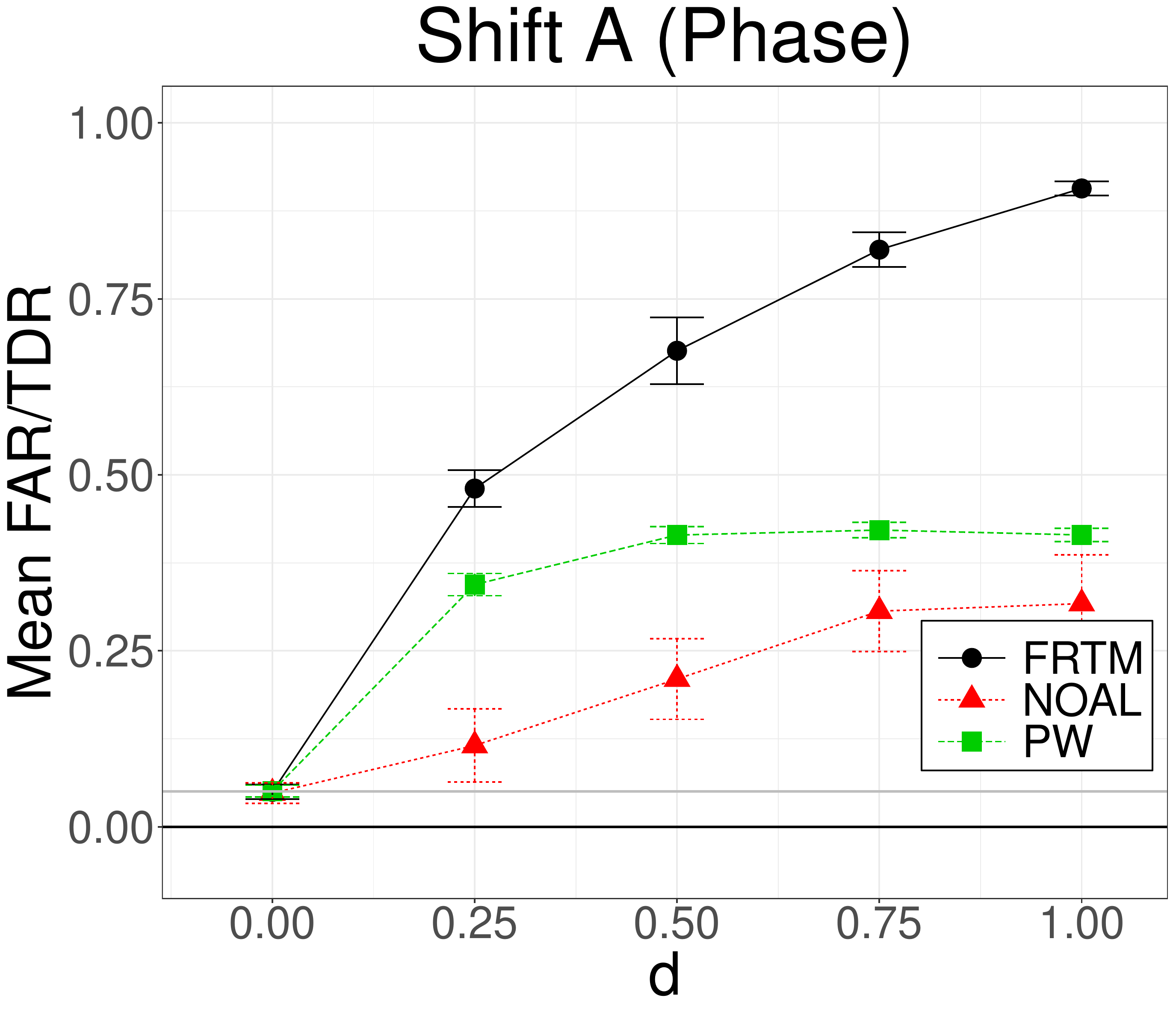}
		&\includegraphics[width=.3\textwidth]{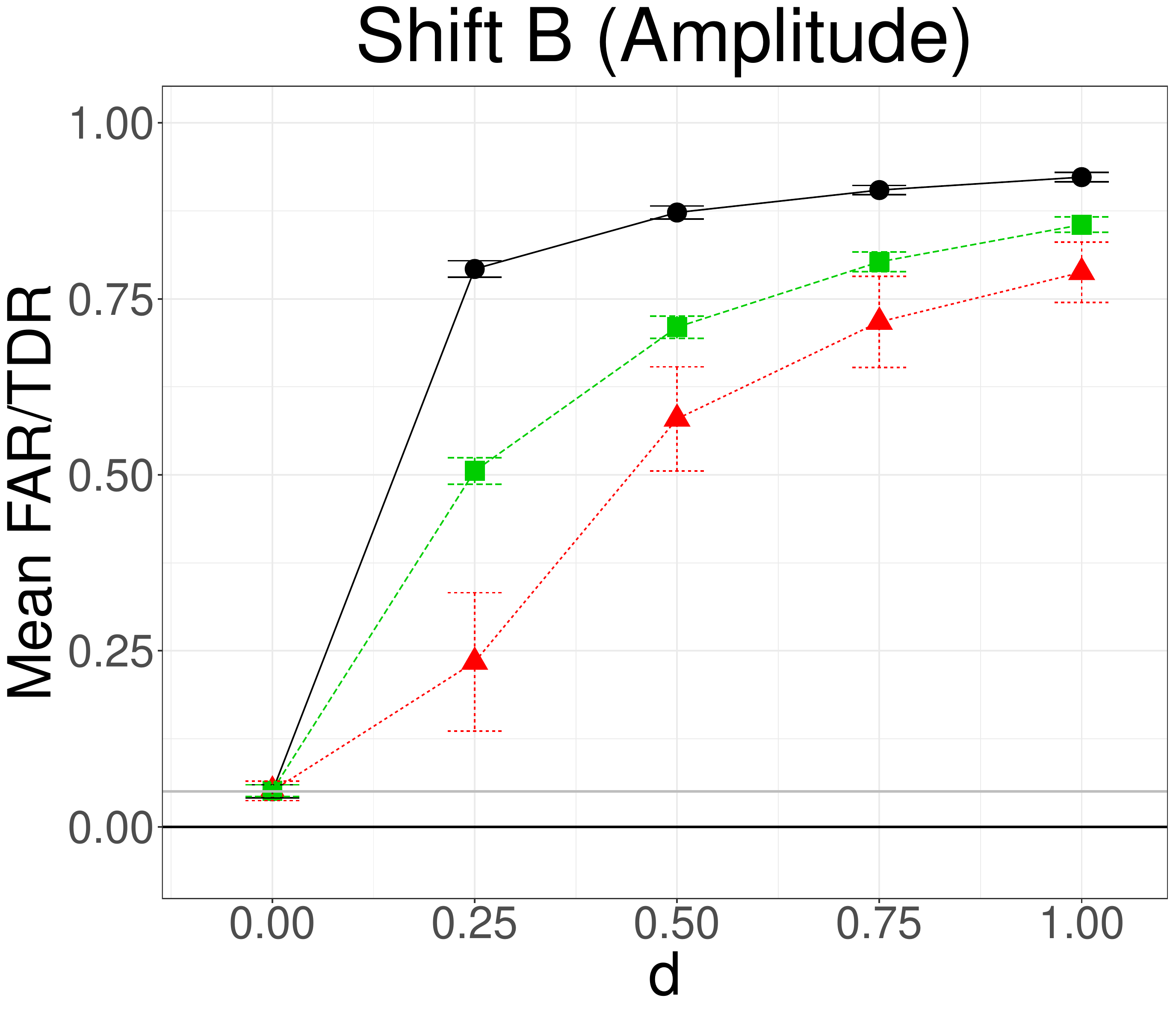}
		&\includegraphics[width=.3\textwidth]{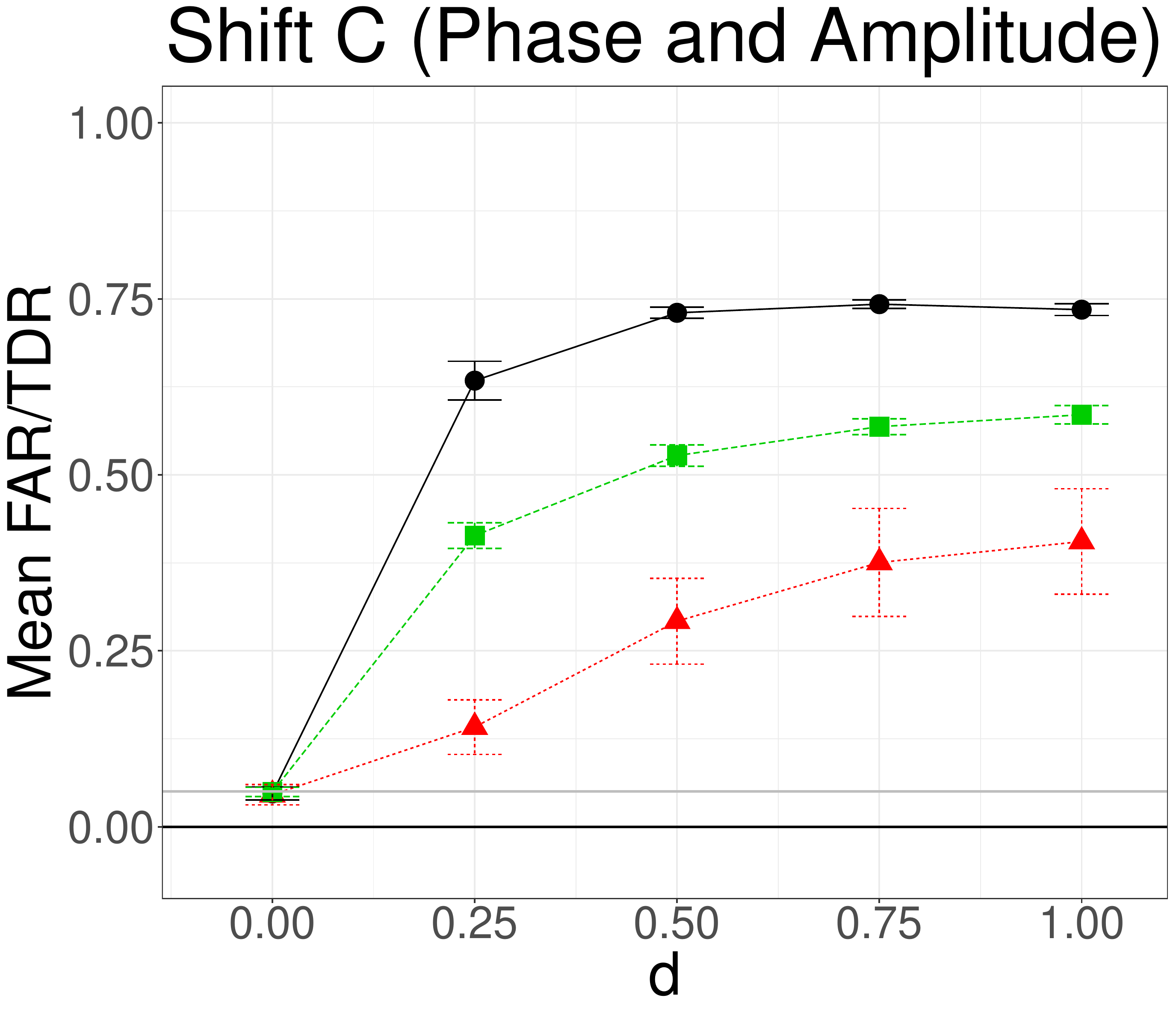}
	\end{tabular}
	\vspace{-.5cm}
\end{figure}

Additional simulations are presented to analyse the performance of  FRTM and the NOAL and PW methods  when OC  behaviours start at the 60\% of the process total duration, i.e., $ x^{*}_{out}=0.6 $.  FRTM is implemented as described in Section 3.
Figure \ref{fi_results_1_06} and Figure \ref{fi_results_2_06} show for Scenario 1 and Scenario 2 and for M1, M2 and M3,  the mean FARs and TDRs  plus/minus one standard error  for each shift type as a function of  the  severity level, with $ x^{*}_{out}=0.6 $. Note that $ d=0 $ corresponds to a Phase II sample that is in an IC state and thus, in this case, the FAR is studied.

By comparing Figure \ref{fi_results_1_06} and Figure \ref{fi_results_2_06} with Figure 2 and Figure 3,  FRTM  confirms itself as the best method in all considered scenarios.
However, we note an overall loss of detection power for  FRTM and the NOAL method, with respect to the case of $ x^{*}_{out}=0.3 $, differently from the PW method, which shows similar performance.
This behaviour is expected, because both  FRTM and the NOAL are functional methods that take into account information about the whole data trajectory of the partially observed functional data.
This feature allows them to be  effective in identifying persistent more than short OC behaviours. Instead,  the PW method,  thanks to its pointwise nature, is more reactive to sudden OCs.
Thus, this setting is less favourable to  FRTM and NOAL method  than the scenarios considered in Section 3, because the OC state arises later in the process.
Nevertheless,  FRTM stands out still as the best method in terms of TDR, because of its ability to successfully combine information from both  amplitude and phase components. 
Note that,  differently from Section 3,   the amplitude variation dominates the data variability and thus, the NOAL method performance is still much worse than that of  FRTM. This comes from the fact that the warping transformation employed through the mFPCA step allows  FRTM to be more sensitive in identifying shifts that affect the amplitude and phase components. 
As in Section 3, the greater complexity of the warping component in Scenario 2 reduces the detection power of  FRTM, especially for a large fraction of phase variation. However,  FRTM shows the best performance  in this case as well. In addition, the NOAL method badly performs also when the amplitude variation dominates the data variability. This happens because the   last part of the process is characterized by a larger amplitude variability, which masks the OC states in the reduced space considered by the NOAL method.  FRTM is clearly less affected by this issue because OC conditions are  detected  in the  phase component space as well.

\section{Additional Plots in the Data Example}
\begin{figure}[h]
	\caption{The 400  IC observations of the O\textsubscript{2}  and the BFR for D1,  D2, D3 and D4.}
	
	\label{fi_csvar}
	
	\centering
		\begin{tabular}{cM{0.45\textwidth}M{0.45\textwidth}}
		D1&	\includegraphics[width=.49\textwidth]{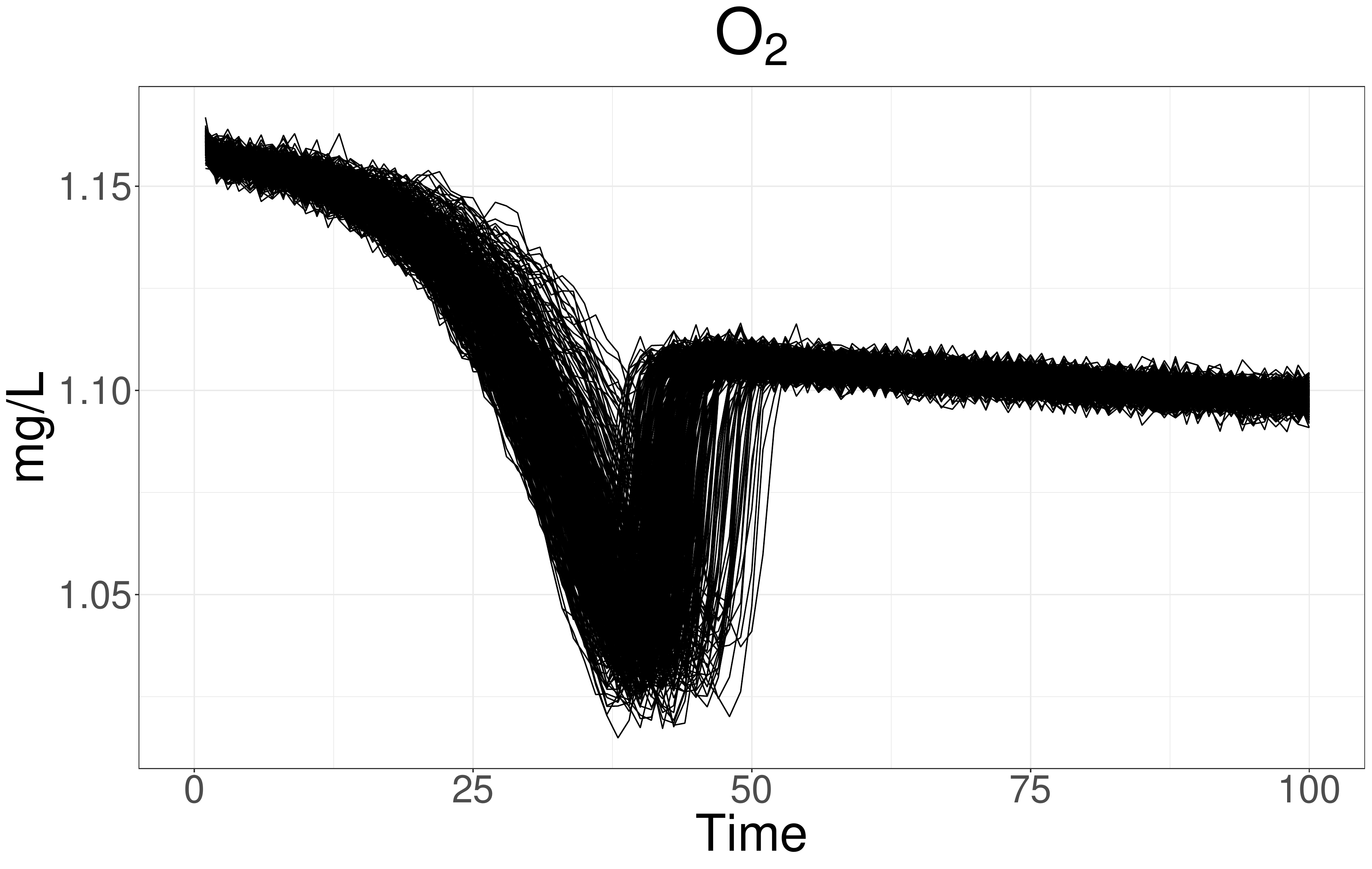}&
	\includegraphics[width=.49\textwidth]{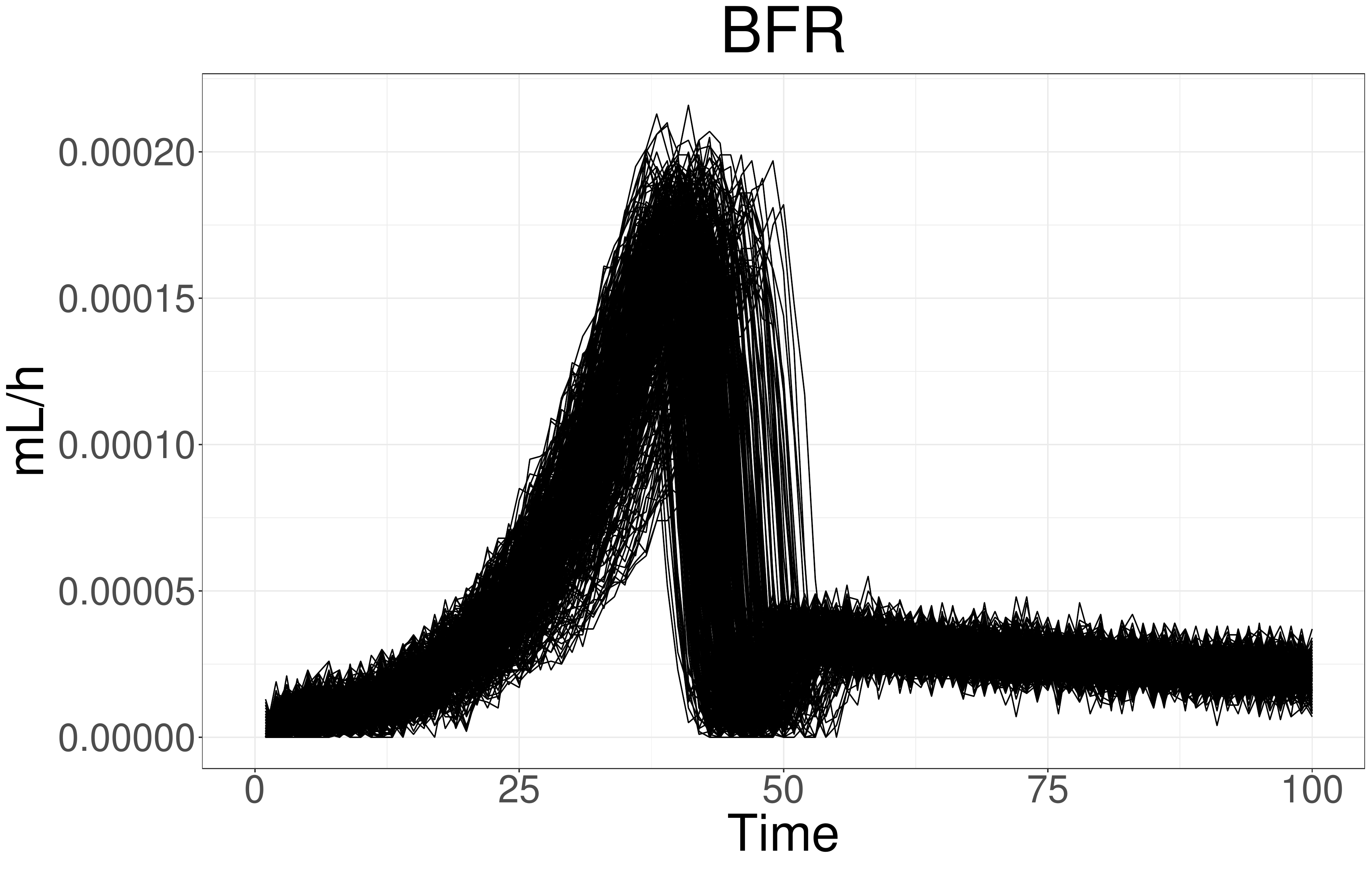}\\
		D2&\includegraphics[width=.49\textwidth]{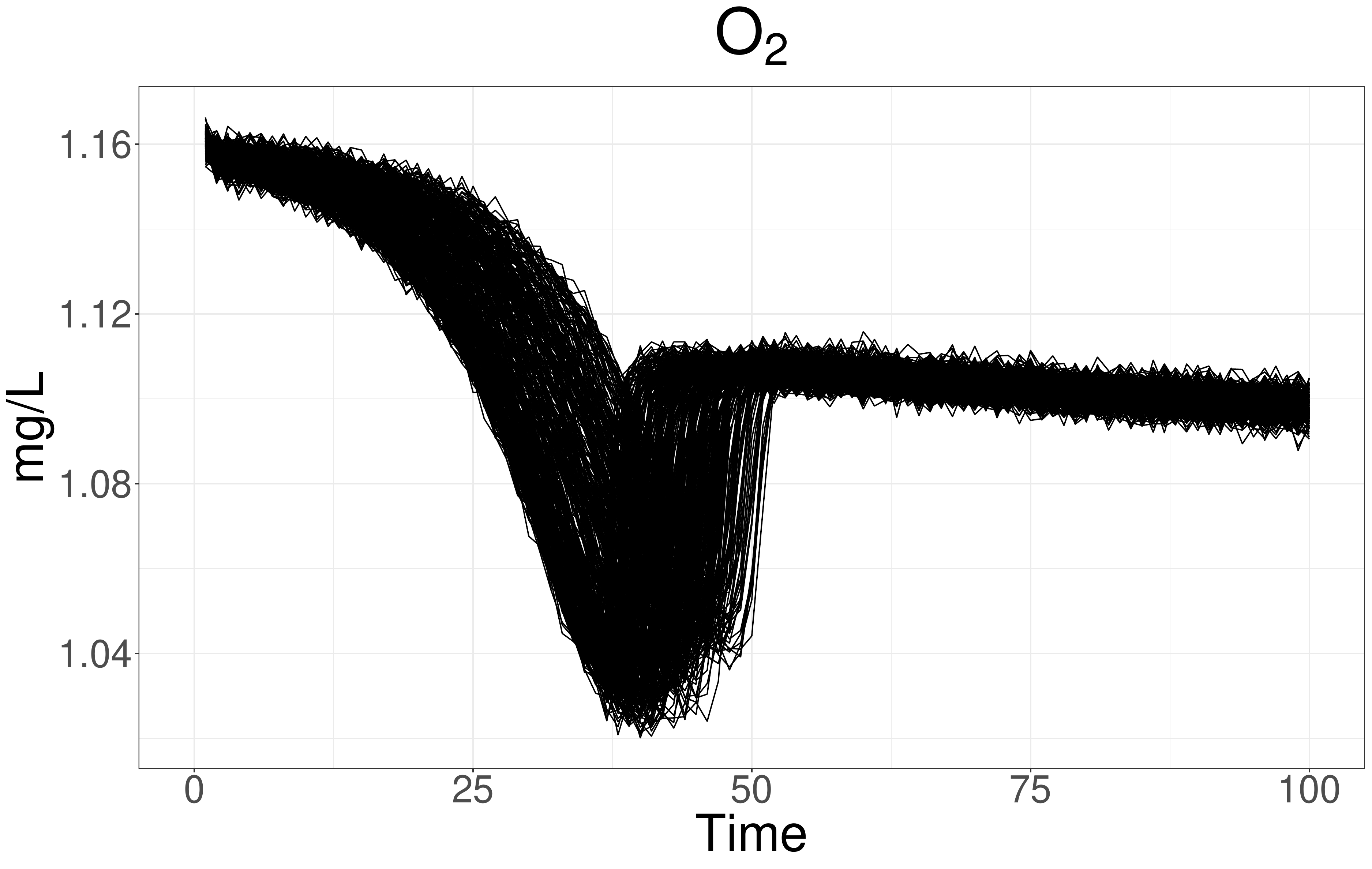}&
	\includegraphics[width=.49\textwidth]{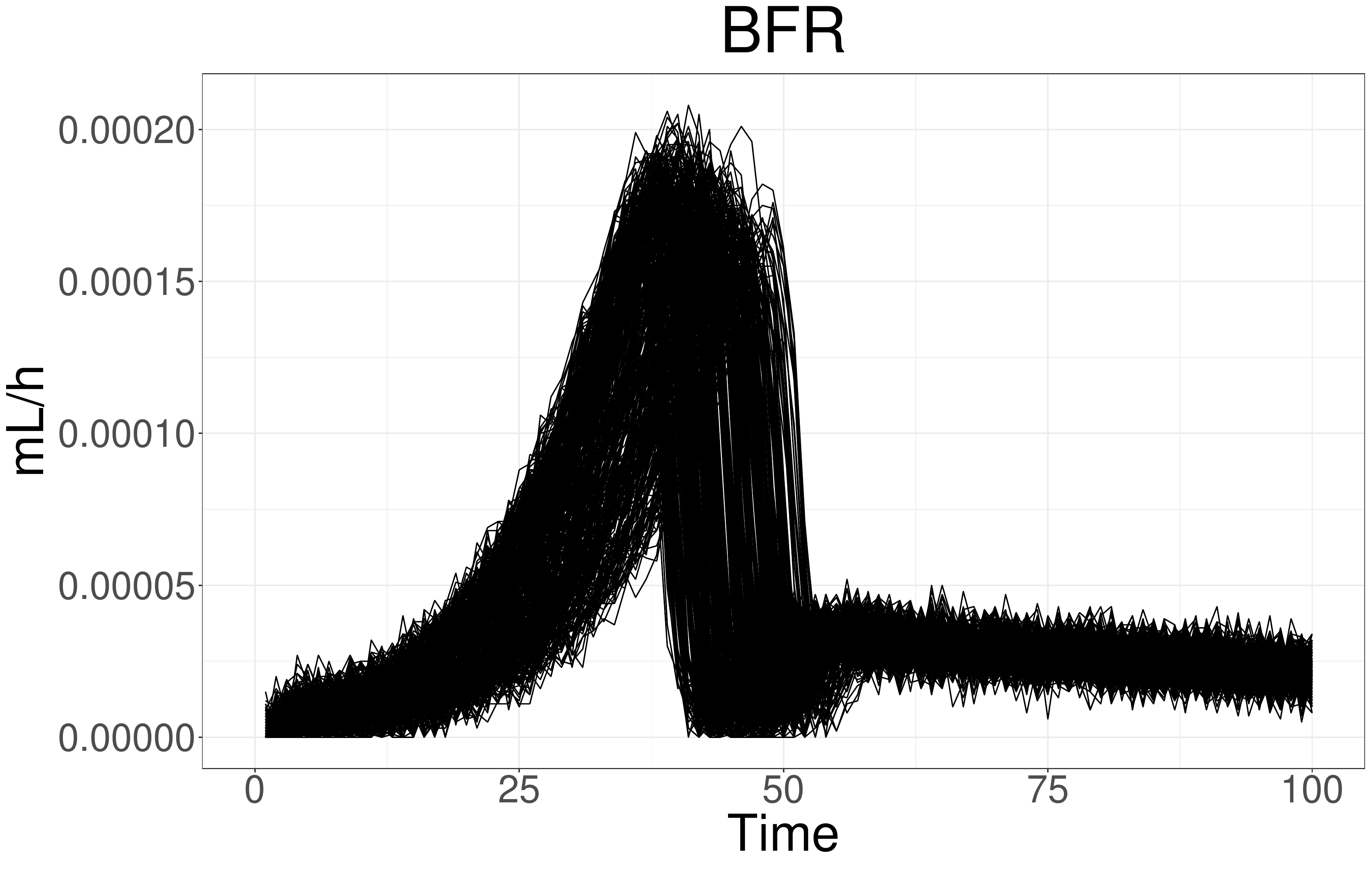}\\
		D3&\includegraphics[width=.49\textwidth]{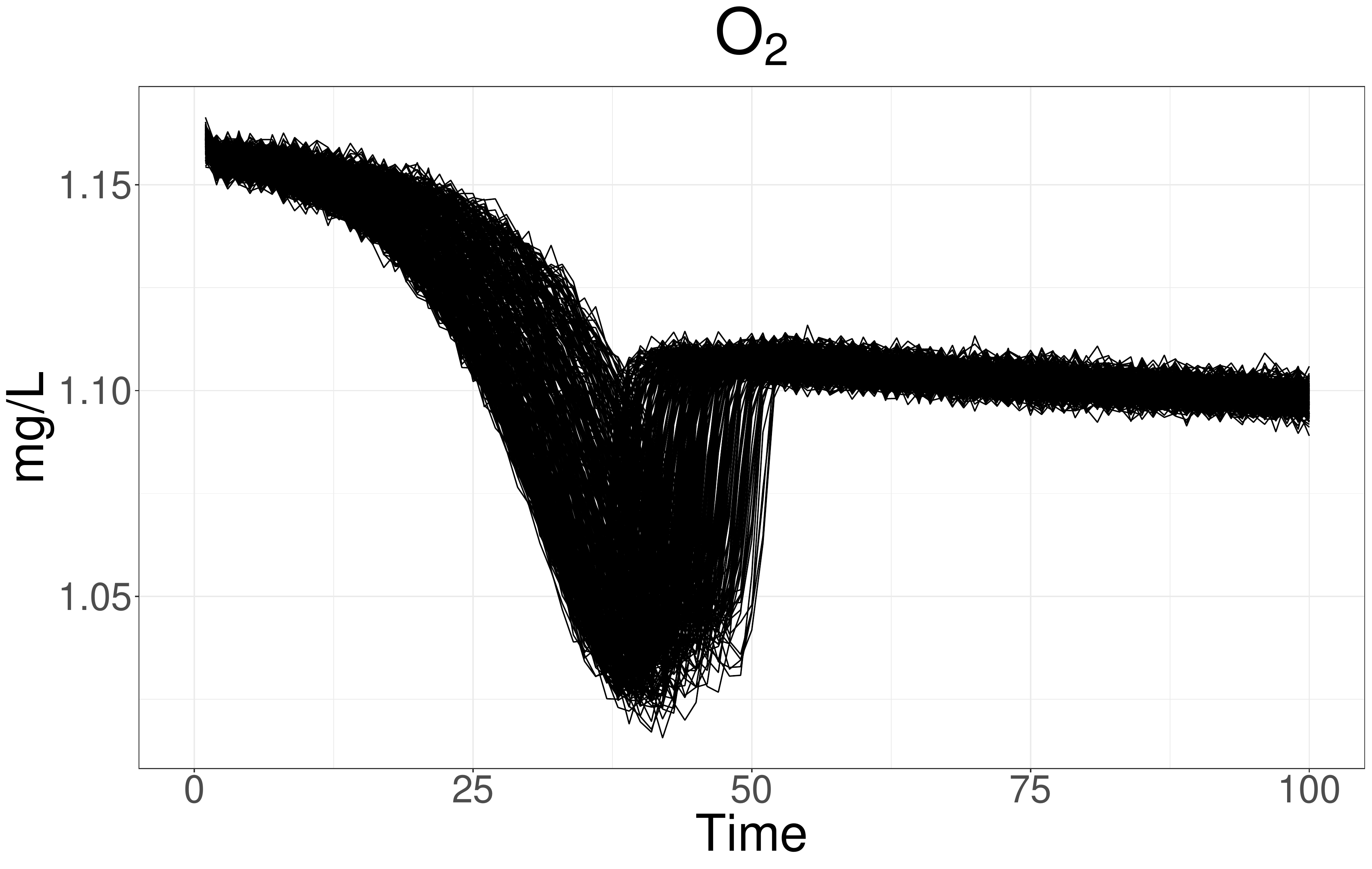}&
	\includegraphics[width=.49\textwidth]{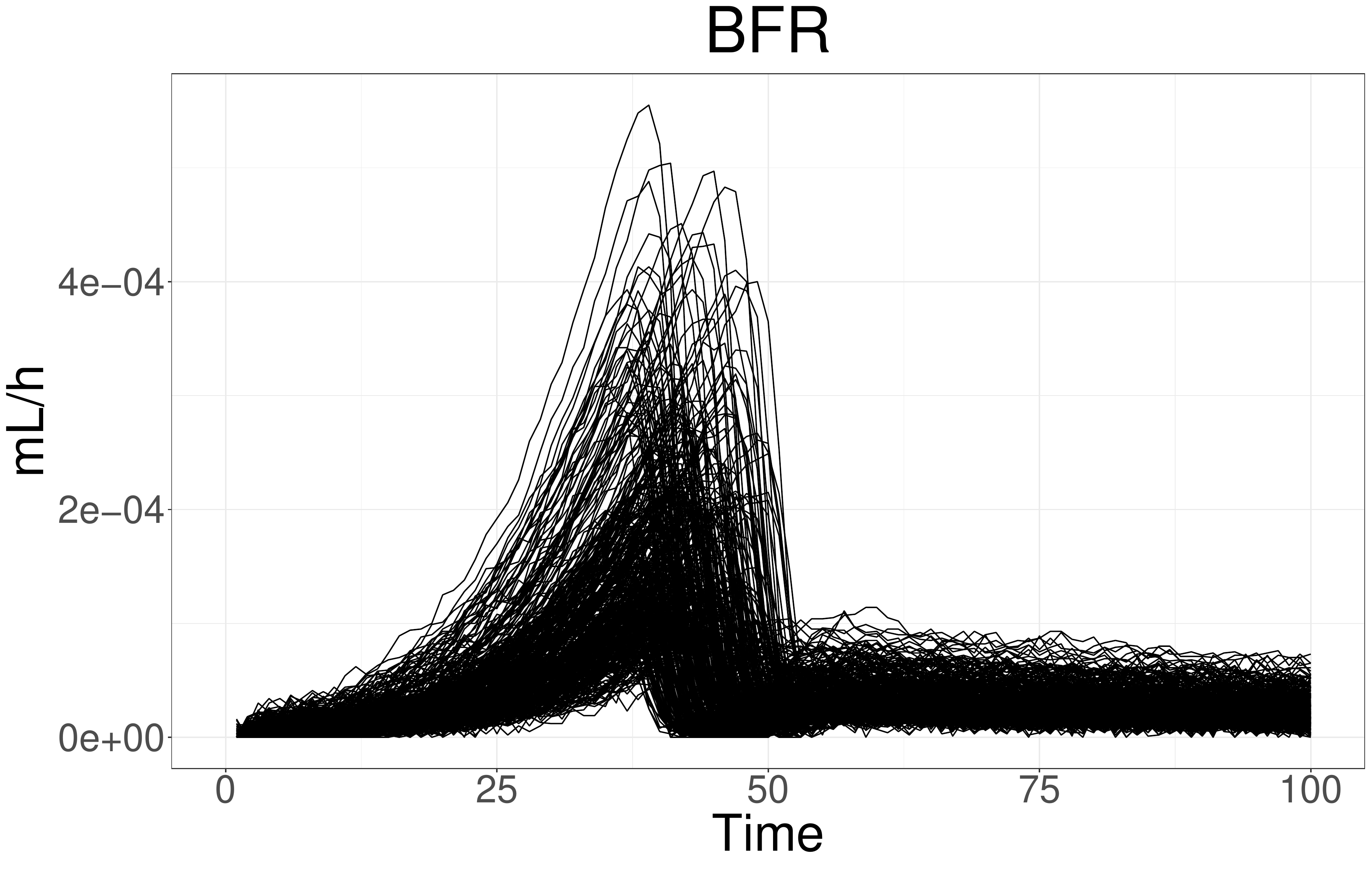}\\
		D4&\includegraphics[width=.49\textwidth]{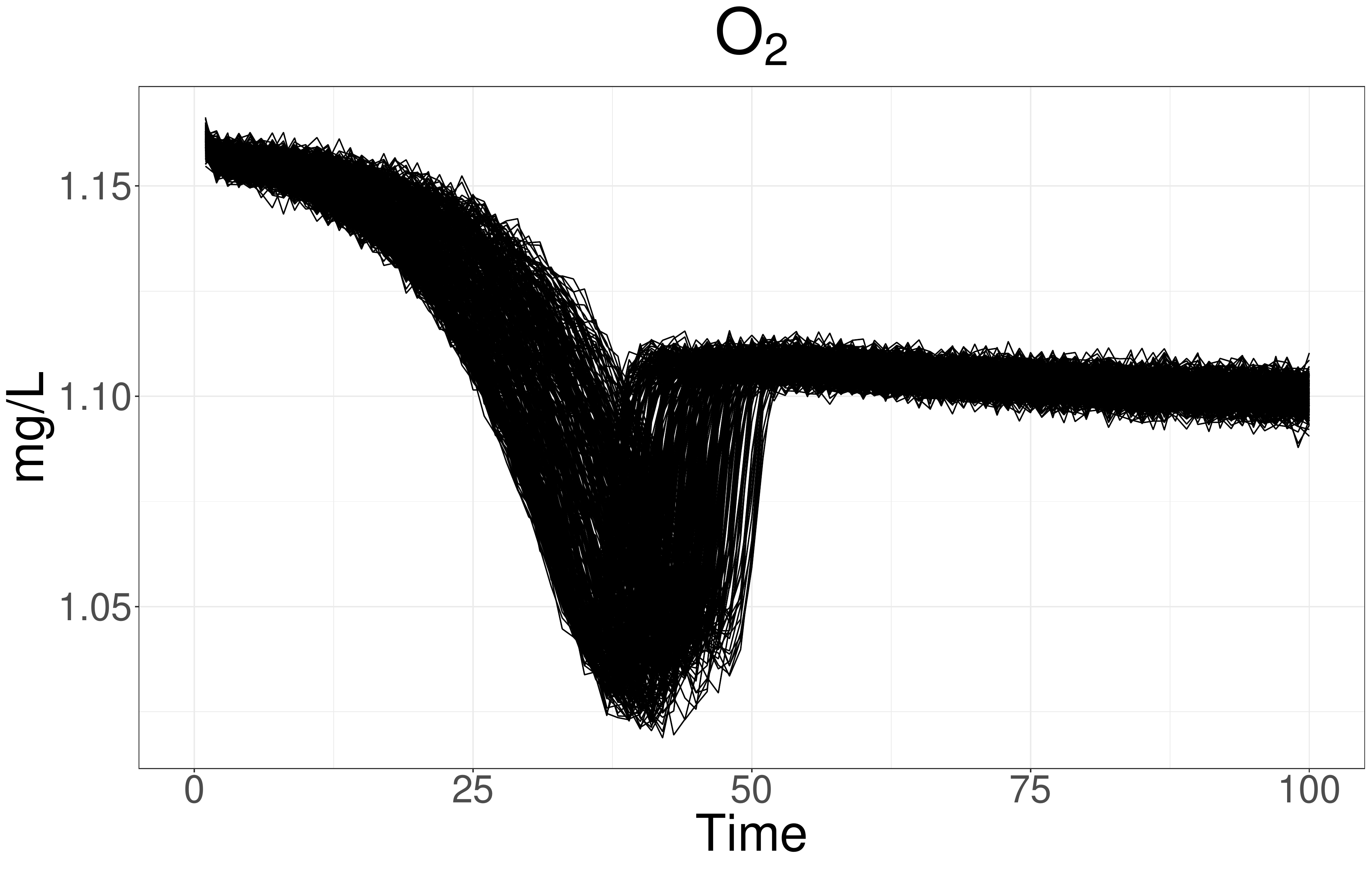}&
	\includegraphics[width=.49\textwidth]{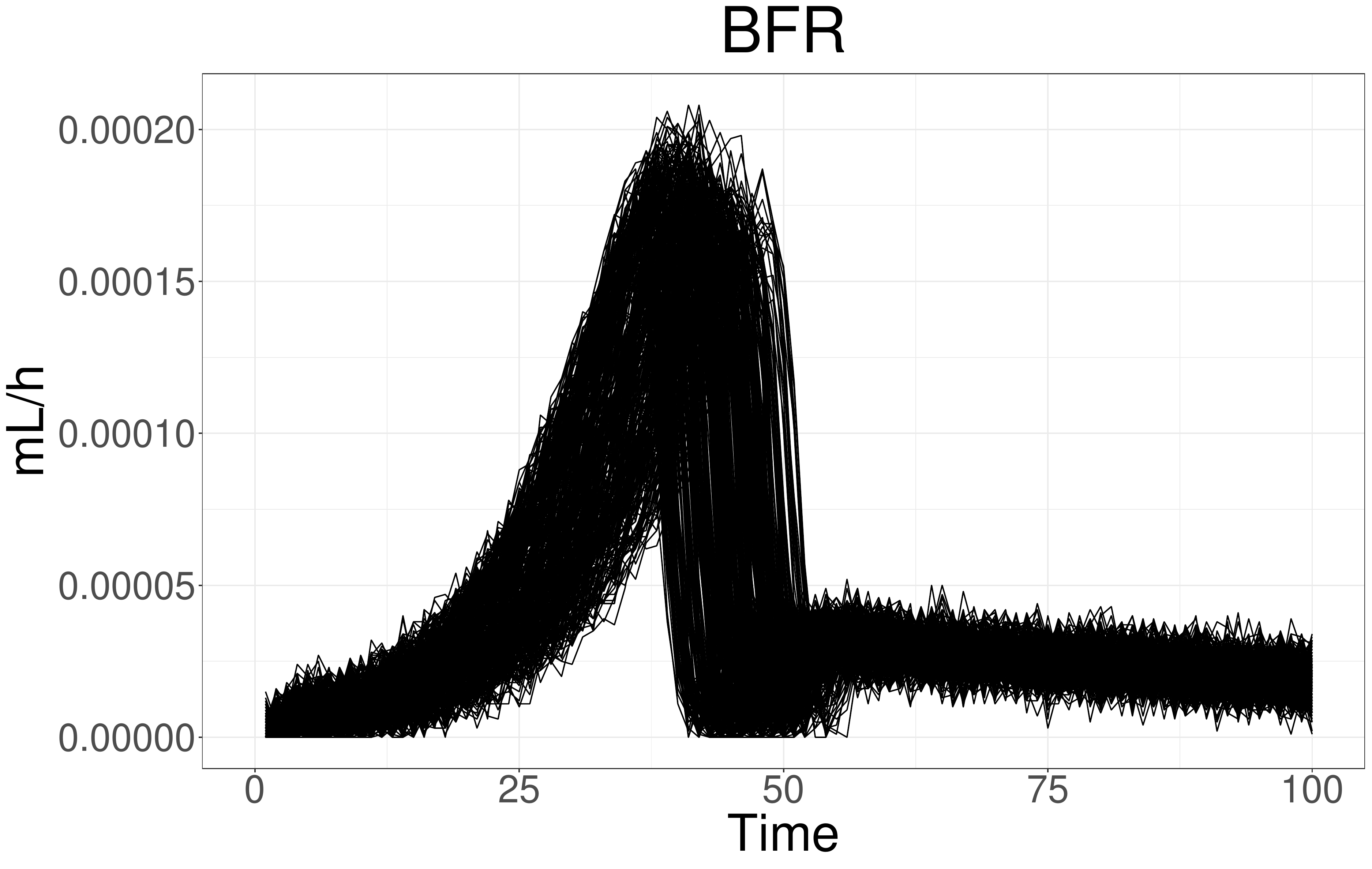}\\
\end{tabular}
	
	\vspace{-.5cm}
\end{figure}

 Figure \ref{fi_csvar} shows the 400 O\textsubscript{2}  and  BFR IC sample in the data example in Section 4, for all the  considered datasets.
 From these plots, a significant proportion of phase variation can be explained and justifies the use of  FRTM. This is particularly true for  the variable BFR in  dataset D3.

\bibliographystyle{chicago}
{\small
\bibliography{References}}
%\begin{description}
%%
%%\item[Title:] Brief description. (file type)
%%
%%\item[R-package for  MYNEW routine:] R-package  containing code to perform the diagnostic methods described in the article. The package also contains all datasets used as examples in the article. (GNU zipped tar file)
%%
%%\item[HIV data set:] Data set used in the illustration of MYNEW method in Section~ 3.2. (.txt file)
%
%\end{description}